\documentclass[sigconf]{acmart}

\AtBeginDocument{%
  \providecommand\BibTeX{{%
    Bib\TeX}}}

\usepackage{array}
\usepackage{arydshln}
\usepackage{amsmath,amssymb,amsfonts,amsthm}
\usepackage{autobreak}
\usepackage{algorithm}
\usepackage{algpseudocode}
\algnewcommand\algorithmicforeach{\textbf{for each}}
\algdef{S}[FOR]{ForEach}[1]{\algorithmicforeach\ #1\ \algorithmicdo}
\usepackage{graphicx}
\usepackage{textcomp}
\usepackage{listings}
\usepackage{xcolor}
\usepackage{multirow}
\usepackage{enumitem}

\theoremstyle{definition}

\usepackage{xcolor}
\usepackage{mdframed}
\usepackage{algorithm}
\usepackage{algpseudocode}

\usepackage{subfigure}
\usepackage{hyperref}
\usepackage{booktabs}

\usepackage{xspace}
\usepackage{soul}
\makeatletter
\usepackage{etoolbox}
\patchcmd{\@makecaption}
  {\scshape}
  {}
  {}
  {}
\makeatother

\usepackage{makecell}

\def\BibTeX{{\rm B\kern-.05em{\sc i\kern-.025em b}\kern-.08em
    T\kern-.1667em\lower.7ex\hbox{E}\kern-.125emX}}
    
\usepackage{url}

\newcommand{\sys}{{\textsf{TRAP}}\xspace}

\newcommand{\delete}[1]{}

\usepackage{tikz}
\usepackage{etoolbox}
\newcommand{\circled}[2][]{%
  \tikz[baseline=(char.base)]{%
    \node[shape = circle, draw, inner sep = 0.3pt]
    (char) {\phantom{\ifblank{#1}{#2}{#1}}};%
    \node at (char.center) {\makebox[0pt][c]{#2}};}}
\robustify{\circled}

\usepackage{pifont}
\usepackage{tabularray}
\usepackage{framed}
\usepackage{xcolor}
\def\BibTeX{{\rm B\kern-.05em{\sc i\kern-.025em b}\kern-.08em
    T\kern-.1667em\lower.7ex\hbox{E}\kern-.125emX}}
\usepackage[normalem]{ulem}

\usepackage{tcolorbox}
\tcbuselibrary{breakable}

\usepackage[title]{appendix}

\makeatletter
\usepackage{etoolbox}
\patchcmd{\@makecaption}
  {\scshape}
  {}
  {}
  {}
\makeatother

\def\BibTeX{{\rm B\kern-.05em{\sc i\kern-.025em b}\kern-.08em
    T\kern-.1667em\lower.7ex\hbox{E}\kern-.125emX}}

\makeatletter 
\newcommand\figcaption{\def\@captype{figure}\caption} 
\newcommand\tabcaption{\def\@captype{table}\caption} 
\makeatother

\renewcommand{\texttt}[1]{%
  \begingroup
  \ttfamily
  \begingroup\lccode`~=`/\lowercase{\endgroup\def~}{/\discretionary{}{}{}}%
  \begingroup\lccode`~=`[\lowercase{\endgroup\def~}{[\discretionary{}{}{}}%
  \begingroup\lccode`~=`.\lowercase{\endgroup\def~}{.\discretionary{}{}{}}%
  \catcode`/=\active\catcode`[=\active\catcode`.=\active
  \scantokens{#1\noexpand}%
  \endgroup
}

\makeatletter
\AtBeginDocument{%
  \pagestyle{empty}% 全局空白
  \thispagestyle{empty}% 首页空白
  \fancyhf{}%
  \renewcommand{\shortauthors}{}%
}
\makeatother

\renewcommand\arraystretch{1.4}

\renewcommand\footnotetextcopyrightpermission[1]{} 
\begin{document}
\begin{sloppypar}

\title{Time Tells All: Deanonymization of Blockchain RPC Users with Zero Transaction Fee (Extended Version)}

%%
% The "author" command and its associated commands are used to define
% the authors and their affiliations.
% Of note is the shared affiliation of the first two authors, and the
% "authornote" and "authornotemark" commands
% used to denote shared contribution to the research.

\author{Shan Wang}
% \author{Ming Yang}
% \authornote{Corresponding Author}
% \email{yangming2002@seu.edu.cn}
\affiliation{%
  \institution{The Hong Kong Polytechnic University}
 \city{Hong Kong}
 \country{China}
}
\affiliation{%
  \institution{Southeast University}
  \city{Nanjing}
  \state{Jiangsu}
  \country{China}
}
\email{shan-cs.wang@polyu.edu.hk}

% \author{Ming Yang}
% \email{shan-cs.wang@polyu.edu.hk}
\author{Ming Yang}
\authornote{Corresponding Author}
% \author{Yu Liu}
\affiliation{%
  \institution{Southeast University}
  \city{Nanjing}
  \state{Jiangsu}
  \country{China}
}
\email{yangming2002@seu.edu.cn}
% \email{liuyu_@seu.edu.cn}

% \author{Ming Yang}
% \authornote{Corresponding Author}
\author{Yu Liu}
\affiliation{%
  \institution{Southeast University}
  \city{Nanjing}
  \state{Jiangsu}
  \country{China}
}
\email{liuyu_@seu.edu.cn}

\author{Yue Zhang}
% \author{Ming Yang}
% \authornote{Corresponding Author}
% \email{yangming2002@seu.edu.cn}
\affiliation{%
  \institution{Shandong University}
  \city{Qingdao}
  \state{Shandong}
  \country{China}
}
\email{zyueinfosec@sdu.edu.cn}

\author{Shuaiqing Zhang}
% \authornote{Corresponding Author}
% \author{Zhen Ling}
\affiliation{%
  \institution{Southeast University}
  \city{Nanjing}
  \state{Jiangsu}
  \country{China}
}
\email{xjzsq@seu.edu.cn}

% \author{Shuaiqing Zhang}
% \authornote{Corresponding Author}
\author{Zhen Ling}
\affiliation{%
  \institution{Southeast University}
  \city{Nanjing}
  \state{Jiangsu}
  \country{China}
}
\email{zhenling@seu.edu.cn}
% \email{zhenling@seu.edu.cn}

% xjzsq@seu.edu.cn

\author{Jiannong Cao}
% \authornote{Corresponding Author}
% \author{Zhen Ling}
% \email{xjzsq@seu.edu.cn}
\affiliation{%
  \institution{The Hong Kong Polytechnic University}
  \city{Hong Kong}
  \country{China}
}
\email{jiannong.cao@polyu.edu.hk}

\author{Xinwen Fu}
% \authornote{Corresponding Author}
% \author{Zhen Ling}
% \email{xjzsq@seu.edu.cn}
\affiliation{%
  \institution{University of Massachusetts Lowell}
  \city{Lowell}
  \state{MA}
  \country{USA}
}
\email{xinwen_fu@uml.edu}

% \affiliation{%


\begin{abstract}
Remote Procedure Call (RPC) services have become a primary gateway for users to access public blockchains. While they offer significant convenience, RPC services also introduce critical privacy challenges that remain insufficiently examined. 
Existing deanonymization attacks either do not apply to blockchain RPC users or incur costs like transaction fees assuming an active network eavesdropper. 
In this paper, we propose a novel deanonymization attack that can link an IP address of a RPC user to this user's blockchain pseudonym. Our analysis reveals a temporal correlation between the timestamps of transaction confirmations recorded on the public ledger and those of TCP packets sent by the victim when querying transaction status. 
We assume a strong passive adversary with access to network infrastructure, capable of monitoring traffic at network border routers or Internet exchange points.
By monitoring network traffic and analyzing public ledgers, the attacker can link the IP address of the TCP packet to the pseudonym of the transaction initiator by exploiting the temporal correlation. This deanonymization attack incurs zero transaction fee.
We mathematically model and analyze the attack method, perform large-scale measurements of blockchain ledgers, and conduct real-world attacks to validate the attack. Our attack achieves a high success rate of over $95\%$ against normal RPC users on various blockchain networks, including Ethereum, Bitcoin and Solana.
\end{abstract}

\vspace{-3mm}
\keywords{Blockchain; RPC service; Wallet; Deanonymization attack}


\maketitle

\vspace{-1mm}
\section{Introduction}

%In the current blockchain ecosystem, 
Remote Procedure Call (RPC) protocols and services have become a primary gateway for users to access blockchain networks. 
Wallet applications such as browser plug-ins rely on RPC services for easy development and seamless interaction with blockchains.
This allows a blockchain user to transact effortlessly through lightweight wallets on their personal devices with no need of running a full blockchain node. 
Ethereum pioneered the use of blockchain RPC services, with Bitcoin and newer blockchains such as Solana following suit. 
For instance, the most popular Ethereum wallet {\em MetaMask} relies exclusively on the RPC service, \textit{Infura}, for blockchain access \cite{infura2}.
% serves 30 million monthly active users and 
\looseness=-1

%While RPC services offer significant convenience, they also introduce critical privacy challenges that remain insufficiently examined. 

%The inherent anonymity properties of blockchain, such as those in Bitcoin, are widely valued for promoting privacy and encouraging adoption. 

Despite various deanonymization attacks against blockchains, 
anonymity of blockchain RPC users remains insufficiently examined.
(i) Some works aim to cluster blockchain addresses belonging to the same user \cite{zhao2015graph, zheng2020identifying, remy2018tracking, meiklejohn2013fistful}, which cannot reveal users' real-world identities.
(ii) Existing deanonymization methods typically link a transaction, along with its initiator's pseudonym, to the IP address of a blockchain node that firstly broadcasts the transaction to the network \cite{biryukov2019deanonymization,biryukov2019transaction,gao2021practical,koshy2014analysis,shen2020transaction,zheng2023ledger}.
This type of attack typically requires the attacker to establish a super node that connects to almost all blockchain nodes.
However, RPC users' personal devices operate off-chain, 
with RPC service nodes acting as intermediaries that broadcast transactions and synchronize the ledger on their behalf.
% and the nodes of RPC services act as intermediaries to broadcast transactions and synchronize ledger on behalf of them.
As a result, this type of work cannot reveal the IP address of RPC users' personal devices.
(iii) Most existing attacks exploiting blockchain RPC services do not focus on compromising user anonymity \cite{li2021strong, li2021deter,hara2020profiling, kim2023etherdiffer}. The only prior study on deanonymization of Ethereum RPC users 
assumes an active network adversary capable of both persistently eavesdropping on and {\em injecting} RPC traffic via routers \cite{wang2024deanonymizing}. 
The injected traffic is used to send transactions through the same router as the victim, enabling estimation of the victim’s transaction propagation timing.
However, this active attack incurs transaction fees as well as makes it more likely to be detected.
Moreover, this work \cite{wang2024deanonymizing} falls short of exposing widespread flaws in user anonymity across various blockchains and wallets.\looseness=-1

% incurs the cost of transaction fees and falls short of exposing widespread flaws in user anonymity across various blockchains and wallets \cite{wang2024deanonymizing}. \red{Moreover, this attack method \cite{wang2024deanonymizing} assumes an active network adversary capable of both eavesdropping on and injecting transaction traffic via routers, making it more likely to be detected.}\looseness=-1

We assume a passive adversary with access to network infrastructure, capable of monitoring traffic at network border routers or Internet exchange points. This threat model is stronger than those typically used in prior blockchain research. For example, front-running attacks \cite{eskandari2019sok}, DoS attacks \cite{li2021strong},  Eclipse attacks \cite{heilman2015eclipse}, flash loan attacks \cite{gan2022understanding}, and deanonymization via super nodes \cite{biryukov2019deanonymization,biryukov2019transaction,gao2021practical,koshy2014analysis,shen2020transaction,zheng2023ledger} generally assume malicious behavior from ordinary nodes or users. In contrast, the adversary in our model possesses capabilities far beyond those of typical blockchain participants.
That said, this paper addresses a less studied problem, with only one prior work \cite{wang2024deanonymizing} directly targeting it. Our threat model is justified by its established use in anonymity research and real-world examples, such as certain countries blocking Tor at border routers \cite{lopes2024flow}.

This paper fills the gap: deanonymizing RPC users 
across various blockchains 
without incurring any transaction fee, under a passive network eavesdropping model.
%, thereby highlighting the severe and widespread risks to blockchain privacy concerning RPC protocols and services.
We identify a temporal correlation vulnerability in the widely deployed "wallet---RPC service---blockchain network" interaction paradigm.
%, which universally exists across platforms. 
While processing a transaction, the RPC service responds to the transaction status query by a wallet {\em shortly} after the transaction is confirmed by the ledger. 
We call the timestamp of such a RPC response as {transaction query timestamp} $\mathcal{T}_q$ and denote the transaction confirmation timestamp as $\mathcal{T}_c$.
An attacker may infer $\mathcal{T}_q$ through monitoring and analyzing encrypted the network traffic between a wallet at a monitored IP address and its associated RPC service.
Therefore, the attacker can estimate $\mathcal{T}_c = \mathcal{T}_q - I_{c,q}$, where $I_{c,q}$ is the interval between these two timestamps
and can be derived via analyzing wallet codes or network traffic.
By searching the ledger for transactions confirmed around the estimated timestamp $\mathcal{T}_c$, the attacker obtains the pseudonyms in the transactions and can then link the monitored IP address to those pseudonyms.
Such a passive way of deanonymization does not incur any transaction fee.

% \red{
% While our deanonymization method works against RPC users without incurring transaction fees, it requires a powerful adversary capability of passive and persistent access to TLS-encrypted TCP traffic.
% Such access is typically limited to privileged entities such as router operators, ISPs, or regulatory authorities, resulting in a relatively narrow threat model.
% Compared to prior work on deanonymizing blockchain nodes by establishing a super node \cite{biryukov2019deanonymization,biryukov2019transaction,gao2021practical,koshy2014analysis,shen2020transaction,zheng2023ledger}, our threat model is stronger, as it assumes network-level eavesdropping capabilities---an ability not all adversaries possess---whereas deploying a super node generally involves no access restrictions.
% Compared to the most relevant work on deanonymizing Ethereum RPC user assuming an {\em active} network attacker \cite{wang2024deanonymizing}, our passive threat model is slightly weaker, as it requires no traffic injection and is inherently more stealthy.
% }

Although the RPC-based communication paradigm exposes temporal correlation vulnerability and opens a new attack surface, 
linking a RPC user's IP address to a {\em unique} pseudonym for deanonymization presents grave challenges.
% deanonymizing a RPC user by linking its IP address to a {\em unique} pseudonym presents grave challenges.
First, encrypted TCP packets in RPC protocol complicate the identification of a specific transacting behavior such as status query and its timestamp $\mathcal{T}_q$.
Second, design diversities across different blockchains, RPC protocols, and wallets pose significant challenges to estimate a reliable $I_{c,q}$. 
Finally, the derived transaction confirmation timestamp $\mathcal{T}_c = \mathcal{T}_q - I_{c,q}$ is a rough estimation, and may correspond to quite many transactions and pseudonyms, thereby concealing the target.
The attacker can link the monitored IP of a wallet user to a set of pseudonyms, 
but may be unable to {\em uniquely} identify the target.
% but may not pinpoint the target uniquely.

To solve these challenges, we propose a novel deanonymization attack method named \underline{\textbf{T}}imestamp \underline{\textbf{R}}eveals \underline{\textbf{A}}ssociated \underline{\textbf{P}}seudonym (\sys), which aims to {\em uniquely} link an IP address of a RPC user to its pseudonym.
(i) By carefully analyzing RPC API calls along with the sizes and sequences of TCP packets generated by these calls in wallets, 
% we design a machine learning model based on features of TCP packet size and sequences for accurate detection of $\mathcal{T}_q$.
we design a machine learning model that leverages TCP packet size and sequence features to accurately detect $\mathcal{T}_q$.
(ii) Thanks to wallet design philosophy for good user experience, a wallet typically queries transaction status automatically and promptly without user intervention to provide timely feedback to users.
This allows the attacker to estimate an upper bound of the interval $I_{c,q}$ (which is short and spans only a few blocks) and derive the range of transaction confirmation timestamp $\mathcal{T}_c$.
(iii) By searching ledgers based on the estimated $\mathcal{T}_c$, the attacker can obtain a set of candidate pseudonyms. To uniquely identify the target, the attacker conducts multiple rounds of traffic and ledger analysis against the same IP address and each round generates a candidate set. Intersecting these sets may reveal the target. \looseness=-1

We mathematically model the attack, conduct large-scale measurements on blockchain ledgers and perform real-world attacks to analyze and validate the \textsf{TRAP} attack across various blockchains including Ethereum, Bitcoin and Solana.
We classify users who transact infrequently based on a threshold as normal users, who represent the overwhelming majority of all users (e.g., 99.97\% in Ethereum).
We find if a normal RPC user performs $3$ or $4$ transactions with their wallet at a monitored IP address, the success rate of {\em uniquely} identifying the user's pseudonym can reach up to $96.80\%$ in Ethereum testnet, 
$95.33\%$ in Ethereum mainnet, 
$97.70\%$ in Bitcoin testnet and $96.58\%$ in Solana testnet.
Furthermore, our results demonstrate that \textsf{TRAP} is universally effective across diverse open-source or closed-source wallets.
The success rate remains consistently high for attackers located in different cities or countries,
regardless of their geographic distance from the victim.

\textbf{Attack impact.} 
Our findings demonstrate that, an abusive entity---such as a big brother---can deanonymize public blockchain users simply by monitoring network traffic, 
without any cooperation from blockchains, wallets, or RPC services, with zero transaction fee and in a stealthy manner. 
\textsf{TRAP}
%could put users’ real identities beyond the control of blockchain systems, 
potentially exposes cryptocurrency users to unexpected censorship.
Moreover, tens of millions of users use RPC services to send trillions of transactions \cite{infura2}. Targeting RPC users has a widespread impact. 
% This poses severe threats to users in regions with strong censorship since the attacker can deanonymize blockchain users without any cooperation of RPC services, which may be located in other regions.
\looseness=-1

% \vspace{1mm}
\textbf{Responsible disclosure}.
We have reported our findings in this paper to 12 affected vendors, including 3 blockchains and their 9 wallets. 
10 of them have acknowledged the vulnerability, including 7 who have confirmed the attack and 3 who are investigating the issue.
2 vendors have not yet responded. 
Notably, the Bitcoin RPC protocol designer {\em Electrum} (also a wallet vendor) assigned CVE-2025-43968 to this issue, and awarded us a bug bounty of \$5,000.
The Ethereum Foundation awarded us an academic grant on the attack and protocol-level countermeasures (No. FY25-2131).
% (Grant No. FY25-2131; Total Funding: \$47,099). 
Moreover, 3 vendors are collaborating with us on the countermeasures, and/or driving changes at other relevant vendors to address the issues more broadly. 
More details can be found in \autoref{appendix:vendorResp}.

% in the Appendix.
% and is collaborating with us on countermeasures, 
% 4 vendors have acknowledged our findings,
% 2 vendors are currently investigating our reports, 
% and 5 vendors have not yet responded.

% We have reported our findings to 12 affected vendors, including 3 blockchains and their 9 wallets. Among them, 
% one vendor awarded us \textbf{a bounty of \$5000} and is collaborating with us on countermeasures, 
% 4 vendors have acknowledged our findings,
% 2 vendors are currently investigating our reports, 
% and 5 vendors have not yet responded.

% \vspace{1mm}
\textbf{Ethical considerations}. 
Our research adheres to the principles outlined in the Menlo Report \cite{bailey2012menlo}.
We only consider a {\em passive} attacker who monitors network packets from a victim user. The attacker will not interfere with normal operations of blockchain networks. 
Our statistics are calculated based on {\em public} ledgers.
% We launched the \textsf{TRAP} attack against our own blockchain accounts, without exposing real-world identities of other blockchain users.
In our evaluation experiments, we strictly log only the network traffic associated with our own devices/IPs with informed consent, ensuring that no data is collected from unaware users/IPs.
Additionally, the transactions generated during our experiments are designed to emulate the transacting patterns and rates observed in public blockchain ledgers. This approach ensures that our experiments do not place any noticeable load on the RPC services or blockchain networks under study.
\looseness=-1

\vspace{-2mm}
\section{Background}
% \vspace{-1mm}

% \vspace{-1mm}
\subsection{Blockchain Ledgers and Wallet Applications}
% \vspace{-1mm}

%\noindent
\textbf{Public blockchain ledgers.}
Public blockchain ledgers record transactions in a sequence of append-only blocks. Once a transaction is packaged into a block and confirmed by the blockchain network, it is permanently stored in the ledger, accessible to all at any time. 
Different blockchains usually adopt unique consensus protocols, leading to distinct block time (the period of generating and confirming a new block).
% and performance measured in TPS (the number of transactions a blockchain can process per second). 
Although each public blockchain features distinct attributes, blocks in their ledgers exhibit some common fields, such as a collection of transactions
% ledgers exhibit common fields. Typically, a block includes a block number, a block hash, the hash of the preceding block, 
and a timestamp marking the {\em confirmation time} of these transactions.
Each transaction includes a blockchain address of its initiator user, which is derived from the user’s public key and acts as a pseudonym, enhancing user anonymity. \looseness=-1

% \subsection{Cryptocurrency Wallets}
% \vspace{1mm}
%\noindent
\textbf{Wallet applications.}
% Due to their decentralization and Transparency nature, 
Public blockchain ledgers serve as a foundational decentralization technology for various applications, particularly supporting the development and operation of cryptocurrency wallets. 
% Cryptocurrency wallets are the mainstream way for normal users to transact in blockchain networks. 
These wallets leverage the blockchain to perform critical functions such as validating and recording transactions and tracking balances. 
Currently, a wallet typically relies on blockchain RPC services to enable easy development without deploying a blockchain full node.
% The wallet can work in a lightweight way by simply sending RPC requests to RPC services for sending transactions or querying ledger data.
% , with the operation results returned in the RPC responses.
This has prompted the emergence of multiple types of wallet applications. They can be browser plug-ins like \textsf{MetaMask}, which provides access to Ethereum; desktop applications such as \textsf{Electrum}, which is used for Bitcoin transactions; or websites like \textsf{Torus}, which facilitates interactions with \textsf{Solana}. 
% Consequently, the developer of the wallet only needs to maintain a server with simple functions for wallet routine maintenance such as updates.
\autoref{table::wallet} provides a short review of various blockchain-based wallets. It is evident that these wallets enjoy significant popularity. 
% \red{Please note that, this paper primarily concentrates on analysis of open-source wallets for examining wallet behaviors and identifying root causes; however, our attack can work against both open-source and closed-source targets as attackers only need to observe network packets.}

% \usepackage{multirow}
\begin{table}[h!]
\vspace{-3mm}
\caption{Blockchain-based cryptocurrency wallets}
\vspace{-3mm}
\label{table::wallet}
\setlength\tabcolsep{2.8pt}
\renewcommand{\arraystretch}{1.1}
\scriptsize
\begin{tabular}{lllllcccc}
\toprule[1.25pt]
\textbf{Wallet} & \textbf{Year} & \textbf{Country} & \textbf{Users} & \textbf{\begin{tabular}[c]{@{}l@{}}Block-\\ chain\end{tabular}} & \textbf{\begin{tabular}[c]{@{}c@{}}Multi-\\ Curr.\end{tabular}} & \textbf{\begin{tabular}[c]{@{}c@{}}Standard\\ RPC Prot.\end{tabular}} & \textbf{\begin{tabular}[c]{@{}c@{}}Open\\ Source\end{tabular}} & \multicolumn{1}{c}{\textbf{\begin{tabular}[c]{@{}c@{}}Sup.\\ Testnet\end{tabular}}} \\ \hline
MetaMask        & 2016          & USA              & 30 M+          & \multirow{3}{*}{ETH}                                            & $\times$                                                        & $\checkmark$                                                         & $\checkmark$                                                  & $\checkmark$                                                                         \\
Enkrypt   & 2015          & USA              & 3 M+           &                                                                 & $\checkmark$                                                    & $\checkmark$                                                         & $\checkmark$                                                  & $\checkmark$                                                                         \\
Taho            & 2021          & USA              & 70 K+           &                                                                 & $\checkmark$                                                    & $\checkmark$                                                         & $\checkmark$                                                  & $\checkmark$                                                                         \\ \hdashline[2pt/2pt]
Electrum        & 2011          & Germany          & 1 M+           & \multirow{3}{*}{BTC}                                            & $\times$                                                        & $\checkmark$                                                         & $\checkmark$                                                  & $\checkmark$                                                                         \\
Green           & 2016          & Canada           & 100 K+         &                                                                 & $\times$                                                        & $\checkmark$                                                         & $\checkmark$                                                  & $\checkmark$                                                                         \\
Sparrow         & 2020          & South Africa     & N/A            &                                                                 & $\times$                                                        & $\checkmark$                                                         & $\checkmark$                                                  & $\checkmark$                                                                         \\ \hdashline[2pt/2pt]
Phantom         & 2020          & USA              & 3 M+           & \multirow{3}{*}{SOL}                                            & $\times$                                                        & $\checkmark$                                                         & $\times$                                                      & $\checkmark$                                                                         \\
Solflare        & 2020          & USA              & 2 M+           &                                                                 & $\times$                                                        & $\checkmark$                                                         & $\times$                                                      & $\checkmark$                                                                         \\
Torus           & 2019          & Singapore        & 1 M+           &                                                                 & $\checkmark$                                                    & $\checkmark$                                                         & $\checkmark$                                                  & $\checkmark$                                                                         \\ 
\bottomrule[1.25pt]
\end{tabular}
\vspace{-2mm}
\end{table}

\vspace{-2mm}
\subsection{Blockchain RPC Protocol}
\label{sec::rpcProtocol}
% \vspace{-2mm}

\begin{figure}[h!]
\vspace{-2mm}
\centering
\includegraphics[width=1.0\columnwidth]{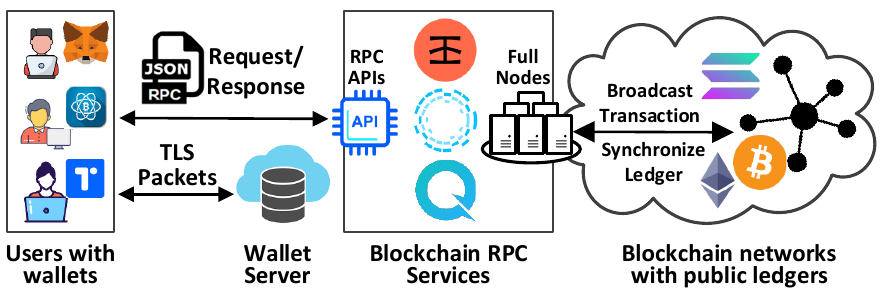}
\vspace{-8mm}
\caption{RPC-based Blockchain communication paradigm}
\label{fig::Ecosystem} 
\vspace{-3mm}
\end{figure}

%\vspace{2mm}
%\noindent
\textbf{RPC-based blockchain communication paradigm.}
% We now turn to a discussion of how the RPC protocol operates in real-world scenarios.
The RPC protocols and services have reshaped the communication paradigm in the blockchain ecosystem, as they provide a lightweight way for users to access blockchain networks. As shown in \autoref{fig::Ecosystem}, users manage their cryptocurrencies using wallets on their personal devices. 
% These wallets rely on RPC services for seamless access to the blockchain networks. A RPC service provider manages multiple full nodes within the blockchain network for ledger synchronization and exposes standard RPC APIs to clients, which play a pivotal role in enabling transactions and retrieving ledger data. 
% Wallets call RPC APIs of various functions by sending RPC requests to the services, and the operation results are returned in RPC responses.
% The communication protocol between the wallet and the RPC service adheres to the JSON-RPC standard, 
% and adopts the TLS protocol to ensure data integrity and confidentiality.
These wallets depend on RPC services to seamlessly access blockchain networks. A RPC service provider operates multiple full nodes for ledger synchronization and exposes standardized RPC APIs that enable clients to submit transactions and retrieve ledger data. Wallets interact with these APIs by sending RPC requests and receiving corresponding responses. Communication between wallets and RPC services follows the JSON-RPC protocol over TLS, ensuring data integrity and confidentiality.
Additionally, wallets may interact with their own servers for routine maintenance such as updates, using TLS-secured communications.

% \vspace{1mm}
% \noindent
\textbf{RPC protocol specification.}
A blockchain RPC service typically employs the JSON-RPC protocol to facilitate the communication between a client and a RPC server. 
The client sends a RPC request to call a specific RPC API associated with a function on a remote RPC server, and the server responds with the results of the function execution.
Each pair of request and response is meticulously structured in JSON, detailing the remote functions, their parameters, and the function execution results. Specifically,
the request JSON object includes four key-value pairs, such as "{\em jsonrpc}" indicating the protocol version, "{\em method}~" indicating the function to be called, "{\em params}" containing parameters and "{\em id}~" uniquely identifying this request. The response JSON object includes "{\em jsonrpc}", "{\em id}~", and "{\em result}" detailing the execution outcomes.
% Adhering to the RPC protocol, clients can efficiently conduct transactions, query ledger data, and interact with smart contracts, streamlining their blockchain operations. \looseness=-1

\begin{figure}[h]
\centering
\includegraphics[width=0.9\columnwidth]{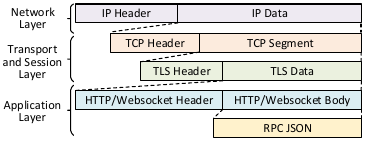}
\vspace{-4mm}
\caption{Network packet format of blockchain RPC protocol}
\label{fig::tcpSegment} 
\vspace{-5mm}
\end{figure}

% \vspace{1mm}
% \noindent
\textbf{Format and layout of RPC network packets.} 
JSON-RPC protocol (or Blockchain RPC Protocol) runs atop a set of low-level network protocols, as shown in \autoref{fig::tcpSegment}. It employs the standard TCP/IP protocol, where each connection between a client and a server is uniquely identified by a quadruple of $\langle${\em source IP, source port, destination IP, destination port}$\rangle$. 
To safeguard the data transmission, the TLS protocol is adopted to encrypt the TCP data. At the application layer, HTTP protocol or Websocket protocol is employed, with a Header and a Body that encapsulates the actual application content.
% of a RPC-JSON object, that includes ledger queries or raw transaction data. 
Central to the HTTP/Websocket body is the RPC JSON object, which holds the actual RPC request or response details.\looseness=-1

%By passively monitoring the network packets, the monitor can obtain the size of the TLS segment but its content is invisible. The TLS segment size is primarily decided by the size of the carried RPC JSON object, as HTTP or Websocket headers typically have relatively fixed fields and size ranges. Fields like IP header, TCP header and TLS header are plaintext.

% A publicly accessible blockchain explorer is maintained, providing users seamless access to the blockchain's public ledger, enhancing transparency in the system.

% \subsection{Blockchain Wallet}

% \section{\textsf{RPCBleed} Attack}
\vspace{-1mm}
\section{\textsf{TRAP} Attack Overview}
\vspace{-1mm}

This section covers the motivation (\S\ref{sec::motivation}), threat model (\S\ref{subsec:threatmodel}), the basic idea (\S\ref{subsec:keyidea}), challenges and methodology of the attack (\S\ref{sec::challenge}).

\vspace{-2mm}
\subsection{Motivation}
\label{sec::motivation}
\vspace{-1mm}

Existing work largely overlooks the critical risks to user anonymity arising from the new communication paradigm reshaped by RPC protocols and services. 
Existing deanonymization methods typically rely on a super node to link a transaction, along with its initiator's pseudonym, to the IP address of a blockchain node that firstly broadcasts the transaction to the network \cite{biryukov2019deanonymization,biryukov2019transaction,gao2021practical,koshy2014analysis,shen2020transaction,zheng2023ledger}.

However, in the context of blockchain RPC services as illustrated in \autoref{fig::Ecosystem}, 
users' personal devices operate off-chain and behind the RPC service. 
% It is the RPC service nodes---not the users' personal devices---that firstly broadcast users' transactions and thus are the targets of these attacks.
It is the RPC service nodes, rather than the users' personal devices, that first broadcast users' transactions and thus are the targets of these attacks.
% As a result, this type of work cannot reveal RPC users' real-world identities.
Consequently, such approaches cannot uncover the real-world identities of RPC users.
Most existing attacks exploiting blockchain RPC services do not focus on compromising user anonymity \cite{li2021strong, li2021deter,hara2020profiling, kim2023etherdiffer}. 
The only prior study on deanonymization of Ethereum RPC users relies on an active network attacker capable of eavesdropping on and injecting traffic via network devices to observe the RPC users' IPs and link them to pseudonyms.
However,
it incurs the cost of transaction fees and falls short of exposing widespread flaws in user anonymity across diverse blockchain networks \cite{wang2024deanonymizing}.

Our work fills the gap: {\em deanonymizing RPC users across various blockchains without incurring any transaction fees under a passive network eavesdropping model, thereby highlighting the severe and widespread risks to blockchain privacy concerning RPC protocols and services.}
% Today, tens of millions of blockchain users, particularly those who are technologically inexperienced, rely on RPC services to
% interact with blockchain networks. 
% % Tens of millions of users depend on RPC services to 
% send trillions of transactions \cite{infura2}.
% Since a large proportion of transactions are facilitated through RPC services \cite{infura2}, targeting RPC users can have a widespread impact. 
Specifically, we propose the \textsf{TRAP} attack which aims to link a RPC user's pseudonym, i.e., its blockchain address, to the IP address of the user's personal device (which may reveal the victim user's real-world identity),
across mainstream blockchains including Ethereum, Bitcoin and Solana, in a passive way and without transaction fees.
% This kind of zero-cost and passive method makes our attack particularly noteworthy,
This kind of passive and zero-cost attack represents an advanced threat model, as the attacker can compromise blockchain users stealthily and without encountering economic constraints.
% which is often a critical barrier to the feasibility and scalability of blockchain-oriented attacks \cite{yaish2024speculative,zhao2024dethna, wang2024deanonymizing}.
\looseness=-1

% since the attacker can deanonymize large number of users without encountering financial constraints.
% Notably, economic cost is often a critical barrier to the feasibility and scalability of blockchain-oriented attacks \cite{yaish2024speculative,zhao2024dethna}, and thus our attack is particularly noteworthy.

\vspace{-2mm}
\subsection{Threat Model}
\label{subsec:threatmodel}
\vspace{-1mm} 

% \red{Our work fills the gap of existing works: The attacker's objective is to link a RPC user's pseudonym, i.e., its blockchain address, to the IP address of the user's personal device (which may reveal the victim user's real-world identity) with zero transaction free.}

% The attacker's objective is to link a RPC user's pseudonym, i.e., its blockchain address, to the IP address of the user's personal device (which may reveal the victim user's real-world identity) with zero transaction free.

\begin{figure}[h!]
% \vspace{-2mm}
\centering
\includegraphics[width=1.0\columnwidth]{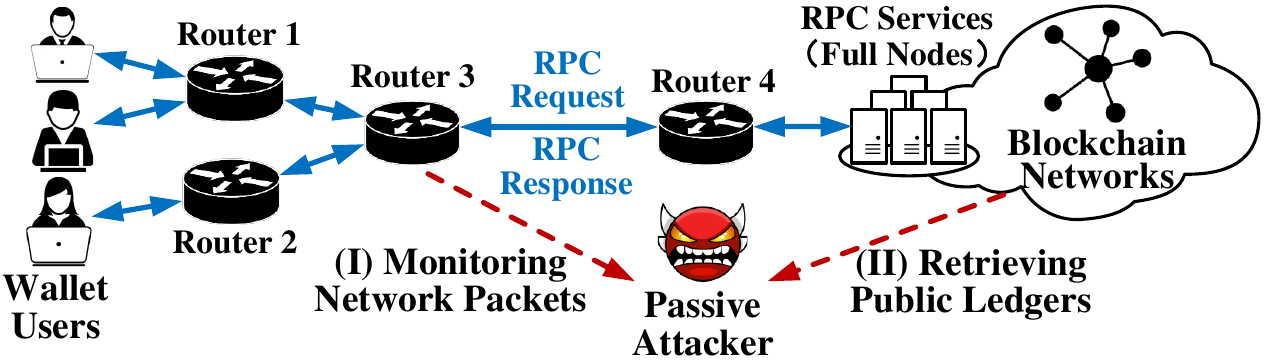}
\vspace{-6mm}
\caption{Threat model}
\label{fig::systemThreatModel} 
\vspace{-5mm}
\end{figure}

In our threat model, illustrated in \autoref{fig::systemThreatModel}, the attacker passively observes network traffic between the victim users' devices and RPC services, via a network device such as a router along the traffic routing path across cities and countries.
An attacker may locate in different region from victims and RPC services.
% The attacker does not manipulate the traffic in any way, nor does it modify, delay, or discard packets.
The attacker collects only the metadata of TCP packets such as the quadruple of $\langle${\em source IP, source port, destination IP, destination port}$\rangle$, timestamps and sizes, which are always unencrypted.
The actual packet contents are encrypted by TLS and remain inaccessible to the attacker.
The attacker can monitor multiple users simultaneously, as the quadruple can uniquely identify a network connection between a wallet application and a RPC service, allowing for the categorization of packets associated with different wallet users and their RPC services.
Potential actors in this scenario include network router owners, internet service providers (ISPs), or regulatory government bodies (big brothers), with network monitoring capabilities via port mirroring or optical splitters \cite{esmail2012physical}. 
% The main difference among different types of attackers is the scale of routers and users they can monitor. Attackers such as ISPs or government entities potentially can compromise a larger number of users without any cooperation of RPC services. 
% \red{This poses severe threats to users in regions with strong censorship since the attack can deanonymize blockchain users without any cooperation of RPC services, which may be located in other regions.}
% \red{
% The attack primarily applies when the attacker is on the natural routing path between wallet users and RPC services, and we do not assume the attacker has routing control capabilities. 
% }
This is a common threat model in network attacks
%as weak security assumptions 
\cite{ling2019novel,shensubverting,tran2020stealthier,lopes2024flow}.
Furthermore, the attacker can access public ledger data from blockchain networks, a capability available to anyone.\looseness=-1
% We also assume that users expect trust their chosen RPC services.
%\looseness=-1
% Users expect RPC providers to not disclose their identities since public blockchain/cryptocurrency users expect anonymity.

In our context, the adversary does not need to identify or control the full routing path between users and RPC services. If the adversary (e.g., an ISP) has access to the network topology and infrastructure, it can monitor traffic at key chokepoints such as border routers or exchange points, where traffic enters or exits the network of interest. An enterprise can monitor its users via internal switches or border routers. Governments can surveil citizens by monitoring national border routers. The computational cost of monitoring traffic at a router depends on inspection depth and traffic volume. In our case, the adversary only needs to capture traffic for specific IPs without deep inspection, enabling low-cost collection with offline analysis.

Our attack is effective against both open-source and closed-source wallets, as the attacker infers wallet behaviors from network traffic.
% only needs to monitor network traffic and access public ledgers. 
To rigorously examine wallet behaviors and identify root causes, this paper will concentrate on analysis of open-source wallets for establishing a theoretical foundation.

% for examining wallet behaviors and identifying root causes; however, our attack can work against both open-source and closed-source targets as attackers only need to observe network packets.

% \vspace{-2mm}
% \subsection{Root Cause and Key Idea}
\vspace{-2mm}
\subsection{Basic Idea}
\label{subsec:keyidea}
% \vspace{-2mm}

\begin{figure}[h!]
\vspace{-3mm}
\centering
\includegraphics[width=1.0\columnwidth]{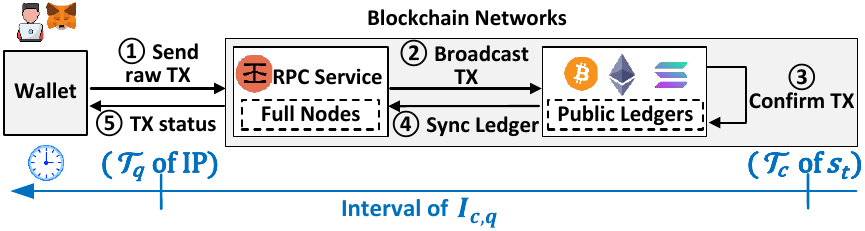}
\vspace{-8mm}
\caption{Timing of processing a transaction}
\label{fig::temporalCorrelation} 
\vspace{-4mm}
\end{figure}

In the "wallet---RPC service---blockchain network" interaction paradigm, processing a single transaction involves coordinated actions and multiple rounds of interactions among these three entities.
%which occur sequentially over the transaction's lifecycle and naturally expose temporal correlations.
As illustrated in \autoref{fig::temporalCorrelation}, 
when a user issues a transaction (TX) through their wallet, the wallet first sends it to the RPC service,
which then broadcasts the transaction to the blockchain network. 
Later, at time $\mathcal{T}_c$ (called transaction {\em confirmation} timestamp), the transaction is confirmed by the ledger and replicated across blockchain nodes including those of the RPC service, 
following a consensus protocol. 
% Assume the transaction is confirmed at time $\mathcal{T}_c$.
Subsequently, the wallet interacts with the RPC service to query the transaction status and receives the response at time $\mathcal{T}_q$ (called transaction {\em query} timestamp), allowing it to verify the transaction's validity and provide feedback to the user.
%We name $\mathcal{T}_c$ and $\mathcal{T}_q$ as transaction {\em confirmation} timestamp and transaction {\em query} timestamp, respectively.
The interval $I_{c,q} = \mathcal{T}_q - \mathcal{T}_c$.

If an attacker can monitor encrypted network traffic between a wallet at a monitored IP address and its RPC service, and is able to infer the fact of transacting and derive $\mathcal{T}_q$, the anonymity of the wallet user will be in grave danger. 
A transaction in public ledgers reveals its confirmation timestamp and its initiator's pseudonym (denoted as $s_t$).
When the attacker obtains $\mathcal{T}_q$ through traffic analysis, they may derive the transaction confirmation timestamp $\mathcal{T}_c = \mathcal{T}_q - I_{c,q}$ by estimating $I_{c,q}$. It then searches the ledger for transactions confirmed around $\mathcal{T}_c$ and obtains the pseudonyms in the transactions.   
Therefore, the attacker can link the monitored IP address with those pseudonyms.  
Such a passive way of deanonymization does not incur any transaction fee.\looseness=-1

% Moreover, this timestamp can be inferred from network packets, allowing the attacker to target diverse wallets.

\vspace{-2mm}
\subsection{Challenges and Methodology}
\label{sec::challenge}
\vspace{-1mm}

% Despite the emergence of the new attack surface in
% Despite the temporal correlation vulnerability in
% the RPC-reshaped communication paradigm in blockchains, deanonymizing the RPC users still remains challenging, as follows.
The basic idea of deanonymization above faces grand challenges.

%\vspace{1mm}
%\noindent
% \textbf{(C-I) Traffic Analysis for Identifying Transaction Query Timestamp $\mathcal{T}_q$.}
\textbf{(C-I) Traffic analysis for the fact of transacting and $\mathcal{T}_q$}.
The TCP traffic between the wallet and the RPC service is encrypted using TLS, preventing the attacker from inferring what is going on between them. 
Only metadata, such as IP addresses, packet sizes, and timestamps, is observable.
% The attacker can only observe metadata such as packet IP addresses, packet size and timestamp and cannot infer what is going on between them from encrypted messages.
Furthermore, a blockchain RPC protocol typically offers tens of APIs of a wide range of functionalities, which facilitate diverse interactions between a wallet and its RPC service, highly concealing the RPC response packet for a transaction status query by the wallet.

% \vspace{-1mm}
% \begin{tcolorbox}[
%   breakable,
%   left=1pt,right=1pt,top=0pt,bottom=0pt,
%   colframe=gray!50,       
%   colback=gray!0,        
%   coltitle=black,         
%   fonttitle=\bfseries,    
%   title=(S-I) Identifying $\mathcal{T}_q$ via TCP packet size and sequence (\S\ref{sec::TxCfmTime}),
%   % sharp corners,          
%   boxrule=0.5mm          
% ]

\textbf{(S-I) Identifying $\mathcal{T}_q$ via TCP packet size and sequence (\S\ref{sec::TxCfmTime}).}
Through systematic analysis of the RPC protocols and wallet business logic, we find that the TCP packet size and sequence features can effectively facilitate the identification of a specific TCP packet that carries $\mathcal{T}_q$ of the target RPC response. 
First, each RPC API has unique parameters and return values, producing RPC-JSON objects of distinct sizes that cause variations in TCP packet sizes.
Second, wallets often use a limited set of APIs, narrowing down the candidates.
Third, the fixed wallet business logic results in relatively consistent and distinct packet sequences surrounding the target.
Together, these features enable accurate detection of the RPC response of interest, which is generated by the transaction status query API call, along with its timestamp.
%\looseness=-1
%\end{tcolorbox}

% which is reflected in the packet sequence features.
% First, each RPC API has unique parameters and return values, 
% % such as a transaction hash, 
% which have theoretical minimum and maximum sizes. Calling different RPC APIs will result in 
% % variations in RPC-JSON object sizes and further lead to 
% distinct sizes of TCP packets.
% % according to the RPC protocol layout in \autoref{fig::tcpSegment}. 
% Second, wallets typically omit the use of certain RPC APIs, such as those for mining and validating blocks, which naturally narrows down the candidates. 
% Third, each wallet operates on fixed business logic, resulting in distinct packet context surrounding the target, which is reflected in the packet sequence features.
% % relatively consistent RPC API calls sequence around the target API, which produce distinct TCP packet sequences. 
% These features allow attackers to detect the TCP packet of interest and its timestamp with high accuracy.

%\vspace{-1mm}
\textbf{(C-II) Estimating $I_{c,q}$ to derive $\mathcal{T}_c$.}
As shown in Figure \ref{fig::temporalCorrelation}, during the interval $I_{c,q}$, significant activities occur among the wallet, RPC service and blockchain system, which may affect $I_{c,q}$.
Each blockchain employs its own consensus protocol, which dictates the time required to confirm a transaction in a block. 
RPC protocols expose different sets of functionalities across blockchains.
Wallets differ in how they manage transactions, such as querying for transaction status.
These variations complicate the estimation of a reliable $I_{c,q}$ with a universal and effective approach.\looseness=-1

%, per the functions provided by RPC protocols.
% RPC protocols expose different sets of functionalities across blockchains, and wallet implementations vary in how they interact with these protocols.
% wallets differ in how they manage transactions, particularly in the timing and strategy for querying transaction status.

% \vspace{-1mm}
% \begin{tcolorbox}[
%   breakable,
%   left=1pt,right=1pt,top=0pt,bottom=0pt,
%   colframe=gray!50,       
%   colback=gray!0,        
%   coltitle=black,         
%   fonttitle=\bfseries,    
%   title=(S-II) Exploiting wallet design to estimate $I_{c,q}$ (\S\ref{sec::walletBehaviorVul}),
%   % sharp corners,
%   boxrule=0.5mm          
% ]
\textbf{(S-II) Exploiting wallet design to estimate $I_{c,q}$ (\S\ref{sec::walletBehaviorVul}).}
To provide users with feedback on transaction status, a wallet must rely on RPC services to query this information as it does not synchronize the ledger like blockchain full nodes.
For a good user experience, wallets perform these queries automatically and promptly, via either periodic queries or notification subscriptions.
Such a wallet design philosophy of efficiency allows a reliable estimation of the upper bound of $I_{c,q}$, which spans only a few blocks (denoted as $k$ blocks).
This indicates that the timestamp $\mathcal{T}_c$ precedes $\mathcal{T}_q$ by no more than $k$ consecutive blocks. 
% \end{tcolorbox}

%\vspace{1mm}
%\noindent
% \textbf{(C-III) Ledger Analysis for Target Pseudonym Identification.}
%\vspace{-1mm}
\textbf{(C-III) Ledger analysis for the identification of the pseudonym making the observed transaction}.
$\mathcal{T}_c$ is a rough estimate, which is within $k$ consecutive blocks away from $\mathcal{T}_q$.
Hundreds of thousands of users conduct millions of transactions daily,
% \footnote{\url{https://etherscan.io/chart/active-address}}, 
potentially appearing in the $k$ blocks alongside the target in a random way. 
By retrieving the ledger, 
the attacker can link the monitored IP address with a set of transactions and their pseudonyms, including the target one, within the $k$ blocks.
However, uniquely and accurately pinpointing the target pseudonym responsible for the observed transaction remains challenging.\looseness=-1
% it is challenging to uniquely and accurately pinpoint the target pseudonym who makes the observed transaction.\looseness=-1
% \red{
% However, uniquely and accurately pinpointing the target pseudonym responsible for the observed transaction remains challenging.
% }\looseness=-1
% it is challenging to uniquely and accurately pinpoint the target pseudonym who makes the observed transaction.

% \vspace{-1mm}
% \begin{tcolorbox}[
%   breakable,
%   left=1pt,right=1pt,top=0pt,bottom=0pt,
%   colframe=gray!50,       
%   colback=gray!0,        
%   coltitle=black,         
%   fonttitle=\bfseries,    
%   title=(S-III) Uniquely identify the target pseudonym via rounds of intersection (\S\ref{sec::intersect}),
%   % sharp corners,          
%   boxrule=0.5mm          
% ]
\textbf{(S-III) Uniquely identify the target pseudonym via rounds of intersection (\S\ref{sec::intersect}).}
% To address this challenge, 
We propose uniquely identifying the target pseudonym via multiple (denoted as $m$) rounds of traffic and ledger analysis.
In each round of the attack, where a transaction is detected by the attacker, it obtains a candidate set $\mathcal{S}_i$ which includes pseudonyms of all transactions within the selected $k$ blocks.  
The target (called $s_t$) is always in $\mathcal{S}_i$.
The attacker then performs the intersection of the $m$ candidate sets:  $\mathcal{S}_m=\bigcap_{i=1}^{m}\mathcal{S}_i$. The target user will be in the intersection $\mathcal{S}_m$. 
We find most blockchain users, referred to as normal users, transact very infrequently. After a few attack rounds (such as $m=3$), the chance that their pseudonyms show up in $\mathcal{S}_m$ along with the target $s_t$ is extremely small.
Active users do exist and show up with $s_t$. We design an efficient filtering strategy to reduce active users' influence so as to uniquely identify the target normal user. 
If the target is an active user, the cardinality $|\mathcal{S}_m|$ of the intersection is not 1, and the target cannot be uniquely identified while we find $|\mathcal{S}_m|$ is small in this case.\looseness=-1

\vspace{-2mm}
\section{Deriving Transaction Query Timestamp $\mathcal{T}_q$}
\label{sec::TxCfmTime}
% \vspace{-1mm}

This section presents the detailed solution (S-I), the traffic analysis method for deriving the transaction query timestamp $\mathcal{T}_q$ and linking $\mathcal{T}_q$ to the IP address of a wallet user. 

% This section presents the first stage (S-I) of the attack in detail.
% An attacker first identifies TCP packets associated with wallets and blockchain RPC services, and determines which wallet, blockchain and RPC service the target user is using. This can be achieved by analyzing domain names of RPC services and wallet servers appearing in the TLS handshakes (\S\ref{sec::domainName}). 
% The attacker exploits features of TCP packet sizes to obtain some packet candidates that potentially are generated by calling the target RPC API (\S\ref{sec::featureSize}). 
% However, relying solely on features of packet sizes may lead to false positives. 
% Therefore, the attacker further extracts features in TCP packet sequences, caused by RPC API calls sequences in the wallet.
% The sequence features provide additional context, such as preceding packets for sending transactions or querying ledger data. By training a machine learning model, the attacker can achieve detection of the specific TCP packet for querying transaction status, and ultimately getting a timestamp anchor of the transaction confirmation (\S\ref{sec::featureSequence}).

\vspace{-2mm}
\subsection{Identify RPC Related TCP Packets}
\label{sec::domainName}
% \vspace{-2mm}
% \vspace{2mm}
% \noindent\textbf{Step-I: RPC Connection Identification via Handshake.} 

% RPC protocols rely on the TCP-based protocol---TLS---for secure network transport.
% The domain names and the IP addresses of RPC services and wallet servers (shown in \autoref{fig::Ecosystem}) are either well known or can be obtained through traffic analysis.
% A passive attacker can analyze the traffic originating from the IP of interest and extract the quadruple of $\langle${\em source IP, source port, destination IP, destination port}$\rangle$ to 
% identify network traffic of interest
% between a wallet and a RPC service. 

An attacker may observe multiple traffic flows simultaneously but can identify RPC packets by inspecting IPs and domain names associated with RPC services. 
A Blockchain RPC protocol relies on TCP/IP and TLS for secure network transport.
Each TCP connection is uniquely identified by a quadruple $\langle${\em source IP, source port, destination IP, destination port}$\rangle$, indicating who is communicating with whom. 
While different applications on the same device share the same source IP, they build distinct network connections with different service servers at different destination IPs. 
The RPC service's IP maps to its domain name, which is either well known or can be extracted from TLS packets. 
A TLS handshake occurs each time a wallet establishes a TCP connection with a RPC server.
During the TLS handshake, the client's "Client Hello" message includes the domain name in plaintext, such as "\textit{sepolia.infura.io}", which corresponds to the destination IP.
% which corresponds to the destination IP of the TCP connection. 
% This indicates that the victim user is connecting to the RPC service named \textit{Infura} for accessing the Ethereum testnet {\em Sepolia}. 
% {\em Sepolia} is an Ethereum testnet. 
Wallets may also connect to their own servers for routine maintenance, and the attacker can similarly obtain the wallet server's domain names and IPs.
Therefore, by analyzing the IPs and domain names in network packets, a network eavesdropper can identify RPC packets and infer which wallet is interacting with which RPC service and blockchain.

% \usepackage{multirow}
\begin{table*}[htbp]
\caption{TCP packet sizes considering practical API usage in wallets. ({\em MetaMask} relies on the RPC service {\em Infura} and adopts HTTP/2. {\em Electrum} relies on service {\em Blockstream} and adopts Websocket. The details on Solana are presented in \autoref{appe::tcpsize}
}
\vspace{-3mm}
\label{table::overlapAPI}
\centering
% \scriptsize
\setlength\tabcolsep{1.0pt}
\renewcommand{\arraystretch}{1.1}
% \begin{tabular}{c|cccccc|lcc}
% \resizebox{\linewidth}{!}{
\begin{tabular}{ccccccp{4.8cm}cc}
\toprule[1.25pt]
\multirow{2}{*}{\textbf{\begin{tabular}[c]{@{}c@{}}Block-\\ chain\end{tabular}}} & \multirow{2}{*}{\textbf{\begin{tabular}[c]{@{}c@{}}Num. of\\ APIs\end{tabular}}} & \multirow{2}{*}{\textbf{Target RPC API}}                                                       & \multirow{2}{*}{\textbf{Wallet}} & \multicolumn{2}{c}{\textbf{TCP Segment Size}}          & \multirow{2}{*}{\textbf{Overlapped APIs in Theory}} & \multicolumn{2}{c}{\textbf{TCP Segment Size}}                                                                                                           \\ 
\cline{5-6} \cline{8-9}
% \cmidrule(lr){5-5}\cmidrule(lr){6-6}  \cmidrule(lr){8-8}\cmidrule(lr){9-9} 
                                                                                 &                                                                                  &                                                                                                &                                  & \textbf{Request}~         & ~\textbf{Response}           &                                                     & \textbf{Request}                                                           & \textbf{Response}                                                          \\ \hline
\multirow{7}{*}{Ethereum}                                                             & \multirow{7}{*}{44}                                                              & \multirow{7}{*}{\begin{tabular}[c]{@{}c@{}}eth\_getTransaction\\ Receipt\end{tabular}}         & \multirow{7}{*}{MetaMask}        & \multirow{7}{*}{193-194}     & \multirow{7}{*}{$\geq$1061} & eth\_getBlockByHash                                 & 192-193                                                                        & $\geq$1437                                                                 \\ \cdashline{7-9}[2pt/2pt] 
                                                                                 &                                                                                  &                                                                                                &                                  &                          &                             & eth\_getTransactionByBlockHash                     & \multirow{3}{*}{\begin{tabular}[c]{@{}c@{}}Without\\ Overlap\end{tabular}} & \multirow{3}{*}{\begin{tabular}[c]{@{}c@{}}Without\\ Overlap\end{tabular}} \\
                                                                                 &                                                                                  &                                                                                                &                                  &                          &                             & eth\_getTransactionByHash                           &                                                                            &                                                                            \\
                                                                                 &                                                                                  &                                                                                                &                                  &                          &                             & eth\_getUncleByBlockHashAndIndex                    &                                                                            &                                                                            \\ \cdashline{7-9}[2pt/2pt] 
                                                                                 &                                                                                  &                                                                                                &                                  &                          &                             & eth\_getProof, eth\_createAccessList, eth\_feeHistory, 
                                                                                 eth\_getFilterLogs,
                                                                                 eth\_getLogs                                        & Unused                                                    & Unused                                                    \\
                                                                                 \hline
\multirow{4}{*}{Bitcoin}                                                             & \multirow{4}{*}{29}                                                              & \multirow{4}{*}{\begin{tabular}[c]{@{}c@{}}blockchain.transaction\\ .get\_merkle\end{tabular}} & \multirow{4}{*}{Electrum}        & \multirow{4}{*}{172-182} & \multirow{4}{*}{$\geq$167}  & blockchain.scripthash.get\_history                  & 170-174                                                                    & $\geq$148                                                                  \\
                                                                                 &                                                                                  &                                                                                                &                                  &                          &                             & blockchain.scripthash.listunspent                   & 170-174                                                                    & $\geq$59                                                                   \\ \cdashline{7-9}[2pt/2pt] 
                                                                                 &                                                                                  &                                                                                                &                                  &                          &                             & blockchain.transaction.get                          & w/o overlap                                                                & w/o overlap                                                                \\ \cdashline{7-9}[2pt/2pt] 
                                                                                 &                                                                                  &                                                                                                &                                  &                          &                             & blockchain.scripthash.get\_mempool                  & Unused                                                                     & Unused                                                                     \\ \bottomrule[1.25pt]

\end{tabular}
% } 
\vspace{-3mm}
\end{table*}

\vspace{-2mm}
\subsection{Identify Packet Candidates via Packet Size}
\label{sec::featureSize}
% \vspace{-2mm}

We first analyze and extract size features of TCP packets to identify packet candidates potentially generated by RPC API calls for querying transaction status.
A RPC protocol provides tens of APIs supporting diverse functions such as sending transactions, querying ledgers, mining and block validation. 
The RPC protocols of Ethereum, Bitcoin and Solana define 44 APIs, 29 APIs and 54 APIs, respectively.  
Among these RPC APIs, one is often used by a client (such as a wallet) to check whether the initiated transaction has been confirmed by the ledger, 
enabling timely verification of transaction validity and balance updates.
% so as to verify transaction validity and update user balance promptly.
For example in Ethereum, the API \texttt{eth\_getTransactionReceipt} is used with a transaction hash as the parameter to query for its status, and the response provides a receipt containing block information that includes this transaction, gas usage, and so on. 
{\em Nil} will be returned for an unconfirmed transaction.
The Bitcoin RPC API \texttt{{blockchain.transaction.get\_merkle}} and Solana RPC API \texttt{{getSignatureStatuses}} are used by wallets for similar functions.

Each RPC API has distinct semantics, unique parameters, and specific return values. 
% We can analyze the size of TCP packets generated by different API calls to identify the packet of interest.
By analyzing the sizes of TCP packets generated by different API calls, we could identify the packet of interest.

% Therefore, if we can identify the specific TCP packets that are generated by calling these APIs, we can infer the confirmation time of a transaction.
% from the source IP address of these TCP packets.

\begin{figure}[h!]
% \vspace{-2mm}
\centering
\includegraphics[width=0.95\columnwidth]{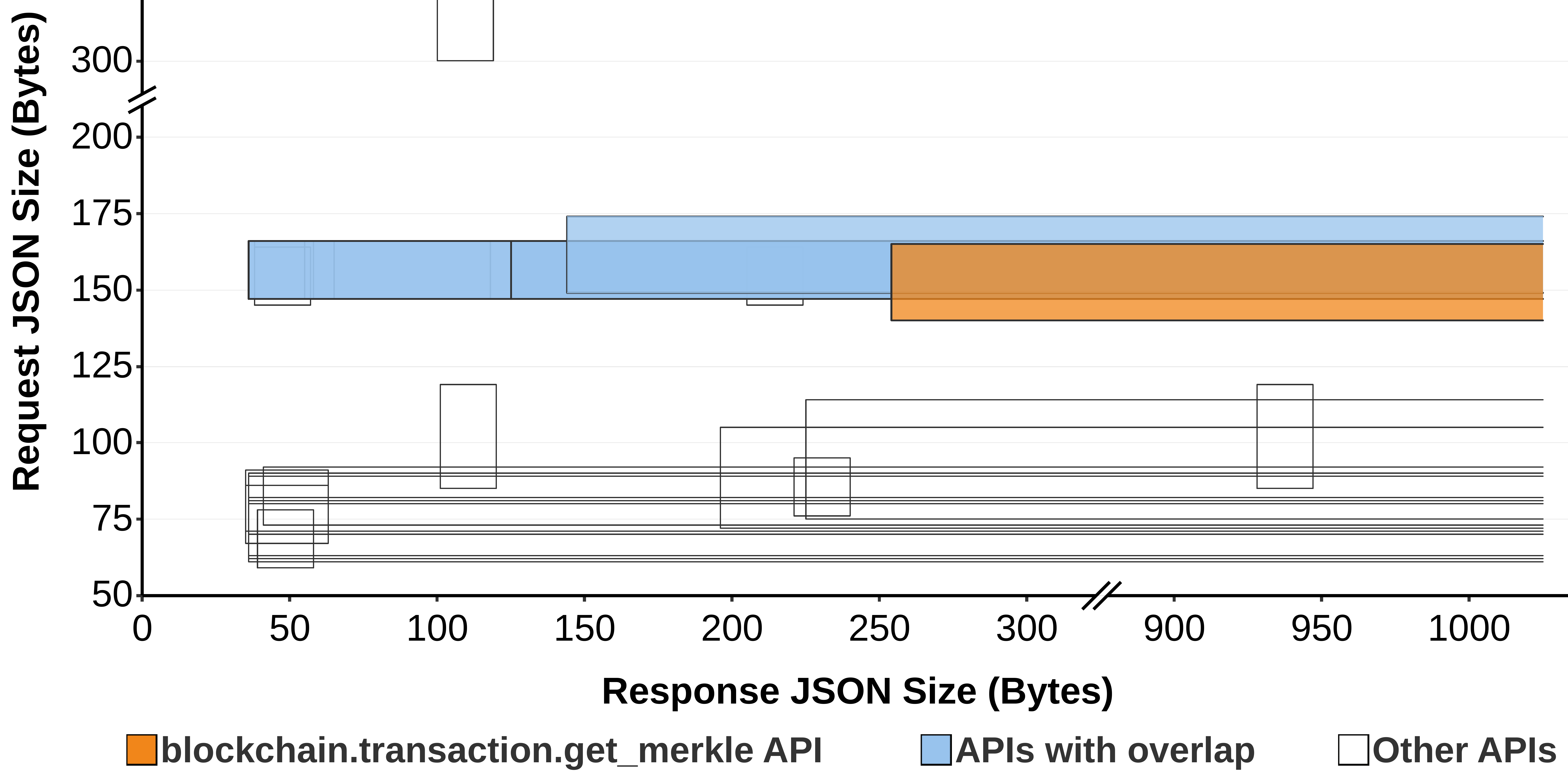}
\vspace{-4mm}
\caption{Size ranges of request/response JSON in Bitcoin}
\label{fig::JSONBTC} 
\vspace{-2mm}
\end{figure}

% \vspace{2mm}
% \noindent
\textbf{(Step-I) Estimating packet size ranges via maximum and minimum sizes of message fields.}
%The first step is to estimate the size range for the packet of interest, which refers to the minimum and maximum possible sizes (typically measured in bytes) that the packet can occupy. 
%This can be achieved by summing the theoretical sizes of each field and its value in the packet. 
When a wallet calls a RPC API, the required parameters are organized in a request RPC-JSON object and the corresponding return values are in a response RPC-JSON object.
These JSON objects contain fixed fields and predictable values according to the API semantics and RPC protocol format in \S\ref{sec::rpcProtocol}. 
% For example, an API for querying transaction status usually specifies a transaction hash as the parameter and returns an object containing transaction status details.
% An API for sending transactions usually takes a raw transaction as the parameter and returns a transaction hash.
% Therefore, the size ranges of a RPC-JSON object can be estimated based on the theoretical maximum and minimum sizes of values of each field. 
For example, a transaction status query API typically takes a transaction hash as input and returns an object containing status details. Conversely, a transaction sending API accepts a raw transaction as input and returns a transaction hash. Therefore, the size range of a JSON-RPC object can be estimated based on the theoretical minimum and maximum sizes of each field's value.
\autoref{fig::JSONBTC} summarizes the analysis results in Bitcoin.
The results for Ethereum and Solana can be found in \autoref{fig::JSONETH} and \autoref{fig::solanaJSON} in the Appendix.
% \autoref{fig::JSONETH} and \autoref{fig::solanaJSON} 
It can be observed that the size ranges of request/response JSON objects generated by an API for transaction status query only overlap with those of a few other APIs.
For example, \autoref{table::overlapAPI} shows that the JSON sizes of Bitcoin RPC API \texttt{blockchain.transaction.get\_merkle} overlap with those of only $4$ other APIs.
% It can be observed that, these mainstream blockchains share similar RPC JSON size features.
% The size ranges of most of APIs do not overlap with that of APIs for querying transaction confirmation status. 
% Ethereum RPC API \texttt{eth\_getTransactionReceipt} overlaps with $9$ out of $44$ APIs, and Solana RPC API \texttt{getSignatureStatuses} overlaps with $13$ out of $54$ APIs.
% It can be observed that, these mainstream blockchains share similar features RPC JSON size.
% The size ranges of request/response JSON objects of a transaction confirmation status API only overlap with that of a few other APIs.

According to the RPC packet format in \autoref{fig::tcpSegment}, the TCP segment contains a TLS header, a HTTP/Websocket header and a HTTP/Websocket body which only carries a RPC JSON object. 
The size of a TLS header and a HTTP/Websocket header are relatively fixed according to the protocol standards. Consequently,
the TCP segment size is mainly decided by the size of RPC JSON object.
Therefore, distinguishability in RPC JSON size ranges will lead to distinguishability in the corresponding TCP packets' sizes.\looseness=-1

% and the RPC JSON object size ranges are decided based on the RPC API semantics. 
% The size of a TLS header and a HTTP header are relatively fixed according to the protocol standards, and the major fluctuations on size are caused by the RPC JSON. 

% \vspace{2mm}
% \noindent
\textbf{(Step-II) Narrowing size ranges via heuristics derived from wallet implementation.}
Several heuristics based on API usage in wallets can further narrow down the candidates of packets carrying the transaction status query result.
% First, a wallet typically does not call RPC APIs for mining or validating blocks.
First, wallets typically do not invoke RPC APIs related to mining or block validation.
For example, Ethereum RPC API \texttt{ eth\_getTransactionReceipt} theoretically overlaps in size with $9$ APIs, but $5$ out of these $9$ APIs are not used by its wallet \textit{MetaMask} as shown in \autoref{table::overlapAPI} and will not introduce noise to the identification.
Second, the RPC JSON objects generated by a wallet usually occupy a limited subset of their theoretical size ranges, resulting in fewer overlaps.
For example, when we calculate the theoretical size ranges, we assume the value of {\em id} field is an integer ranging from $1$ byte to $20$ bytes.
However, \textit{MetaMask} only uses a 15-byte or 16-byte random number as the value of the {\em id}~ field in JSON. 
Considering the practice, packets generated by the RPC API \texttt{eth\_getTransactionReceipt} potentially overlap in size with those of {\em only one API}, for \textit{MetaMask}.
Note: When a wallet calls an API for transaction status, {\em nil} will be returned if the transaction has not been confirmed. Our size-based detection result can exclude packets with {\em nil}, since their size is small.
%}}
\looseness=-1

With the size-based strategies above, 
the identified packet candidates may still include noise, leading to false positives.
% For example, in Ethereum, calling \texttt{eth\_getBlockByHash} may generate packets with the same sizes as the transaction status query API, leading to false positives.
\looseness=-1

\vspace{-2mm}
\subsection{Identify Target Packet via Packet Sequence}
\label{sec::featureSequence}
% \vspace{-1mm}

\begin{figure}[h]
\vspace{-2mm}
\centering
\includegraphics[width=0.9\columnwidth]{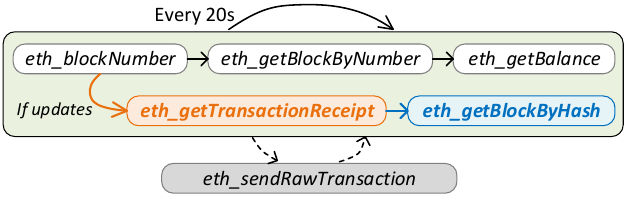}
\vspace{-2mm}
\caption{Ethereum RPC API calls sequence in {\em MetaMask}}
\label{fig::ETHAPISequence} 
\vspace{-2mm}
\end{figure}

% We find that features in the TCP packets sequence can be utilized to remove the false positives.
% Blockchain wallets generally adhere to a standardized business logic, resulting in predictable and relatively fixed RPC API call sequence and TCP packet sequence. Prior to invoking an API to check transaction status, a wallet typically initiates an API call to send the transaction, followed by ledger queries to receive latest information.
We find that features in the TCP packet sequence can be leveraged to eliminate false positives. Blockchain wallets typically follow standardized business logic, leading to predictable and relatively consistent sequences of RPC API calls and corresponding TCP packets. For example, before querying a transaction’s status, a wallet usually first calls an API to send the transaction, followed by additional ledger queries to retrieve the latest information.
\autoref{fig::ETHAPISequence} shows the business logic of the Ethereum wallet {\em MetaMask}. It first initiates a transaction by calling API \texttt{eth\_sendRawTransaction}. Simultaneously in an asynchronous task, it cyclically calls API \texttt{eth\_blockNumber} to query the latest block number every $20$ seconds. 
If the block number increases, it queries for the transaction status by calling our target API \texttt{eth\_getTransactionReceipt}, followed by calling \texttt{eth\_getBlockByHash} for verifying the transaction validity. 
At the same time, it also calls APIs to retrieve the latest block details and the current account balance.
Bitcoin and Solana wallets have similar features according to \autoref{fig::BTCAPISequence} and \autoref{fig::SolanaAPISequence} 
in the
% the extended version of the paper.
Appendix. 
\looseness=-1

% First, although RPC service provides rich APIs for providing comprehensive functions, a wallet may not call all APIs. 
% Second, 

% Features in RPC API call sequences per the wallet business logic can be utilized to improve the detection accuracy.
The target API and the noise APIs (those that could have size overlaps with the target) have different contexts in the API call sequence, as they have different functionalities. 
As shown in  \autoref{fig::ETHAPISequence},
% \autoref{fig::BTCAPISequence} and \autoref{fig::SolanaAPISequence}, 
the surrounding APIs, invoked before and after, differ for the target and noises.
This distinction results in unique TCP packet sequences around the target packet. Consequently, the sequence features can enhance detection accuracy. 
\looseness=-1

% \red{Given the asynchronous nature of periodic and other RPC calls, manually defining rules to identify the target packet across multiple patterns would be cumbersome and inflexible. Rule-based methods also struggle with selecting the appropriate sequence length, weighting features, and handling inter-arrival time variations. Therefore, we adopt a basic ML model to automatically and efficiently learn the data features for performing the identification task.}
% A machine learning (ML) based method can be used to efficiently identify the packet sequence of interest. 

Given the asynchronous nature of periodic and other RPC calls, manually defining rules to identify the target packet across multiple patterns would be cumbersome and less effective. 
Rule-based methods typically struggle with determining an appropriate packet sequence length, weighting features, and handling all possible sequences.
Therefore, we adopt a basic ML model to automatically and efficiently learn the data features and rules for performing the identification task better, as discussed and validated in \autoref{sec::rule_comparison}.
A feature vector as defined in Equation (\ref{equ::feature}) is constructed to represent a sequence of TCP packets, for training a ML model for detection.
% The first step is to train a ML model using TCP packets with ground truth. 
% To achieve this, we construct a feature vector, as defined in Equation (\ref{equ::feature}), to represent a sequence of TCP packets. 
Specifically, each RPC request is paired with a RPC response, meaning that each API call generates two TCP packets. Assume the middle two TCP packets correspond to the target API. 
% These two packets with $r$ packets preceding them and $r$ packets following them form a sequence together. 
Together with $r$ packets preceding them and $r$ packets following them, these $2r+2$ packets form a complete sequence.
% To form the complete sequence, the attacker selects $r$ packets preceding and $r$ packets following the two target packets.
% Totally, the sequence contains $2r+2$ packets.
Each packet is described by three properties including its packet size $l$, its direction $d$ (request or response) and the inter-arrival time interval $t$ to the former.
%The direction property is a boolean value to distinguish RPC request and response.
The feature vector is shown as follows.
% and $1$ indicates it is a JSON request TCP packet and $0$ indicates it is a JSON response TCP packet.
% \vspace{-2mm}
\begin{equation}
\begin{aligned}
    \langle ~&(l_1,d_1,t_1),...,(l_{r+1},d_{r+1},t_{r+1}),(l_{r+2}, d_{r+2}, t_{r+2}), ...,(l_{2r+2},\\&d_{2r+2},t_{2r+2})~ \rangle
\label{equ::feature}
\end{aligned}
% \vspace{-2mm}
\end{equation}
Based on this feature vector, training a basic machine learning model, such as decision tree, support vector machine (SVM), and Random Forest, for binary classification would be sufficient to achieve high detection accuracy.
%(Please refer to \S\ref{subsec:realworldresult}), due to the distinct features in TCP packet sizes and sequences as analyzed above.\looseness=-1
% Additionally, these models are computationally efficient and scalable, making them practical for real-time or large-scale network analysis without the high resource demands of complex models like neural networks. 

%\vspace{2mm}
%\noindent
% \textbf{(Step-II) Identifying Target TCP Packet:} 
The attacker uses the trained ML model to classify a sequence of $2r+2$ packets in the form of the feature vector.
% for identifying the timestamp of interest.
% To be more specific, when the attacker identifies a pair of TCP packets with sizes that fall within the estimated range for the target API, as outlined in \autoref{table::overlapAPI}, they select $n$
% packets preceding and $n$ packets following them to form a sequence and compute its feature vector. 
% The attacker then applies the model to classify this feature vector. 
If the feature vector is classified as a positive instance, the timestamp of the response TCP packet in the middle is the desired transaction query timestamp $\mathcal{T}_q$.
% \red{used as the anchor} for the transaction confirmation time. \looseness=-1

%Finally, when the attacker detects a pair of TCP packets whose sizes fall in the estimated size ranges of the target API, as summarized in Table \ref{table::overlapAPI}, the attacker picks up $n$ packets before and $n$ packets after the potential packets and calculate the feature vector of this sequence. Then the attacker uses the trained machine learning model to classify it. If it is a positive instance, the attacker takes the timestamp of the response TCP packet as the anchor of the transaction confirmation time.\looseness=-1

\vspace{-2mm}
\section{Estimation of Interval $I_{c,q}$ to Derive $\mathcal{T}_c$}
\label{sec::walletBehaviorVul}

% We identify a vulnerability in wallet behaviors: When a user issues a transaction (TX) through their wallet, the interval between subsequent transaction confirmation in the ledger and the RPC service's response to the query of transaction status by the wallet is short. Therefore, if an attacker monitors the user's IP and network traffic and can identify the RPC service's response packets at a timestamp $\mathcal{T}_q$, they may link pseudonyms in blocks created around $\mathcal{T}_q$ in the ledger to the IP, and thus the user.

This section presents the detailed solution (S-II), a universally effective 
% theoretically sound 
approach to derive the interval $I_{c,q}$, measured in $k$ blocks, 
so as to further derive the transaction confirmation timestamp $\mathcal{T}_c$.

\vspace{-2mm}
\subsection{Transaction Status Query Methods}

The detailed interaction pattern among a wallet, a RPC service and a blockchain network for processing a single transaction is illustrated in \autoref{fig::walletBehavior}. 
The interaction is triggered by a user manually sending a transaction through their wallet.
All subsequent actions introduced below are then automatically performed by the wallet, RPC service and blockchain network.
First, the wallet sends a RPC request containing the raw transaction to the RPC service at the time $\mathcal{T}_s$ in Step \circled[2]{\small 1}. 
The RPC service checks and broadcasts the received transaction to the blockchain network in
Step \circled[2]{\small 2}. It also returns a transaction hash to the wallet as the acknowledgment of the receival of the transaction in Step \circled[2]{\small 3}.
To monitor the transaction status, a wallet has two options as follows.

%\begin{itemize}[leftmargin=5pt]
%    \item 
    \textbf{Periodic query}. As shown in Figure \ref{fig::walletBehavior_query}, a common way is to send a periodical query to the RPC service with the transaction hash in Step \circled[2]{\small 4}. 
    The RPC service then retrieves the ledger in Step \circled[2]{\small 5} and returns a {\em nil} if the transaction has not yet been confirmed in Step \circled[2]{\small 6}. 
    Once the transaction is confirmed and included in a new block (Step \circled[2]{\small 7}), all blockchain nodes including the nodes operated by the RPC service synchronize the latest ledger in accordance with a consensus protocol in Step \circled[2]{\small 8}.
    In the consensus protocol, the combined duration of Step \circled[2]{\small 7} and Step \circled[2]{\small 8} constitutes the block time.
    The transaction confirmation time is denoted as $\mathcal{T}_c$.
    The RPC service then responds to the wallet's query with a transaction receipt in Steps \circled[2]{\small 9} and \circled[2]{\small 10}, and the periodic query then stops. 
    The time of the response at Step \circled[2]{\small 10} is $\mathcal{T}_q$.
    Ethereum wallets and Solana wallets typically adopt this method to check the transaction status, as their RPC protocols lack a notification mechanism.
    
%    \item
    % \vspace{1mm}
    % \noindent \textbullet~ 
    \textbf{Notification subscription}. Alternatively, as illustrated in Figure \ref{fig::walletBehavior_notify}, if the RPC service supports the notification subscription, it can notify the wallet once the transaction is confirmed (Step \circled[2]{\small 6}). The wallet will then immediately initiate a query in Step \circled[2]{\small 7} and receive a response at the time $\mathcal{T}_q$ in Step \circled[2]{\small 9}.
    Bitcoin wallets usually adopt this method, as the Bitcoin RPC protocol defines the notification function.\looseness=-1
%\end{itemize}

According to Figures \autoref{fig::walletBehavior_query} and \autoref{fig::walletBehavior_notify}, the transaction confirmation delay---the period between $\mathcal{T}_s$ and $\mathcal{T}_c$---occurs prior to the transaction confirmation and does not affect the temporal correlation between $\mathcal{T}_c$ and $\mathcal{T}_q$.

\vspace{-2mm}
\subsection{Estimation of the Interval to Derive $\mathcal{T}_c$}
\label{subsec::estimationK}

We measure the upper bound of the interval between $\mathcal{T}_c$ and $\mathcal{T}_q$, denoted as $I_{c,q}^{max}$, in units of blocks (specifically, $k$ blocks). 
The value of the parameter $k$ is primarily determined by the query method adopted by a wallet and the block time $z$, i.e., the time required to create a new block and add it to the ledger in a blockchain.
% \looseness=-1

\begin{figure}[ht!]
\vspace{-2mm}
\centering
% \vspace{-2mm}
\subfigure[Obtain the TX status via periodic query]
{
\begin{minipage}[t]{1\linewidth}
    \vspace{-3mm}
    \centering
    \includegraphics[width=1.0\columnwidth]{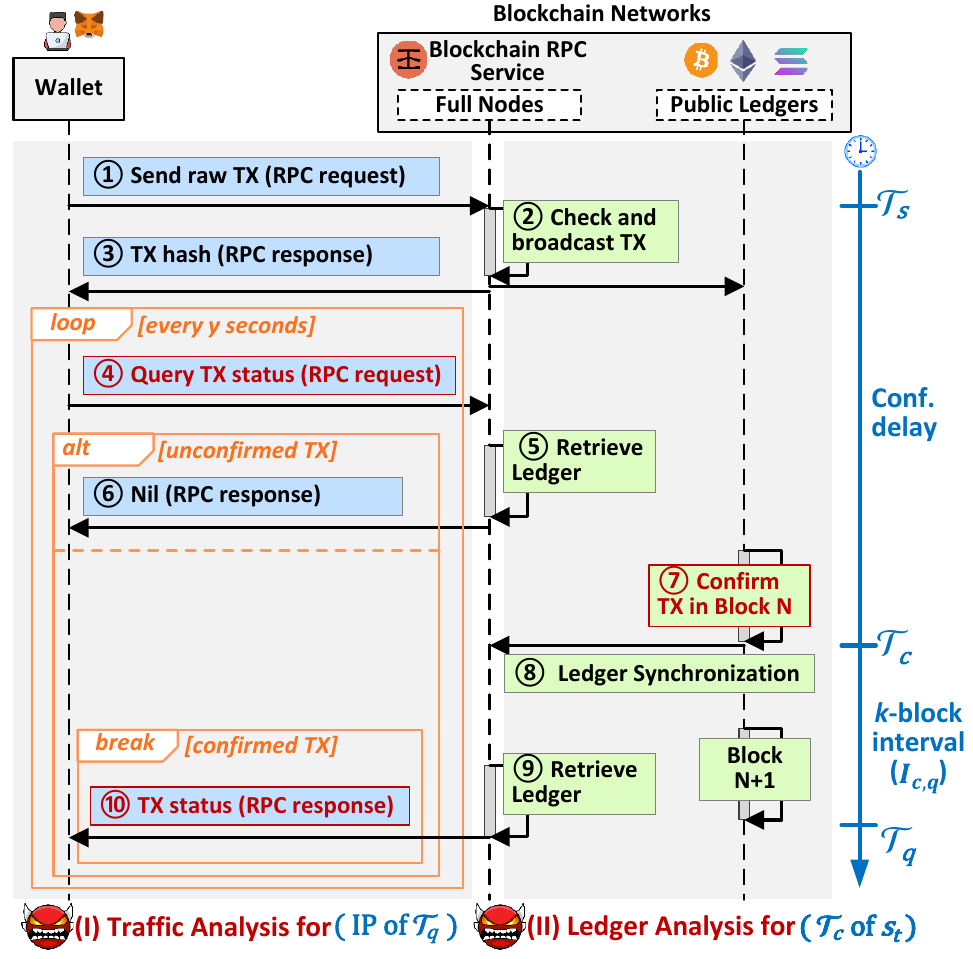}
    \label{fig::walletBehavior_query}
\vspace{-3mm}
\end{minipage}
}
\\
\vspace{-3mm}
\subfigure[Obtain the TX status via notification subscription]{
\begin{minipage}[t]{1\linewidth}
% \vspace{-3mm}
    \centering
    \includegraphics[width=1.0\columnwidth]{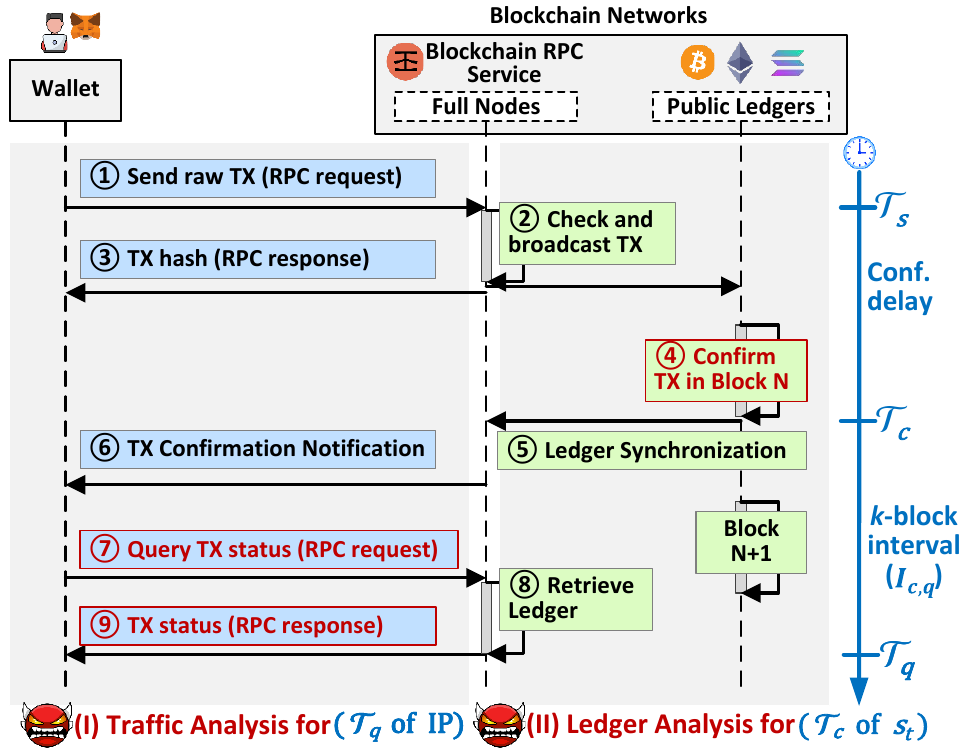}   
    \label{fig::walletBehavior_notify}
\vspace{-3mm}
\end{minipage}
}
\vspace{-5mm}
\caption{The interaction pattern for processing a single transaction among wallets, RPC services and blockchains}
\label{fig::walletBehavior} 
\vspace{-5mm}
\end{figure}

In the case of {\em periodic query}, the interval $I_{c,q}$ is between Steps \circled[2]{\small 7} and \circled[2]{\small 10} as depicted in Figure \autoref{fig::walletBehavior_query}.
Assume the periodic query cycle is $y$ seconds. 
The worst case is that a query cycle just ends right before the transaction is confirmed and the ledger is synchronized. Then the wallet needs to wait for $y$ seconds to initiate the next query. 
% The network RTT (Round-Trip Time) is typically within $0.4$ seconds \cite{RTT}. 
% Therefore, we estimate $k$ as follows,
Considering the network RTT (Round-Trip Time), we estimate $k$ as follows,\looseness=-1
\vspace{-1mm}
\begin{equation}
\vspace{-1mm}
\label{eqn::k-periodicalQuery}
    k = \lceil \frac{y+RTT}{z} \rceil, 
\end{equation}
where RTT is incurred in Step \circled[2]{\small 10} and typically within $0.4$ seconds \cite{RTT}.
The value of the periodic query cycle $y$ is typically hardcoded, and can be obtained by inspecting the open-sourced wallets. For closed-source wallets, $y$ can be derived through examining their network traffic.
The block time $z$ is well-known for each blockchain.
Take the Ethereum wallet {\em MetaMask} as an example, where $z = 12$ and $y=20$, thus $k = 2$ in theory.
In the case of the Solana wallet {\em Torus}, where $z=0.4$ and $y = 20$, we have $k = 51$ in theory.
Considering the practical fluctuations in the consensus process, setting $k=3$ in Ethereum and $k=60$ in Solana is  robust enough.\looseness=-1

% \red{In practice, setting $k=60$ proves to be reliable in practice, considering its block time is relatively small comparing with potential consensus and network fluctuations}.
% Nevertheless, \red{due to the very short block time}, setting $k=60$ proves to be more reliable in practice.

In the case of {\em notification subscription}, we estimate $k$ as follows,
\vspace{-1mm}
\begin{equation}
\vspace{-1mm}
\label{eqn::k-subscription}
k = \lceil \frac{2RTT}{z} \rceil,
\end{equation}
where RTT is incurred in Steps \circled[2]{\small 6}, \circled[2]{\small 7} and \circled[2]{\small 9}. 
% Bitcoin RPC protocol defines a notification function, allowing Bitcoin wallets to operate through notification subscriptions.
For example, for Bitcoin wallets such as {\em Electrum}, $z = 120$, thus $k = 1$. 
Setting $k=2$ is reliable enough to the practical fluctuations in blockchain networks.\looseness=-1
% in practice. proves to be more reliable in practice
% \red{Since $120$ seconds are significantly larger than the RTT, $k=1$ is reliable enough}.

% It can be observed from Equations (\ref{eqn::k-periodicalQuery}) and (\ref{eqn::k-subscription}) that RTT has a minimal influence on the selection of $k$, since RTT may be negligible compared to the block time.
% These analyses have been verified by our experiments in realistic blockchain networks in \S\ref{subsec:realworldresult} and \autoref{table::k}, which also provide more details on other wallets.
% Consequently, an attacker can aprioristically derive the upper bound of the interval, i.e., $I_{c,q}^{max} = k ~~ blocks$.
As shown in Equations (\ref{eqn::k-periodicalQuery}) and (\ref{eqn::k-subscription}), the RTT has minimal impact on the selection of $k$, as it is typically negligible compared to the block time. This observation is further supported by our experiments on realistic blockchain networks in \S\ref{subsec:realworldresult} and \autoref{table::k}, which also include results for other wallets. Consequently, an attacker can aprioristically derive an upper bound on the interval, i.e., $I_{c,q}^{max} = k ~~ blocks$. 
Combining with the identified timestamp $\mathcal{T}_q$ from traffic, the attacker can obtain a transaction confirmation timestamp within the range of $\mathcal{T}_q-I_{c,q}^{max}\leq\mathcal{T}_c\leq \mathcal{T}_q$. 
Compared to the vast ledger, a small value of $k$ significantly reduces the pool of candidates.\looseness=-1

% The vulnerability---specifically, the temporal correlation between transaction confirmation (at time $\mathcal{T}_c$) and the subsequent query of transaction status (at time $\mathcal{T}_q$)---is an inherent and widespread issue across diverse wallets, as it arises from fundamental design philosophy of wallets.
% Following common business logic, after a user sends a transaction via a wallet application, the wallet should provide timely feedback on the transaction status.
% As wallets cannot synchronize ledgers like a blockchain node, it must rely on RPC services to query the transaction status.
% To ensure good user experiences, wallets typically perform these queries automatically and promptly, via either periodic query or notification subscription.
% The business logic and the periodic query cycle are hardcoded and executed automatically by the wallet. 
% These behaviors have been verified by examining open-source wallets listed in \autoref{table::wallet}.
% We find that the Bitcoin RPC protocol defines a notification function, allowing Bitcoin wallets to operate through notification subscriptions. In contrast, the Ethereum RPC protocol and Solana RPC protocol do not offer notification services; their wallets usually rely on periodic queries.

\vspace{-2mm}
\section{Identification of Target Pseudonym}
\label{sec::intersect}
% \vspace{-2mm}

This section details the solution (S-III) to uniquely identify the target pseudonym from the ledger, presents the mathematical models underlying the method, and proposes an optimized method. \looseness=-1

% \vspace{-2mm}
% \subsection{Pseudonym Candidates via Ledger Analysis}
% based on Estimated $\mathcal{T}_c$}
\label{sec::ledgerAnalysis}
% \subsection{Exploiting the Interval to Estimate $\mathcal{T}_q$} 

% (e.g., $k=3$ for Ethereum wallet {\em MetaMask}).

\vspace{-1mm}
\subsection{Identify Target Pseudonym via Intersection}

%In a \textsf{TRAP} attack, 
Identifying the target pseudonym involves multiple rounds of network traffic and ledger analysis and final deanonymization. 
%\looseness=-1

% \vspace{1mm}
% \noindent
\textbf{Step-I: Traffic and ledger analysis rounds}.
In each round, e.g., the $i^{th}$ round, the attacker monitors network packets from the IP address of interest. They identify packets related to transaction queries and obtain the query response timestamp $\mathcal{T}_q$, as in \S\ref{sec::TxCfmTime}.

The attacker then estimates the transaction confirmation timestamp $\mathcal{T}_c$, which is within a range: $\mathcal{T}_q-I_{c,q}^{max}\leq\mathcal{T}_c\leq \mathcal{T}_q$ and $I_{c,q}^{max} = k ~~ blocks$. The attacker can now link the monitored IP with the $k$ consecutive blocks timestamped before $\mathcal{T}_q$, retrieve the pseudonyms of all transaction initiators in the $k$ blocks, and form a candidate set, $\mathcal{S}_i=\{s_{i,1}, s_{i,2}, ..., s_{i,n}\}$. The target $s_t\in\mathcal{S}_i$.

% \red{Note: the observation of transaction query packets implies the user sent a transaction.}
% {\em Ledger analysis}. 
% The attacker then retrieves $k$ consecutive blocks before $\mathcal{T}_q$ from the public ledger,
% % and retrieves the pseudonyms of senders of all transactions in the $k$ blocks to form 
% and obtains a candidate set $\mathcal{S}_i$, as in \S\ref{sec::ledgerAnalysis}.
% , $\mathcal{S}_i=\{s_{i,1}, s_{i,2}, ...\}$. 
%Obviously, a single attack alone is insufficient to uniquely identify the target pseudonym.

% {\em Intersection}. 
The attacker then performs the set intersection as follows,
\begin{equation}
\label{eqn::Intersection}
\mathcal{S}_i= \mathcal{S}_i \cap \mathcal{S}_{i-1},  
\end{equation}
where $\mathcal{S}_i$ is the set of potential target pseudonyms at the $i^{th}$ round. $\mathcal{S}_0=\mathbb U$, where $\mathbb U$ is the universal set including all blockchain users. Note: $\mathcal{S}_1 \cap \mathcal{S}_0=\mathcal{S}_1 \cap \mathbb U = \mathcal{S}_1$.
% \begin{equation}
%     \mathcal{S}_1 \cap \mathcal{S}_0=\mathcal{S}_1 \cap \mathbb U = \mathcal{S}_1
% \end{equation}

% \vspace{1mm}
% \noindent
\textbf{Step-II: Deanonymization}.
The attacker performs multiple rounds (e.g., $m$) of Step-I until the cardinality of the potential target pseudonym set $\left| \mathcal{S}_m \right|=1$ or no longer changes. 
The pseudonyms in $\mathcal{S}_m$ link corresponding users to the IP address being monitored.
If $\left| \mathcal{S}_m \right|=1$, a unique user is associated with the IP address.
It is important to emphasize that our attack goal is ambitious, as it aims to {\em uniquely} link a single IP address to one pseudonym, rather than to a set of pseudonyms.
We consider the attack successful only when $\left| \mathcal{S}_m \right|=1$ and the single pseudonym in $\mathcal{S}_m$ is the target.
% we say our attack succeeds.
\looseness=-1

\vspace{-2mm}
\subsection{Modeling the Attack}
\label{sec::attackModel}
% Based on the attack steps, we derive the probability $P$ (the success rate) of uniquely identifying the target user after $m$ rounds of traffic and ledger analysis as follows,

Based on the attack steps, we can derive the probability $P$ (the success rate) that the target user is uniquely identified from $m$ rounds of traffic and ledger analysis as follows,
\begin{equation}
% \vspace{-1mm}
\label{eqn::P}
P=P_t\cdot P_{t'},
% \vspace{-1mm}
\end{equation}
where $P_t$ is the probability that the target pseudonym appears in all the $m$ candidate sets, and $P_{t'}$ is the probability that non-target pseudonyms are excluded from the candidates.

%\vspace{2mm}
%\noindent
\textbf{(Phase-I) Modeling the target inclusion probability \boldmath $P_t$.} 
If the ML-based timestamp detection method in \S\ref{sec::TxCfmTime} correctly identifies the packet of interest and the timestamp $\mathcal{T}_q$, the target user's pseudonym $s_t$ will appear in the chosen $k$ blocks. Denote the probability of detecting $\mathcal{T}_q$ as $\alpha$,
\begin{equation}
% \vspace{-1mm}
\label{eqn::Pt}
P_t = \alpha^m.
% \vspace{-1mm}
\end{equation}

%\vspace{2mm}
%\noindent
\textbf{(Phase-II) Modeling the non-target exclusion probability \boldmath{$P_{t'}$}.} 
We model users making transaction as a multinomial process. 
Assume there are $N$ users with pseudonyms in the blockchain system, i.e., $\mathbb U=\{ s_i | 1 \le i \le N \}$. 
%\sim Multinomial(x; p_1, p_2, ..., P_{t'})$, as 
This multinomial process is a sequence of independent transactions where a transaction may be made by any of the $N$ users/pseudonyms. The probability of $s_i$ making the transaction is $p_i$. 
% In our scenario, the fixed number of outcomes are the $N$ pseudonyms, i.e., $\{s_1, s_2, ..., s_N\}$, and each trial is a transaction leading to a pseudonym per its initiator.
%The parameter vector $\boldsymbol{p} = (p_1, p_2, ..., p_N)$ describes the probability of a trial leading to each outcome.
%The $x$ denotes the total number of trials, and $X_i$ is the number of trials leading to outcome $s_i$, where $x=\sum_{i=1}^nX_i$.  
%
We model an individual user $s_i$ making transactions 
%user $s_i$ in $k$ consecutive blocks can be modeled by 
as a Poisson process with rate $\lambda_i$.
% , which typically describes the number of events occurring in a fixed interval of time. $\lambda_i$ is an average rate denoting the number of transactions per unit time. 
%(iii) 
Therefore, the number of transactions initiated by $N$ users follows the Poisson distribution with a rate of $\lambda=\sum_{i=1}^N \lambda_i$.
%The relationship of their rates should satisfy {\boldmath$\lambda_i = \lambda p_i$}. 
%Moreover, Gamma distribution can be used as a prior for $\lambda$, allowing for additional variance in overdispersed transaction count data in reality.
% (iv) The variable $\boldsymbol{n}$ also follow Poisson distribution, similar to $\boldsymbol{x}$.\looseness=-1

% Assume $s_i$ is a pseudonym in set $S_1$.
Denote $f_i$ as the probability that a non-target pseudonym $s_i$, where $i\neq t$, is excluded after $m$ rounds of intersection.
Since $s_i$ can be excluded once it fails to appear in any of the sets during the intersection process (Refer to Appendix B
% \autoref{appendix::models}
for more details in the Appendix),
$f_i$ can be derived in Equation (\ref{eqn::f}).\looseness=-1

\begin{equation}
% \vspace{-2mm}
    f_i 
    = 1-\prod\nolimits_{j=2}^{m}(1 - (1 - p_i)^{x_j}),  
%\vspace{-1mm}
\label{eqn::f}
\end{equation}
where
$x_j$ is the number of transactions in the $k$ consecutive blocks in the $j^{th}$ round of the attack.

% The parameters $\boldsymbol{p}$ and $\boldsymbol{x}$ are random variables modeled by stochastic processes. 
% The $p_i$ may take any value of variable $\boldsymbol{p}$, and $x_j$ may take any value of variable $\boldsymbol{x}$ where each value corresponds to a different probability.
% As the parameter $\boldsymbol{p}$ and $\boldsymbol{x}$ follow stochastic processes, the $\boldsymbol{f_i}$ is also a random variable, whose each possible value occurs with a distinct probability.

Assume $s_t$ is the target user in $\mathcal{S}_1$ and $\left| \mathcal{S}_1 \right|=n$. The probability that the $n-1$ non-target users in set $\mathcal{S}_1$ are excluded after $m$ rounds of intersection is derived as follows,
\begin{equation}
\vspace{-1mm}
\label{eqn::Pn}
P_{t'} = \prod\nolimits_{i=1, i \ne t}^{n}f_i,
%\vspace{-1mm}
\end{equation}
where $n$ is the number of pseudonyms in $\mathcal{S}_1$, which is collected from $k$ consecutive blocks.
%This also implies that, the probability of one or more non-targets appearing alongside the target in all $m$ attacks, resulting in false positives, can be given by $h = 1 - P_{t'}$.

\vspace{1mm}
%\noindent
\textbf{(Phase-III) Modeling the success rate:} 
Based on the above equations, we can derive $P$ in Equation (\ref{eqn::P_Final}). 
% which is the identification accuracy.
% \begin{equation}
% \begin{aligned}
%      P = P_t \cdot P_{t'}.
% \end{aligned}
% \label{equ::F}
% \end{equation}
% Based on the above equations.
% Equations (\ref{equ::Ft}) $\sim$ (\ref{equ::F}), 
% Substitute (\ref{eqn::Pt}), (\ref{eqn::f}) and (\ref{eqn::Pn}) into (\ref{eqn::P}), we derive Equation (\ref{equ::F_factor}). showing $P$ is a random variable as its parameter $\boldsymbol{p}$, $\boldsymbol{x}$, and $\boldsymbol{n}$ are random variables.
% a function of five parameters. Among them, $\boldsymbol{p}$, $\boldsymbol{x}$, and $\boldsymbol{n}$ are random variables modeled by stochastic processes. 
The parameters $x_j$, $n$ and $p_i$ are decided by the blockchain user activity  intensity in the ledger.
Parameters $\alpha$ and $m$ are determined by the attack strategy. \looseness=-1
% Overall, the accuracy $P$ is actually a random variable. 

%\vspace{-2mm}
\begin{small}
\begin{equation}
\vspace{-1mm}
%\begin{aligned}
    P = \alpha^m \prod\nolimits_{i=1, i \ne t}^{n} \left(1-\prod\nolimits_{j=2}^{m}(1 - (1 - p_i)^{x_j}) \right).
% \vspace{-1mm}
\label{eqn::P_Final}
\end{equation}
\end{small}

% \red{
% \textbf{(Phase-IV) Estimation of $P$:} 
% Denote $x$ as number of transactions in $k$ blocks. $x$ is a random variable that follows Poisson distribution. 
% $x_j$ in (\ref{eqn::P_Final})  is the number of transactions in the k blocks of the $j^{th}$ round of an attack. To estimate $P$, we use the mean of $x$, i.e., $\bar{x}$, for $x_j$, substitute it into (\ref{eqn::P_Final}), and derive the estimated $P$ as follows,
% \begin{small}
% \begin{equation}
%     P = \alpha^m \prod_{i=1, i \ne t}^{n} \left(1-\left(1 - (1 - p_i)^{\bar{x}} \right)^{m-1}\right).
% \label{eqn::P_Estimation}
% \end{equation}
% \end{small}
% }

% \vspace{-2mm}
\subsection{Analysis and Optimization}
\label{sec::model_Analysis}
% \vspace{-2mm}

According to Equation (\ref{eqn::P_Final}), $P$ decreases as $p_i$ increases, where
$p_i$ quantifies a user's transaction-sending intensity. 
Therefore, if non-target users transact infrequently, there is a better chance to uniquely identify the target user $s_t$.
However, we observe highly active users in real-world ledgers. 
Such users may often appear alongside with the target in candidate sets in each round of the attack, causing false positives. 
To address this issue, we categorize users as normal users and active users and propose an optimized identification method based on a transacting rate threshold.

% different identification strategies accordingly.

\begin{figure*}[t!]
\begin{minipage}[c]{0.49\columnwidth}
\centering
\includegraphics[width=0.99\textwidth]{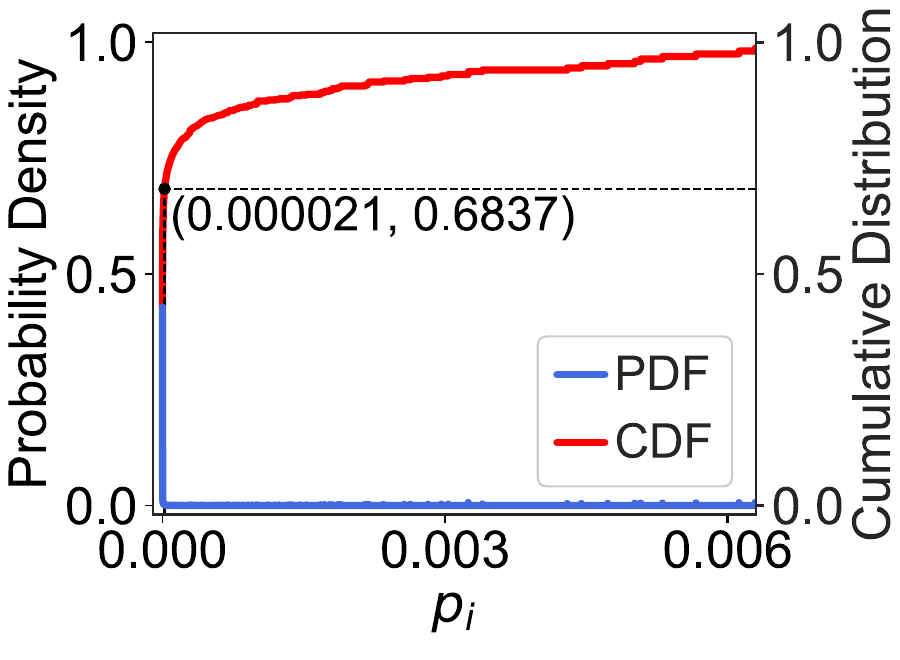}
\vspace{-8mm}
\caption{Distribution of $p_i$ (the probability of a transaction being created by $s_i$)}
% that a transaction is initiated by a pseudonym $s_i$}
\label{fig::distributionOfb} 
\vspace{-3mm}
\end{minipage}
\hspace{0.1cm}
\begin{minipage}[c]{0.49\columnwidth}
\centering
\includegraphics[width=0.99\textwidth]{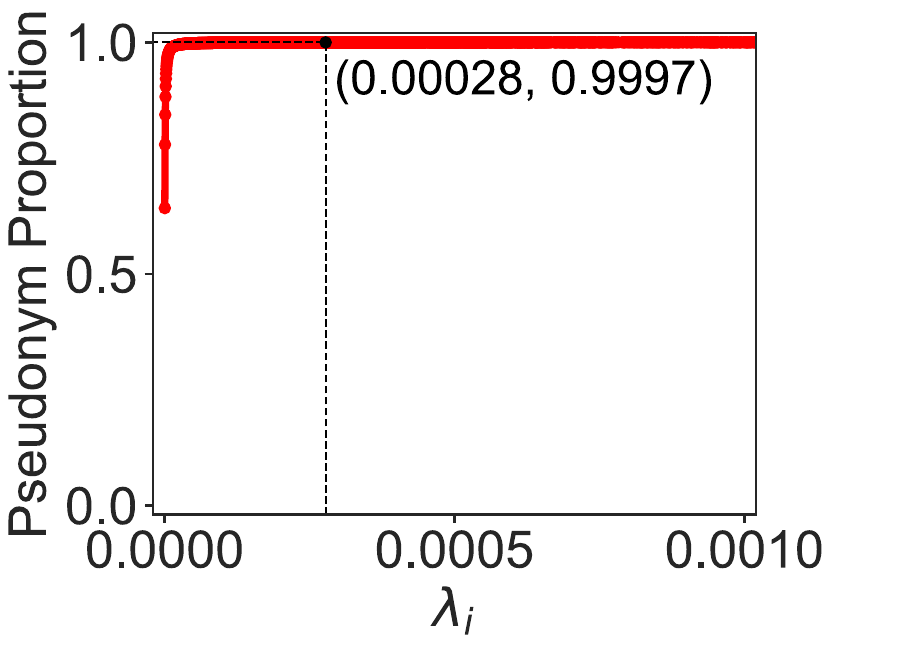}
\vspace{-8mm}
\caption{The proportion of users within each transacting rate $\lambda_i$}
\label{fig::distributionOfLamda}  
\vspace{-3mm}
\end{minipage}
\hspace{0.1cm}
\begin{minipage}[c]{0.49\columnwidth}
\centering
\includegraphics[width=0.99\textwidth]{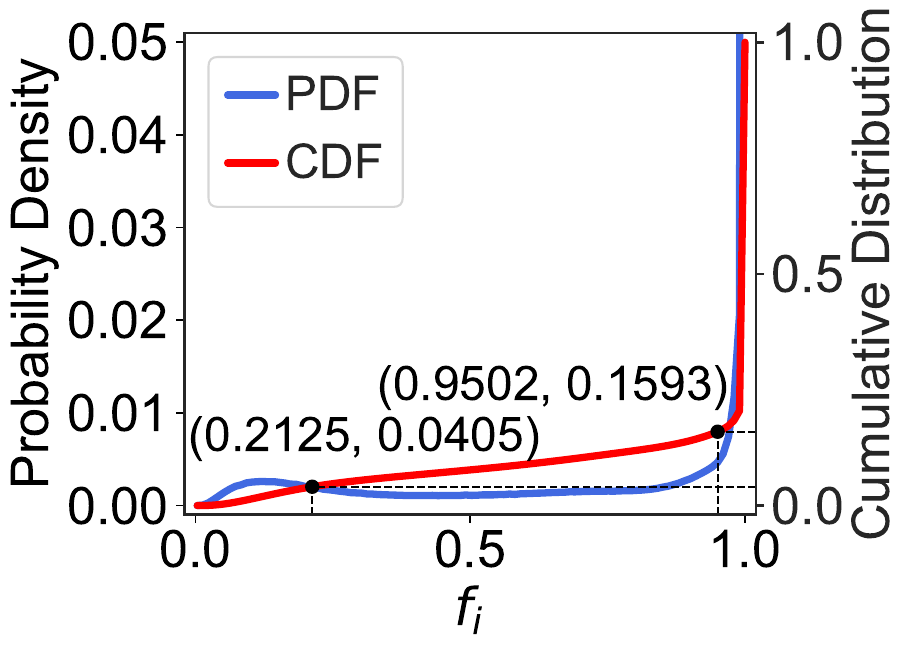}
\vspace{-8mm}
\caption{Distribution of $f_i$ without adopting optimization ($m = 3$)}
\label{fig::F(m,f)} 
\vspace{-3mm}
\end{minipage}
\hspace{0.1cm}
\begin{minipage}[c]{0.49\columnwidth}
\centering
\includegraphics[width=0.99\textwidth]{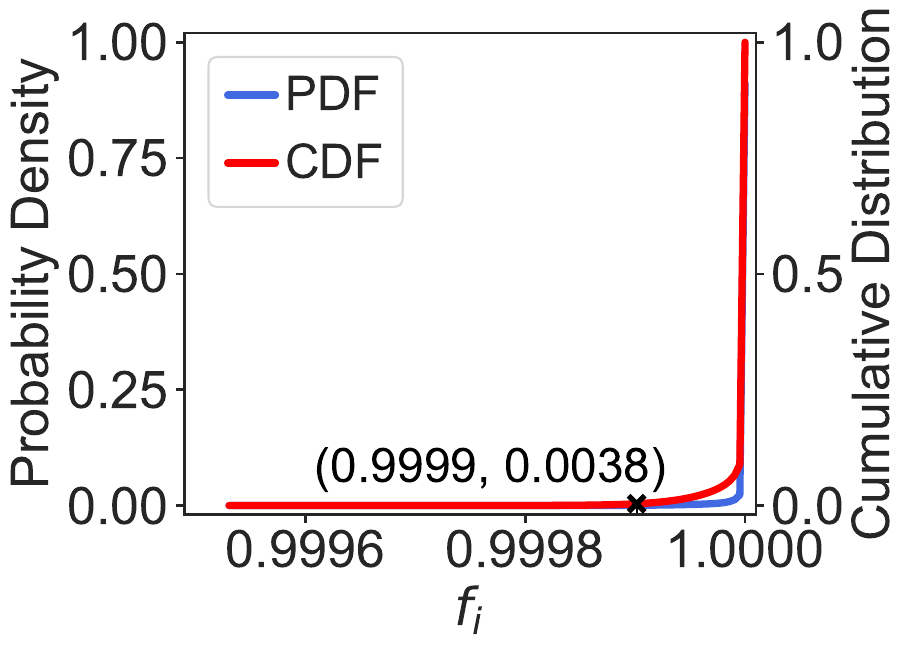}
\vspace{-8mm}
\caption{Distribution of $f_i$ in the optimized identification method ($m = 3$)}
\label{fig::F(m,f)filter_b}  
\vspace{-3mm}
\end{minipage}
\end{figure*}

\begin{figure}
% \hspace{0.2cm}
\begin{minipage}[c]{0.48\columnwidth}
\centering
\includegraphics[width=0.99\textwidth]{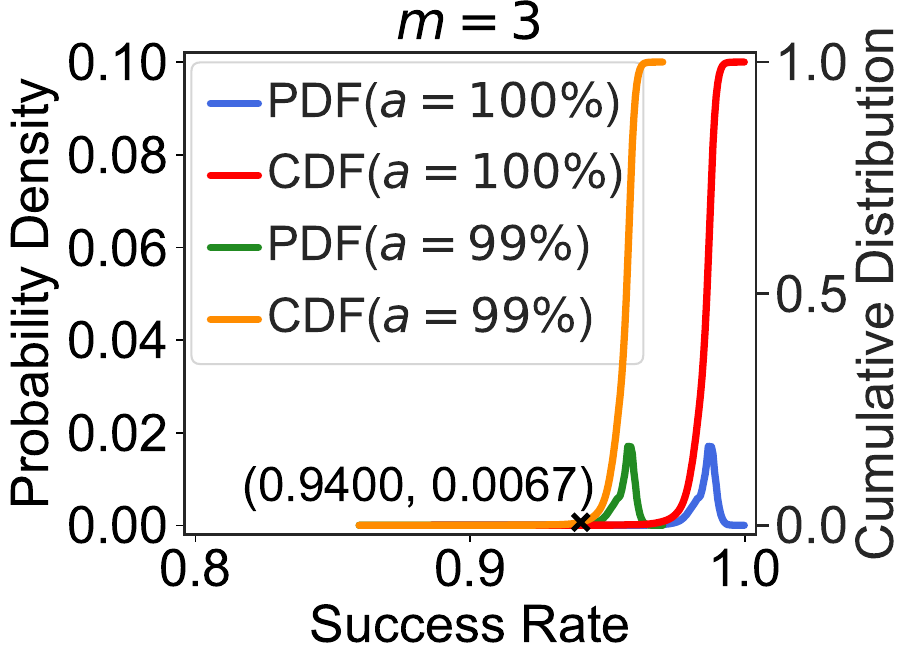}
\vspace{-7mm}
\caption{Distribution of the success rate $P$ ($m = 3$)}
\label{fig::F(P)m=3}  
\end{minipage}
\vspace{-2mm}
\hspace{0.1cm}
\begin{minipage}[c]{0.48\columnwidth}
\centering
\includegraphics[width=0.99\textwidth]
{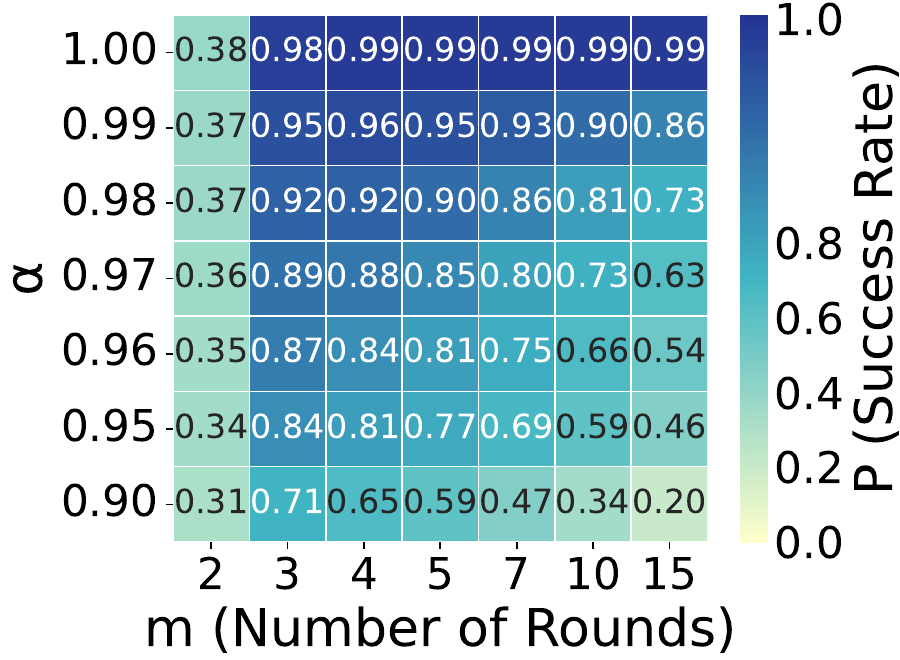}
\vspace{-7mm}
\caption{The success rate $P$ with varying $\alpha$ and $m$}
\label{fig::heatmap}  
% \vspace{-1mm}
\end{minipage}
\vspace{-2mm}
\end{figure}

\textbf{Classifying users based on a transacting rate threshold}.
% \textbf{Transacting Rate.}
An attacker can utilize {\em transacting rate}, i.e., the rate $\lambda_i$ in the Poisson distribution model of each user, as a metric to classify users.
Note that, the Poisson distribution is typically used to model the number of events in a fixed interval of time.
% of this user's activity level, which is named as . 
% $\lambda_i$ denotes the number of transactions initiated by user $s_i$ within an unit time, such as within $1$ second.
The probability of user $s_i$ sending one or more transactions within the period of $k$ blocks can be derived in Equation (\ref{equ::lamda}).
\begin{equation}
    P(X_i \geq 1) = 1 - e^{-\lambda_ikT},
\label{equ::lamda}
\end{equation}
where $X_i$ denotes the number of transactions from $s_i$, and $T$ is the block time. 
%\textbf{Transacting Rate Threshold.}
The attacker can distinguish active users and normal users (inactive users) via a threshold $\theta_{\lambda}$ of transacting rate. We define $\theta_{\lambda}$ by setting $P(X_i \geq 1) = 1\%$, as $1\%$ is generally considered small in probability theory.
Our experiment results are consistent with this threshold setting.
% In different blockchains, $\theta_{\lambda}$ is different with different $k$ and $T$. 
% For example in Ethereum, $k = 3$ and $T = 12s$, then $\theta_{\lambda} \approx 0.00028$.
% It means that, if a user sends over $0.00028$ transactions per second, the  an active user.
% This aligns with the intuition, as $\theta_{\lambda} = 0.00028$ indicates that the user sends around one transaction per hour ($\frac{1}{0.00028 * 60 *60}\approx0.99h$).
% In reality, very few Ethereum users send transactions every hour.
% which occupy only $**\%$ according to our measurements in \autoref{}.

%\vspace{-2mm}
\begin{algorithm}[h]
%\vspace{-1mm}
\caption{Optimized Target Pseudonym Identification Method}
% \label{alg:cap}
\label{algorithm::optimizedIden}
\begin{algorithmic}
\Require $\mathcal{S}_m$ \Comment{The intersection result after $m$ rounds of attacks}
% \Ensure $\mathcal{S}_t$
\State $\mathcal{S}_t \gets \mathcal{S}_m$
\For{$s_i \in \mathcal{S}_m$}
    \If{$\lambda_i \geq \theta_{\lambda}$} \Comment{Its transacting rate exceeds the threshold}
        \State $\mathcal{S}_t = \mathcal{S}_t\setminus\{s_i\}$ \Comment{Filtering out active users}
    \EndIf
\EndFor
\If{$\left| \mathcal{S}_t \right|=1$} \Comment{The target is a normal user}
    \State \textbf{Output} $\mathcal{S}_t$ \Comment{Output the target pseudonym}
% \Else
\ElsIf{$\left| \mathcal{S}_t \right|=0$} \Comment{The target is an active user}
    \State \textbf{Output} $\mathcal{S}_m$ \Comment{Fail (output a pseudonym candidates set)}
\Else
    \State \textbf{Output} $\mathcal{S}_t$ \Comment{Fail (output a pseudonym candidates set)}
\EndIf
\end{algorithmic}
%\vspace{-1mm}
\end{algorithm}
%\vspace{-2mm}

\textbf{Optimized identification method}.
As outlined in Algorithm \ref{algorithm::optimizedIden},
after obtaining the intersection set $\mathcal{S}_m$ via $m$ attack rounds, the attacker calculates the transacting rate for each user in $\mathcal{S}_m$.
Users with $\lambda_i$ exceeding the threshold $\theta_{\lambda}$ are active users and excluded from $\mathcal{S}_m$.
There are three cases per the size of the final set, denoted as $\mathcal{S}_t$, without active users.
(Case I): When $\left| \mathcal{S}_t \right|=1$, the only one pseudonym in $\mathcal{S}_t$ is output as the target.
(Case II): If $\left| \mathcal{S}_t \right|=0$, implying that the target is an active user, the initial $\mathcal{S}_m$ is output and gives a set of pseudonym candidates.
(Case III): If $\left| \mathcal{S}_t \right|>1$, indicating false positives introduced by normal users, the final $\mathcal{S}_t$ gives a set of pseudonym candidates.
% Our attack goal is ambitious, and we treat the last two cases which give a candidate set as failures, as they cannot give the unique target.
Our attack goal is ambitious; therefore, we consider the last two cases, where the result is a candidate set rather than a unique target, as failures.
Fortunately, according to our measurements in \S\ref{sec::measurements_section}, the occurrence probability of the Case III is only $1.47\%$ when $m=3$ in Ethereum. 
Moreover, only a small portion of users are active based on the transacting rate threshold (e.g., only around $0.03\%$ in Ethereum), indicating that our attack method can uniquely identify overwhelming majority of users.
\looseness=-1

\vspace{-2mm}
\section{Measurements-Based Numerical Evaluation}
\label{sec::measurements_section}
\vspace{-1mm}
% Empirical

In this section, we conduct large-scale measurements of public ledgers from real-world blockchains, including Ethereum, Bitcoin and Solana, so as to derive numerical results of the attack based on mathematical models in \S\ref{sec::attackModel}, since the user activity intensity in ledgers may impact the attack results as indicated by Equation (\ref{eqn::P_Final}). 
These numerical results validate the intersection-based attack method, demonstrating the root cause of \textbf{TRAP} lies in the inherent characteristics of blockchain users and ledgers.
The numerical evaluation answers the following two questions.
\begin{itemize}[leftmargin=10pt]
%\vspace{1mm}
%\noindent 
%\textbullet~ 
\item \textbf{Q1}: 
%Hundreds of thousands users transact over a blockchain network every day \footnote{\url{https://etherscan.io/chart/active-address}}. 
%Intuitively, non-target users may appear alongside the target in the candidate sets obtained from $m$ rounds of attacks, and lead to false positives. 
What is the false positive rate before and after adopting the optimized identification method in Algorithm \ref{algorithm::optimizedIden}
(\S\ref{sec::infulenceRegularUser})?

%\vspace{1mm}

%\noindent 
%\textbullet~  
\item \textbf{Q2}: 
%In realistic blockchains, users' transaction behaviors follow a stochastic process. The number of transactions ($\boldsymbol{x}$) and the number of of pseudonyms ($\boldsymbol{n}$) in $k$ consecutive blocks, and the probability of a transaction trial leading to each pseudonym ($\boldsymbol{p}$) may impact the final result in a random way. 
What is the success rate according to the statistics of public ledgers (\S\ref{subsec::Expectation})?\looseness=-1
\end{itemize}

We assume the victim user is a normal user, who transacts infrequently. If the victim user is an active user, the attacker cannot uniquely identify the active user. The latter case is evaluated in 
% at the end of 
\S\ref{subsec:realworldresult}
. \looseness=-1
% Refer to \S\ref{sec::model_Analysis} for the definition of active users.

\vspace{-2mm}
\subsection{Datasets}
% \label{subsec:experimentsetup}
% \vspace{-2mm}

% \textbf{Ledgers Data Collection and Processing.} We collect public ledgers data on Ethereum, Bitcoin, and Solana networks. For mainnet data analysis, we primarily leverage Google Cloud's {\em BigQuery} service. We retrieve the testnet data via RPC services, and store and analyze it in {\em MongoDB}.\looseness=-1

We collected the public around $30.61$ GB of ledger data from August 1st to August 31st, 2024 in Ethereum mainnet. The dataset contains 221,842 blocks and 33,964,686 transactions. 
These transactions are initiated by 6,769,835 users. 
Please note that, this section primarily presents the measurements from the Ethereum mainnet as a representative example to demonstrate our models and findings. 
The datasets and measurements from other $5$ blockchain networks, including the Ethereum testnet, Bitcoin mainnet and testnet, Solana mainnet and testnet, exhibit similar patterns,
which are presented in 
Appendix E and Figures 26-55 in the Appendix.
% \autoref{sec::measureAppendix} and Figures \ref{fig::distributionOfbETHtest}--\ref{fig::F(P)m=4_SOLtest} 
% in the extended version of the paper.
% Appendix. 
In total, we processed about \textbf{3.10 TB} of ledger data.
\looseness=-1

%88.73 GB = 0.08665 TB 
% 1.29TB + 1.72TB

% (0.80,0.0002668949)
% (0.0000000294,0.1279618778)
% (0.0064308559,0.0064308559)

\vspace{-2mm}
\subsection{False Positive Rate}
\label{sec::infulenceRegularUser}

We derive the false positive rate, denoted as $R$ where $R = 1-P_{t'}$, based on large-scale measurements.
% caused by non-target appearing in the final intersection result. Specifically, $R = 1-P_{t'}$.
% To derive the false positive rate based on large-scale measurements, 
According to Equations (\ref{eqn::f}) and (\ref{eqn::Pn}), and the estimation method in 
% \autoref{sec::estimationOfE(P)} 
Appendix C, we empirically measure the probability density functions (PDFs) of the three key parameters, the number of transactions in $k$ consecutive block---$x$, the number of pseudonyms in $k$ consecutive blocks---$n$, and the probability of a user making a transaction---$p_i$. 
% They characterize the user's activity intensity in a ledger.
The measurement results and analysis are presented in \autoref{fig::distributionOfb} and \autoref{fig::distributionOfLamda}, and 
Appendix D,
through which we make \textit{Finding~I}. \looseness=-1
% from \autoref{fig::distributionOfb} and \autoref{fig::distributionOfLamda}.
% According to the analysis in \S\ref{sec::measureActivityLevel} of Appendix, 

% \vspace{1mm}
% \begin{mdframed}[backgroundcolor=yellow!4]  
% \noindent\textit{\textbf{Finding~I.}{
{\em \textbf{Finding~I.}
Only $0.03\%$ of Ethereum users are marked as active users based on the threshold of transacting rate.
Filtering out the $0.03\%$ of users who are active can significantly reduce the range of values for $p_i$. }
% }}
% \end{mdframed}
% \vspace{1mm}
\looseness=-1

% \vspace{1mm}
%\noindent
\textbf{False positive rate without optimization:} 
We first check the distribution of $f_i$, and then derive the mathematical expectation of the false positive rate $R$.
The calculation methods based on the measured PDFs of $x$, $n$ and $p_i$ are presented in 
% \autoref{sec::estimationOfE(P)} 
Appendix C.
Recall $f_i$ is the probability that a non-target user is excluded after the intersection as discussed in \S\ref{sec::attackModel}.
When $f_i$ approaches $0$, the user $s_i$ most likely introduces the false positive.
When $f_i$ approaches 1, the opposite holds true.
% With the measured $F(\boldsymbol{x})$ and $F(\boldsymbol{p})$, we derive 
The distribution $F(f_i)$ is shown in \autoref{fig::F(m,f)} when $m = 3$. 
It can be observed that some values of $f_i$ approach $0$.
% The expectation of $f_i$ is $\mathbb{E}[f_i]=92.05\%$.
Based on $F(f_i)$, we further derive the expectation of false positive rate $R$.
We have $\mathbb{E}[R] = 97.66\%$, indicating the false positive rate is very high without optimization.
% Obviously, the false positive rate is very high without optimization.
% $E(h) = 97.66\%$
% Based on the analysis, we conclude the \texttt{Finding (II)}.
% as follows.

% \vspace{1mm}
%\noindent
\textbf{False positive rate of optimized method:} 
\autoref{fig::F(m,f)filter_b} shows the distribution $F(f_i)$ when the attacker employs the optimized identification method described in Algorithm \ref{algorithm::optimizedIden}, which excludes active users (only $0.03\%$ of all users) from the intersection result.
The optimized method significantly alters the distribution $F(f_i)$, shifting from \autoref{fig::F(m,f)} to \autoref{fig::F(m,f)filter_b}, where nearly all values of $\boldsymbol{f_i}$ exceed $99.99\%$.
Accordingly, the expectation of false positive rate is changed to $\mathbb{E}[R]=1.47\%$.
Based on this, we make \textit{Finding II}.
%\looseness=-1

% \vspace{2mm}
% \begin{mdframed}[backgroundcolor=yellow!4]  
% \noindent\textit{\textbf{Finding II.}{~
{\em \textbf{Finding II.} The optimized identification method, which excludes only a small portion of users who are active, significantly reduces the false positive rate, from $97.66\%$ to only $1.47\%$ in a $3$-round intersection attack, highly optimizing the attack result.}
% which indicates that daily user activity rarely introduces false positives.
% he probability of non-target users appearing alongside the target in the candidate sets of $3$ attack attempts ($m=3$) 
% }}
% \end{mdframed}
% % \vspace{2mm}

% (0.99,0.0000011188)
% (0.0000000294,0.6419927517)
% (0.0064308559,0.0000001477)

\begin{table}[ht!]
\vspace{-2mm}
\caption{Numerical success rate across blockchains}
\vspace{-3mm}
\centering
% \label{table::expectation}
\setlength\tabcolsep{4pt}
\renewcommand{\arraystretch}{1}
\resizebox{\linewidth}{!}{
\begin{tabular}{lccccc}
\hline
\multirow{2}{*}{Blockchain} & \multirow{2}{*}{\begin{tabular}[c]{@{}c@{}}Number of \\ Attack Rounds\end{tabular}} & \multicolumn{2}{c}{Testnet} & \multicolumn{2}{c}{Mainnet}  \\ \cdashline{3-6}[2pt/2pt] 
                            &                                                                          & $\alpha$ = 0.99 & $\alpha$ = 1.00 & $\alpha$ = 0.99 & $\alpha$ = 1.00 \\ \hline
Ethereum                    & m=3                                                                      & 96.97\%         & 99.94\%         & 95.61\%         & 98.53\%         \\
Bitcoin                     & m=4                                                                      & 93.88\%         & 97.73\%         & 95.98\%         & 99.92\%         \\
Solana                      & m=4                                                                      & 95.32\%         & 99.23\%         & 90.16\%         & 93.86\%         \\ \hline
\end{tabular}
}
\label{table::expectationP}
\vspace{-3mm}
\end{table}

\vspace{-1mm}
\subsection{Success Rate Based on Measurements}
\label{subsec::Expectation}

Based on the measured probability density functions of $x$, $n$ and $p_i$, along with the calculation methods in
\autoref{sec::estimationOfE(P)}, we derive the expected value of $P$, i.e., $\mathbb{E}[P]$, which gives the success rate versus the timestamp detection probability ($\alpha$) and the number of attack rounds ($m$). 
\autoref{fig::F(P)m=3} shows the distribution of success rate $P$ using the optimized identification method. 
When $m=3$ and $\alpha=100\%$, $\mathbb{E}[P]=98.53\%$, 
indicating that variations in blockchain user activity cause minimal interference.
% indicating minimal interference from variations in blockchain user activity.
When $m=3$ and $\alpha=0.99$, almost all values of $P$ exceed $94\%$ and $\mathbb{E}[P]=95.61\%$, suggesting a consistent and high success rate.
The heatmap in \autoref{fig::heatmap} shows the expectations of $P$ versus varying $\alpha$ and $m$.
These two parameters are decided by the attacker's capability and strategy.
This enables us to assess the risks to user anonymity without launching real-world attacks on blockchain networks.
The expectations of $P$ in other blockchains can be derived similarly, which are summarized in \autoref{table::expectationP}.
These results demonstrate that the attacker can achieve a high success rate across blockchains, with a small $m$ and a large $\alpha$.
We then make \textit{Finding III}. \looseness=-1

% \vspace{2mm}
% \begin{mdframed}[backgroundcolor=yellow!4]  
% \noindent\textit{\textbf{Finding III.}{
{\em \textbf{Finding III.} Although the number of transactions and the number of pseudonyms in $k$ consecutive blocks and the user transacting rates are random, they rarely interfere with the final attack result in our optimized identification method.
The attacker can achieve high deanonymization success rates across various blockchains with only a limited number of attack rounds.}
% and high transaction timestamp detection probability. 
% }}
% \end{mdframed}
% % \vspace{-2mm}

\vspace{-1mm}
\section{Evaluation of Real-World Attacks}
\label{subsec:realworldattack}

This section presents the results of real-world attacks on realistic blockchain networks. 
% In our experiments, the attackers locate along the realistic route traces across cities and countries.
These results align with the numerical results in \S\ref{sec::measurements_section}, validating the correctness of the mathematical models.\looseness=-1
%while showing that our attack results are representative rather than isolated cases.

% In this section, we first discuss the experiment setup (\S\ref{subsec:experimentsetup}). 
% Then, we examine the measurements of real-world blockchain networks and derive the theoretical expectation of the deanonymization accuracy (\S\ref{subsec::measurements}).
% Next, we derive the theoretical expectation of the deanonymization accuracy (\S\ref{subsec:Theoreticalaccuracy}).
% Finally, we evaluate the performance of ML models and our attack method across different blockchains. The real-world evaluation results further validate the correctness of the theoretical results (\S\ref{subsec:realworldattack}).
% We now present our attack results against real-world blockchains, which validate the attack effectiveness and the correctness of our theoretical models and results.\looseness=-1

\vspace{-2mm}
\subsection{Experiment Setup}
\label{subsec:experimentsetup}
% \vspace{-2mm}

% \vspace{1mm}
%\noindent
\textbf{Platforms and tools.}  
We use browser-based {\em MetaMask} v12.6.0 for Ethereum, Windows application {\em Electrum} v4.5.5 for Bitcoin, and web-based {\em Torus} for Solana as representatives to demonstrate the attack effectiveness across different blockchains.
% as these wallets are popular, operate with standard RPC protocols and support both blockchain testnet and mainnet.
The generality of the attacks against other wallets is discussed in \S\ref{subsec::attackGenerality}.
% For each wallet, we develop an automation tool built atop {\em Selenium} or {\em pywinauto} to mimic normal user operations on wallets such as sending transactions.
For each wallet, we develop an automation tool built atop {\em Selenium} or {\em pywinauto} to emulate manual click operations on wallets such as sending transactions.
These tools generate transactions based on timings, intervals and frequency observed in realistic ledgers.
By design, each wallet keeps periodical query for network-adjusted gas price from its own server or from the RPC service, which estimates the gas price according to the network conditions in real time.
When a user sends a transaction, the wallet will automatically provide the latest queried gas price as the default. In our experiments, the transactions use the wallets’ default and network-adjusted gas price,
which typically ensures timely confirmation.
\looseness=-1

%\vspace{1mm}
%\noindent

\begin{figure}[t!]
\centering
\includegraphics[width=1.0\columnwidth]{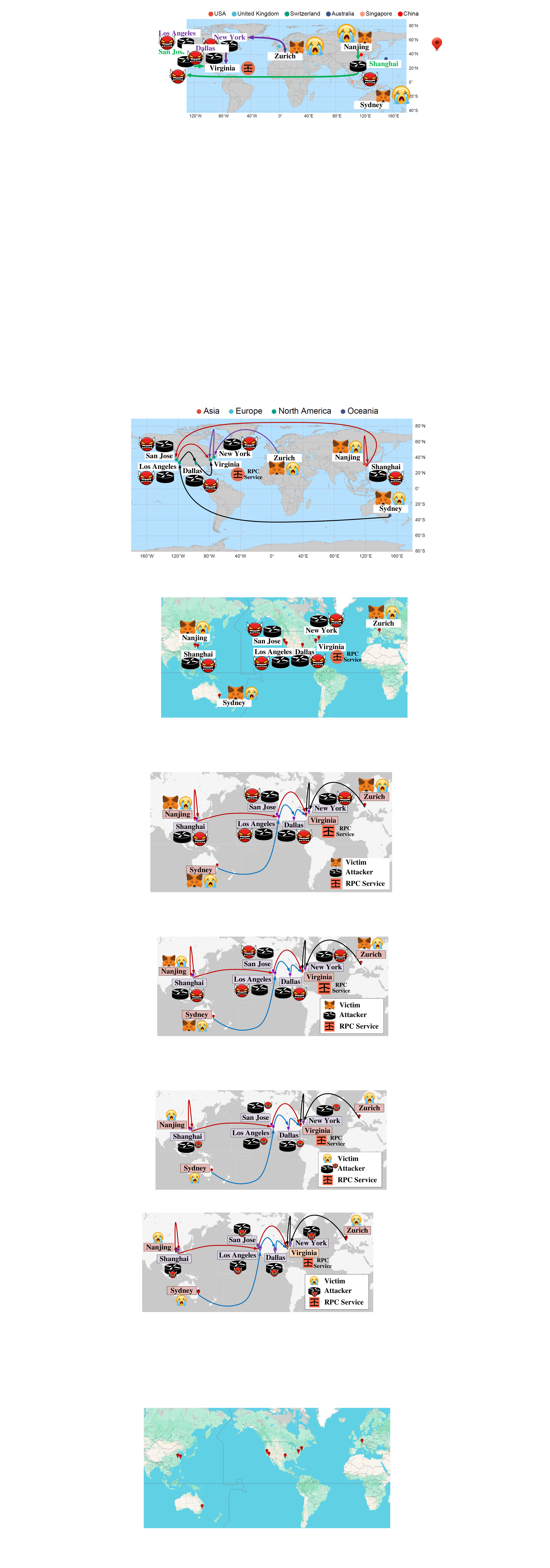}
\vspace{-7mm}
\caption{Attackers and Victims' Geographical Distribution}
\label{fig::attackRoute} 
\vspace{-5mm}
\end{figure}

% \vspace{1mm}
% \noindent
\textbf{Victim user settings.} 
As illustrated in \autoref{fig::attackRoute}, 90 victim users are located in three regions, including {\em Sydney}, {\em Zurich} and {\em Nanjing} respectively, with 30 users in each region.
To simulate realistic transaction behaviors, we randomly select the $90$ normal users (inactive users) from the ledger, and the 90 victim users replicate their transaction patterns, including transaction timings, intervals and frequency as recorded in the ledger. 
During a two-week attack period, these $90$ victims sent $1099$ transactions in total.

\textbf{Attacker settings.}
For Ethereum and its wallet {\em MetaMask}, the RPC service {\em Infura} is hosted in Virginia, USA.
The routing paths from victims in different regions to {\em Infura} vary widely, spanning the globe.
To emulate this scenario, we artificially deploy software routers on cloud servers at various geographic locations along real-world routes identified using traceroute-like tools (e.g., NextTrace), as illustrated in Figure \ref{fig::attackRoute}.
We further employ the Point-to-Point Tunneling Protocol (PPTP) to ensure RPC traffic passes through our routers without modifying TCP packet characteristics.
Each router runs Ubuntu 24.04 with 1 vCPU and 2GB of memory. Captured traffic is analyzed offline using a machine with a 12th Gen Intel Core i5-12500H (3.10GHz) and 32GB of RAM.
This setup aims to demonstrate that network delay has minimal impact on deanonymization accuracy, regardless of victim or malicious router location. 
It is not intended to show how an adversary identifies or controls routing paths. 
Note that, we also set up attackers at the victims' locations.
% To reflect real-world attack scenarios, we deploy attackers along these routing paths as illustrated in \autoref{fig::attackRoute}. 
% \red{Note that, these are real routing paths obtained using traceroute-like tools.}
% We set cloud severs in different geographic locations as \red{software routers}.
% % network traffic relays.
% \red{These software routers run Ubuntu 24.04 and are configured with 1 vCPU and 2GB of memory.}
% Through PPTP (Point-to-Point Tunneling Protocol), {\em iproute} and {\em iptables}, the network traffic between victims and the RPC service goes through these \red{software routers}.
% \red{Note that, the PPTP protocol does not alter the features of TCP packets, allowing us to collect sufficient data to evaluate attack effectiveness versus varying $m$ and attack locations.}
% The attackers control these relays and monitor victim traffic. 

\textbf{Network data collection and processing.} 
The attackers employ the Linux packet analyzer tool {\em tcpdump} to capture network traffic going through the controlled relays.
A {\em Python} library, {\em mitmproxy}, operating in the transparent mode, is used to execute a man-in-the-middle attack,
enabling us to log TLS keys for decrypting TCP packets and label them with their corresponding RPC APIs, which serve as ground truth. Note: actual attacks work on encrypted traffic.
During a two-week period of attacks against the 1099 transactions from the 90 victims, we collected in total of \textbf{26.70 GB} of network traffic related to RPC services.\looseness=-1
% among victims and the RPC service.

\textbf{Generalization to real world Attack.}
Although our experiments are not conducted through real-world ISPs due to ethical and legal constraints, the primary impact of this limitation is restricted to missed transaction observations
(see \S\ref{sec::limitation} for details). Nonetheless, our results based on real traffic and transaction data can be generalized to reflect real-world attack effectiveness, 
given sufficient transaction observations per target IP.

% \textbf{Victim Settings.} 

\vspace{-2mm}
\subsection{Real-World Attack Results}
\label{subsec:realworldresult}
\vspace{-1mm}

%We now present the real-world attack results for each attack step.\looseness=-1

\begin{figure*}[t!]
\begin{minipage}[t]{0.66\columnwidth}
\centering
\includegraphics[width=0.96\textwidth]{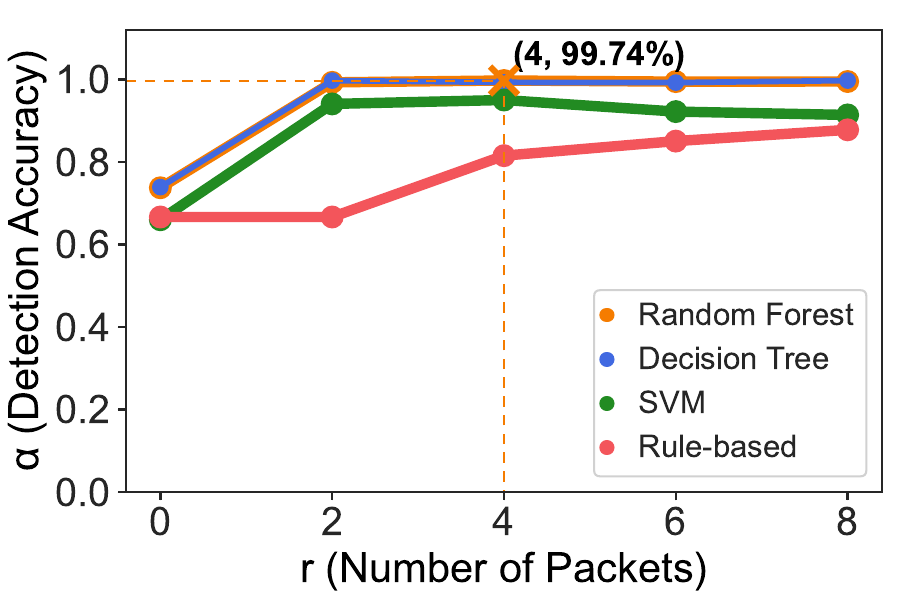}
\vspace{-4mm}
\caption{Detection accuracy of transaction query timestamps in Ethereum}
\label{fig::mlETH} 
\vspace{-2mm}
\end{minipage}
\hspace{0.02cm}
\begin{minipage}[t]{0.66\columnwidth}
\centering
\includegraphics[width=0.96\textwidth]{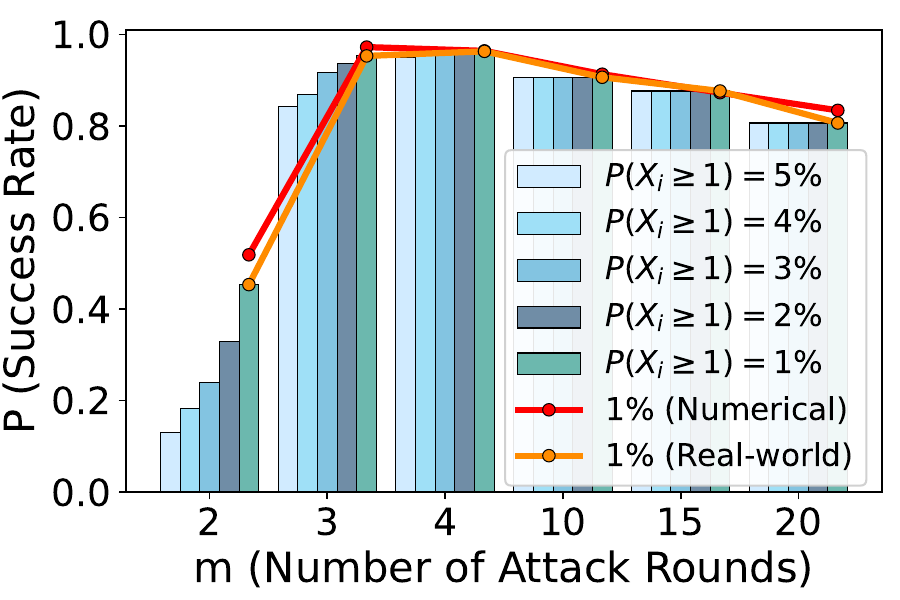}
\vspace{-4mm}
\caption{Success rate of real-world attacks in Ethereum mainnet}
\label{fig::accuracyETHMain} 
\vspace{-2mm}
\end{minipage}
\hspace{0.02cm}
\begin{minipage}[t]{0.66\columnwidth}
\centering
\includegraphics[width=0.96\textwidth]{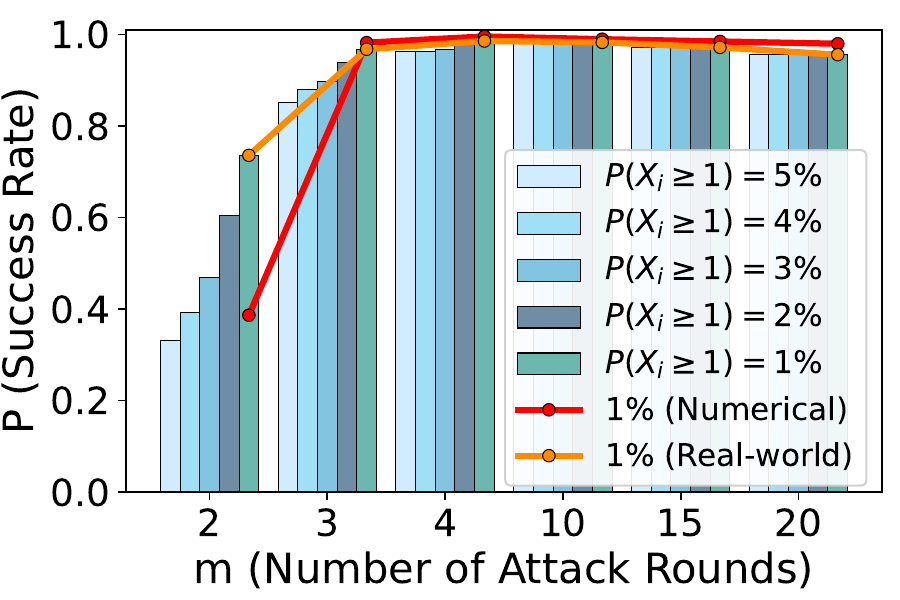}
\vspace{-4mm}
\caption{Success rate of real-world attacks in Ethereum testnet}
\label{fig::accuracyETHTest}  
\end{minipage}
\vspace{-2mm}
\end{figure*}

%\vspace{2mm}
%\noindent
\textbf{Detection accuracy of transaction query timestamp:}  
The attacker trains the machine learning (ML) model to identify the TCP packet of interest and obtain the timestamp $\mathcal{T}_q$. 
The feature vector in Equation (\ref{equ::feature}) is derived from a sequence of TCP packets of length $2r+2$. 
If the middle two packets in the sequence are generated by the RPC API for transaction status query, this sequence is labeled as positive; otherwise, it is labeled as negative.
In Ethereum, the training dataset contains $2000$ positive instances related to target API \textit{eth\_getTransactionReceipt}, $2000$ negative instances related to the noise APIs, and another $2000$ randomly selected negative instances that are not generated by noise APIs but share similar packet sizes.
\autoref{fig::mlETH} shows the training results.
When $r=0$, the accuracy is low, indicating that high accuracy cannot be achieved without incorporating features in packet sequences. 
Using the random forest model with $r=4$, the attacker can achieve a high detection accuracy of $99.74\%$, i.e., high $\alpha$ in Equation (\ref{eqn::Pt}).
Moreover, we define rule-based identification method (in \autoref{sec::rule_comparison}) and compare its effectiveness with ML models. As shown in \autoref{fig::mlETH}, the ML model is better.
\looseness=-1

\begin{figure}[h!]
\vspace{-3mm}
    \centering
    \includegraphics[width=1.0\linewidth]{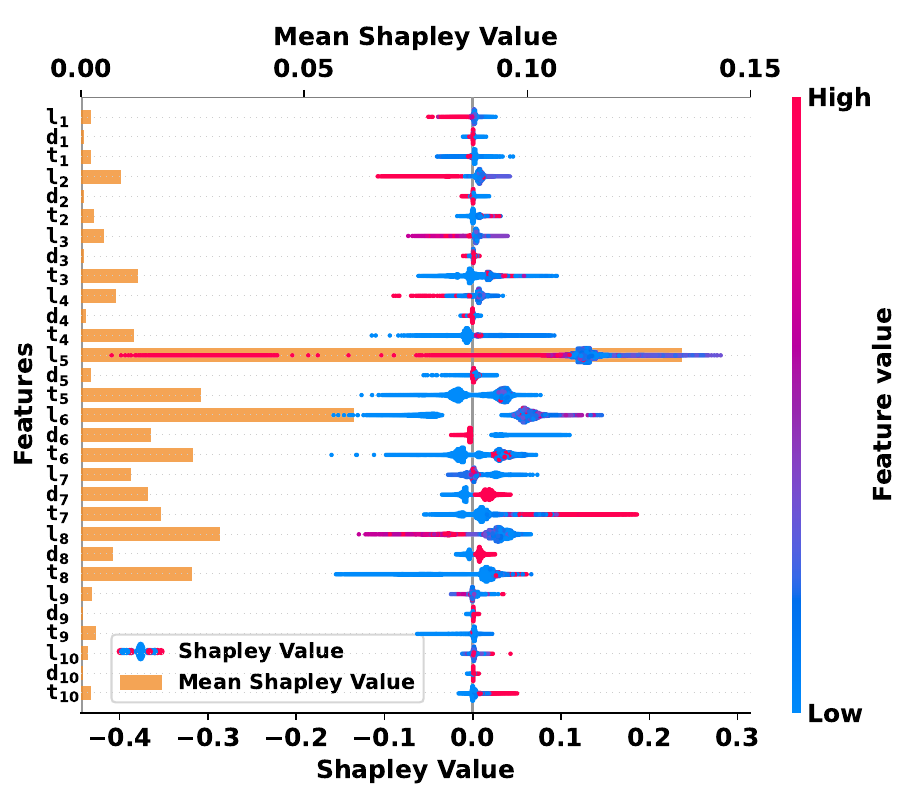}
    \vspace{-4mm}
    \caption{Feature Importance Analysis}
    \label{fig::featureImportance}
     \vspace{-5mm}
\end{figure}

\textbf{Analysis of feature importance}.
Moreover, we utilize Shapley value to evaluate the contribution of each feature to the ML model's prediction.
When $r=4$, the feature vector contains $30$ features in total, with each packet profiled by three features.
Figure \ref{fig::featureImportance} shows the feature importance of each feature. 
It can be observed that, the packet size of the middle packet contributes the most, and the features of its surrounding packets also contribute more. This aligns with our analysis in \S\ref{sec::featureSize} and \S\ref{sec::featureSequence}.

% \usepackage{multirow}
\begin{table}[ht!]
% \vspace{-1mm}
\vspace{-2mm}
\tabcaption{Proportion of target transactions included in $k$ blocks}
\vspace{-2mm}
\setlength\tabcolsep{2.8pt}
\renewcommand{\arraystretch}{1.0}
\scriptsize
\resizebox{\linewidth}{!}{
\begin{tabular}{clccccc}
\hline
\textbf{Blockchain}       & \textbf{Wallet} & \textbf{Periodic Q.} & \textbf{Sub. Notif.} & \textbf{k=1}  & \textbf{k=2}  & \textbf{k=3}  \\ \hline
\multirow{3}{*}{Ethereum} & MetaMask        & 20s                 & $\times$             & 63.30\%       & 99.48\%       & 99.69\%       \\
                          & Enkrypt         & 10s                 & $\times$             & 98.98\%       & 100\%         & 100\%         \\
                          & Taho            & 2s                  & $\times$             & 99.49\%       & 99.49\%       & 100\%         \\ \hdashline[3pt/3pt]
\multirow{3}{*}{Bitcoin}  & Electrum        & $\times$            & $\checkmark$         & 99.85\%       & 100\%         & 100\%         \\
                          & Green           & $\times$            & $\checkmark$         & 100\%         & 100\%         & 100\%         \\
                          & Sparrow         & $\times$            & $\checkmark$         & 99.59\%       & 100\%       & 100\%          \\ \hline
\textbf{Blockchain}       & \textbf{Wallet} & \textbf{Periodic Q.} & \textbf{Sub. Notif.} & \textbf{k=40} & \textbf{k=50} & \textbf{k=60} \\ \hline
\multirow{3}{*}{Solana}   & Torus           & 10s                 & $\times$             & 76.13\%       & 93.62\%       & 99.78\%       \\
                          & Phantom         & 0.5s                & $\times$             & 97.72\%       & 97.72\%       & 98.48\%       \\
                          & Solflare        & 1s                  & $\times$             & 97.91\%       & 97.91\%       & 98.95\%       \\ \hline
\end{tabular}}
\label{table::k}
%\vspace{-2mm}
\end{table}

\textbf{Effectiveness of the $k$-block interval:} 
Upon detecting a transaction query timestamp $\mathcal{T}_q$ from network traffic, the attacker retrieves $k$ consecutive blocks preceding $\mathcal{T}_q$ to capture the target transaction.
\autoref{table::k} presents the proportion of target transactions successfully captured by the $k$ blocks, based on 1000 instances per wallet.
These results validate our estimation of $k$ in \S\ref{subsec::estimationK}, and confirm that a small $k$ is effective.
For example, $99.69\%$ of target transactions are included in the selected $3$ blocks for Ethereum wallet {\em MetaMask}.
Accordingly, we set $k=3$ for Ethereum wallets, $k=2$ for Bitcoin wallets, and $k=60$ for Solana wallets.

%\vspace{2mm}
%\noindent
\textbf{Success rate in real-world attacks:}
Using the trained ML model and the selected parameter $k$, the attacker launches attacks against Ethereum RPC users from the {\em MetaMask} wallet.
These attacks target $1099$ transactions initiated by the $90$ victim users across three regions, from March 23, 2025 to April 05, 2025, on Ethereum testnet.
Additionally, the attacker performs attacks against $100$ transactions on Ethereum mainnet.
Please note that, attack results on other blockchains and wallets are presented in \S\ref{subsec::attackGenerality}.
\autoref{fig::accuracyETHMain} and \autoref{fig::accuracyETHTest} show the results on Ethereum mainnet and testnet with varying $m$ and transacting rate threshold $\theta_{\lambda}$ (determined by setting $P(X_i \geq 1)$ to different values to exclude active users).
We also compare the real-world results with the numerical results in \S\ref{subsec::Expectation} when  $\theta_{\lambda}$ is derived by $P(X_i \geq 1) = 1\%$.
We make the following observations.
(i) Our attacks achieve a high success rate.
With $m=3$ and $P(X_i \geq 1) = 1\%$, the success rate reaches up to $95.33\%$ on Ethereum mainnet and $96.80\%$ on Ethereum testnet.
(ii) The real-world attack results closely align with their mathematical expectations in both trends and values, validating the correctness of our mathematical models of the attack.
We observe deviations between the real-world and numerical results when $m=2$. This is reasonable, as active users are identified over a relatively long period (e.g., one month) in our method, while some normal users may exhibit high transaction intensity within a shorter time frame such as during the attack period. 
This issue can be easily mitigated by performing one more attack, increasing $m$ from 2 to 3.
(iii) A small $m$ can yield a high success rate. This demonstrates that observing a limited number of transactions already allows the attacker to accurately link a user's IP address to its pseudonym.\looseness=-1
% such as just $3$ instances.

% \red{
% \textbf{Impact of Transaction Confirmation Delays.} 
% Transaction confirmation delays do not affect the attack accuracy. This is because the estimate of the transaction confirmation timestamp relies on the status query/response traffic generated 
% % when wallets check the transaction confirmation status 
% \textbf{after} this transaction has been confirmed on-chain, regardless of how long the confirmation takes.
% For unconfirmed transactions or failed transactions,
% the ML model will classifies the corresponding status query/response packets as negative instances, as the responses contain {\em nil} with a small size, preventing false positives. 
% To validate this, victim users send 100 additional transactions per Bitcoin and Ethereum wallet using the minimum self-defined gas price allowed by the wallets.
% Note that, Solana wallets do not allow user-defined gas price.
% Those transactions would have longer-than-normal confirmation delays.
% % For example, the average confirmation delay of the Ethereum transaction via {\em MetaMask} is around 20.40 seconds.
% Despite this, the corresponding ML-based timestamp detection accuracy and the deanonymization success rate remain high, as shown in Figure \ref{fig::ML_lowFee}, Figure \ref{fig::successLowFee} and Table \ref{table::attackMultiWallet}.
% }

\textbf{Attack results against active users:}
We also carry out attacks against active users by configuring the victims to send $1000$ transactions at a rate exceeding the defined threshold.
In this case, the attacker links an IP address to a small set of pseudonyms, rather than a single one.
On average, this set contains around $17$ pseudonyms, with a probability $99.60\%$ of including the target, when $m=3$.\looseness=-1

%\vspace{2mm}
%\noindent

% \begin{table}[h!]
% \tabcaption{Success rate of \textsf{TRAP} attack across different wallets}
% \label{table::attackMultiWallet}
% \vspace{-3mm}
% % \vspace{2mm}
% \setlength\tabcolsep{2.0pt}
% \renewcommand{\arraystretch}{1.1}
% \resizebox{\linewidth}{!}{
% \begin{tabular}{c|ccc|ccc|ccc}
% \hline
% Blockchain  & \multicolumn{3}{c|}{\textbf{Ethereum}}               & \multicolumn{3}{c|}{\textbf{Bitcoin}}                 & \multicolumn{3}{c}{\textbf{Solana}}                   \\ \hdashline[2pt/2pt]
% Wallet      & \textbf{MetaMask} & \textbf{Enkrypt} & \textbf{Taho} & \textbf{Electrum} & \textbf{Green} & \textbf{Sparrow} & \textbf{Torus} & \textbf{Phantom} & \textbf{Solflare} \\ \hline
% Default-fee & 96.80\%           & 95.83\%          & 94.40\%       & 97.70\%           & 96.85\%        & 97.95\%          & 96.58\%        & 96.42\%          & 95.62\%           \\ 
% Low-fee     & 95.05\%           & 95.36\%          &  94.10\%       & 96.21\%           & 95.89\%        & 97.68\%          & N/A            & N/A              & N/A               \\ \hline
% \end{tabular}
% }
% \end{table}

\begin{table}[h!]
\vspace{-2mm}
\tabcaption{Success rate of \sys attack across different wallets}
\label{table::attackMultiWallet}
\vspace{-4mm}
\setlength\tabcolsep{2.0pt}
\renewcommand{\arraystretch}{1.1}
% \scriptsize
\resizebox{\linewidth}{!}{
\begin{tabular}{ccc|ccc|ccc}
\hline
\multicolumn{3}{c|}{\textbf{Ethereum}}               & \multicolumn{3}{c|}{\textbf{Bitcoin}}                 & \multicolumn{3}{c}{\textbf{Solana}}                   \\ \hdashline[2pt/2pt]
\textbf{MetaMask} & \textbf{Enkrypt} & \textbf{Taho} & \textbf{Electrum} & \textbf{Green} & \textbf{Sparrow} & \textbf{Torus} & \textbf{Phantom} & \textbf{Solflare} \\ \hline
96.80\%               & 95.83\%          & 94.40\%           & 97.70\%               & 96.85\%        & 97.95\%              & 96.58\%            & 96.42\%          & 95.62\%           \\ \hline
\end{tabular}
}
\vspace{-3mm}
\end{table}

\vspace{-2mm}
\subsection{Attack Generality}
\label{subsec::attackGenerality}

%We discuss the generality of our theoretical models and deanonymization attack methods across multiple dimensions.\looseness=-1

\textbf{Attack generality across different blockchains:}  
The attack results in Bitcoin and Solana are shown in 
Figure 58 and Figure 59
% \autoref{fig::accuracyBTCtest} and \autoref{fig::accuracySOLtest} 
% in the extended version of the paper,
in Appendix, 
exhibiting high success rate and alignment with numerical results.
When $m = 4$ and $\theta_\lambda$ is defined by $P(X_i \geq 1) = 1\%$, the success rate can reach up to $97.70\%$ in Bitcoin testnet.
% m=3  $97.45\% m=4  $97.70\%$
When $m = 4$ and $\theta_\lambda$ is defined by $P(X_i \geq 1) = 5\%$, the success rate can reach up to $96.58\%$ in Solana testnet.
% m=3  $96.23\%   m=4  $96.58\%$
Although we do not evaluate real-world attacks in Bitcoin mainnet and Solana mainnet, the numerical results in \autoref{table::expectationP} can suggest that a high success rate can be achieved in these two mainnets.\looseness=-1

\textbf{Attack generality across different wallets:}
\autoref{table::attackMultiWallet} shows that our deanonymization attack consistently achieves high success rates across different wallets. 
This is reasonable as our attack exploits universal vulnerabilities across applications and platforms that follow similar RPC-based communication paradigm.
Notably, the attack is effective against both open-source and closed-source wallets (e.g., {\em Solflare} and {\em Phantom}),
as the attacker can infer wallet behaviors via network monitoring, regardless of code transparency.

\begin{figure}[h!]
\vspace{-1mm}
\centering
\includegraphics[width=1.0\columnwidth]{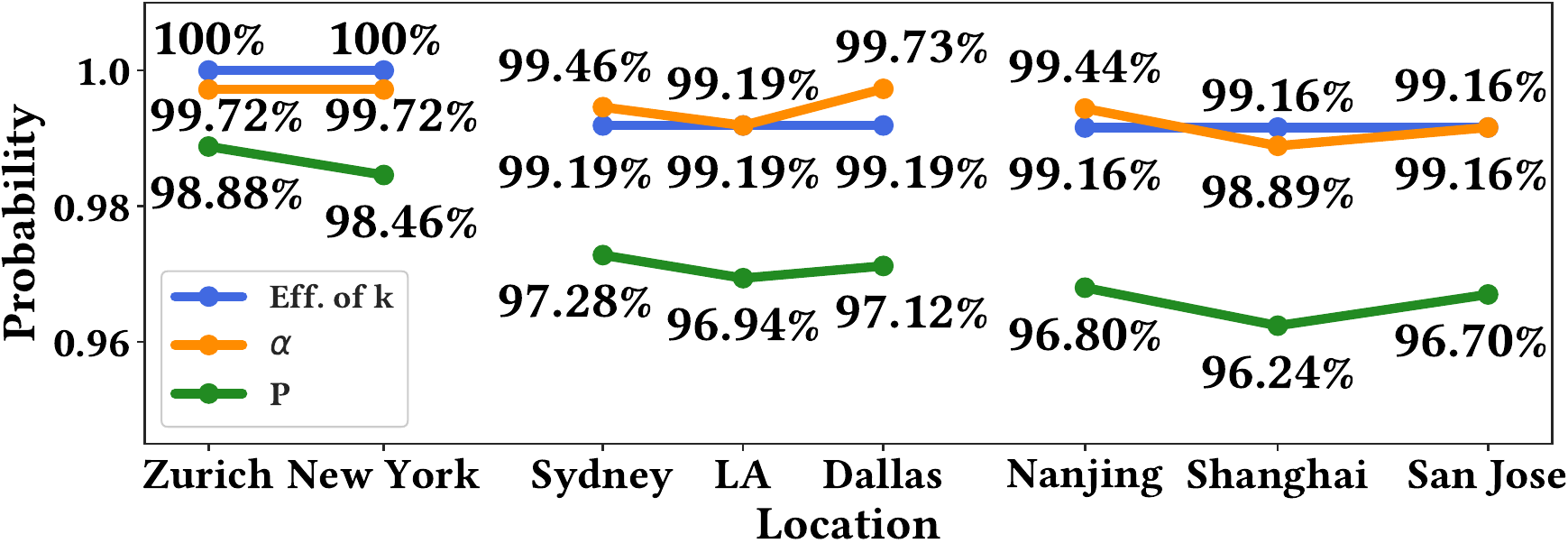}
\vspace{-7mm}
\caption{Attack effectiveness vs. locations of attackers}
\label{table::accuracyLocation}
\vspace{-3mm}
\end{figure}

\textbf{Attack generality across different attacker locations:}
Our experiments cover multiple scenarios in terms of locations of the attacker and victim: in the same city; in different cities but within the same country; in different countries.
As presented in \autoref{table::accuracyLocation}, against Ethereum testnet, the effectiveness of the selected $k=3$ blocks that capture the target transaction, the timestamp detection accuracy $\alpha$ and the deanonymization success rate $P$ remain consistently high at different attack locations, whether across cities or countries along the routing paths. 
This is reasonable because network packets exchanged between victim users and the RPC service typically preserve consistent sizes and sequences along the routing path at the TCP layer.
% Moreover, variations in packet intervals across different locations are ignorable, as network delays typically stay within 400 ms \cite{RTT}, which is small compared to the block time.

% This also demonstrate that out attack resistant to daily network delays, although they may introduce slight variations in packet intervals across different attack locations. 

% Notably, our experiments cover multiple scenarios, where the attacker locates in the same city as the victim, in different cities but within the same country, and in different countries altogether.
% These results demonstrate the widespread impact of our attack, as attackers from around the world may compromise blockchain user anonymity.

% \input{samples/sections/sec7.2_discussion}
\vspace{-2mm}
\section{Limitations in the Experimental Evaluation}
\label{sec::limitation}

Our experiments use artificially positioned users and emulated software routers deployed at various locations to collect data.
However, real-world network conditions may differ significantly from our experimental setup, potentially resulting in insufficient transaction observations (i.e., a small $m$), or failed deanonymization attempt, as discussed below:
% which can reduce the effectiveness of the attack, as discussed below:

\textbf{Network jitters and packet loss.} 
Our attack method is robust to normal network jitters, as discussed in \S\ref{sec::discussion}. However, if the network jitter is extremely high, deanonymization may fail due to inaccurate estimation of the transaction confirmation timestamp based on the query response timestamp.

\textbf{Dynamic IPs.} 
The IP of a wallet user may change over time, particularly for mobile users, and an attacker may not observe enough (fewer than 3 or 4) transactions from a specific IP. 
As a result, the success rate will decrease. 
% For example, as shown in Figure 16 in the paper, the success rate is only about $73.60\%$ when $m=2$ in the Ethereum testnet. 
If an attacker observes only 1 transaction, they can only link the IP to a set of pseudonyms within the selected $k$ blocks, rather than a single pseudonym.

\textbf{Geographic routing diversity.}
Geographic routing diversity can introduce variations in network jitters. It may also lead to partial traffic visibility. For instance, traffic from a particular IP address may not consistently traverse a malicious router or ISP, causing the attacker to miss some transaction observations. If the attacker fails to capture a sufficient number of transactions, the effectiveness of the attack may degrade significantly.
\looseness=-1

\textbf{NAT.} 
Users behind NAT have distinct private IP addresses but may share the same public IP. If the NAT router is the adversary, our attack remains effective. We now consider scenarios where the adversary operates on the Internet, external to the NAT. 
In this case, especially in the enterprise environments, different users may share a single public IP while
using different pseudonyms. 
If the observed transactions originate from multiple users behind the same NAT, the intersection step in our attack yields an empty set, causing the deanonymization attempt to fail.
\looseness=-1

\textbf{Feasibility of persistent network visibility.}
The hierarchical architecture of the global Internet ensures that attackers located at certain positions, such as a border router or an exchange point, could persistently observe sufficient traffic to deanonymize blockchain RPC users.
A border router of a local/regional ISP or an enterprise network that connects the wallet users to the Internet can persistently monitor their traffic and track user IPs.
%, making the attack feasible and robust against dynamic IPs and/or NAT. 
A border router of a local/regional ISP that serves the RPC server can monitor RPC traffic from users at scale, although dynamic IPs and NAT may reduce the attack effectiveness. 
For a national/global ISP or a government interested in surveilling its citizens, they can monitor border routers that connect the national network to the global Internet. Those border routers typically have stable visibility into traffic of users within their service scope.

% \red{
% \textbf{Generalization to Real World Attack.}
% Based on the above analysis of realistic network conditions, the primary impact of deviations from our experimental setting is limited or missed transaction observations. Our experimental results based on real traffic and transaction data can be generalized to real-world attack effectiveness given the number of observed transactions (i.e., the value of $m$) targeting an IP.
% }

\vspace{-2mm}
\section{Discussion}
\label{sec::discussion}

% \vspace{1mm}
% \noindent
\textbf{Countermeasures}. 
Defeating the \sys attack proposed in this paper by enhancing RPC services and wallets is challenging, as it requires protocol-level adjustments and coordinated actions between RPC services and wallets.
While potential mitigation like packet padding and random delays could help, they face significant challenges considering long-term security, practical integration, and user experience. 
First, although padding can obscure the size features, attackers may exploit the packet frequency to infer sensitive transaction behaviors. 
Second, the varying sizes of current RPC data make standardizing their size difficult, as it may disrupt the established data flow and introduce compatibility issues.
Third, although introducing delays can obscure the temporal features, there is a tradeoff between delays and user experience. 
% Moderate delays may prove ineffective, as attackers could adapt their observation strategies over time.
We leave the details of countermeasures as our future work.

\textbf{IP-masking tools for defense.} 
Using IP-masking tools like Tor/VPN can enhance wallet user anonymity and may defeat our attack.
In Tor, wallet traffic passes through an entry node, a middle relay, and an exit node before reaching the RPC service. At the exit, Tor encryption is removed while RPC's TLS encryption remains, allowing adversaries to observe traffic similar to what we study. However, grand challenges remain: (i) exit nodes change over time and vary by location, (ii) multiple users may share the same exit, and (iii) advanced attacks like confirmation attacks \cite{lopes2024flow,ling2019novel}
are needed to reveal the user’s true IP even if the exit IP is linked to a pseudonym using our method. 
Monitoring traffic at the entry node is also limited by Tor’s layered encryption and fixed-size cells, though traffic analysis/fingerprinting techniques \cite{mitseva2024stop,li2022foap,oh2023appsniffer} may offer insights.
VPNs can be analyzed similarly.
% using a single-hop circuit and requiring trust in the provider.
How to deanonymize blockchain RPC users protected by Tor/VPN remains an open problem.\looseness=-1

\textbf{Impact of transaction confirmation delays.} 
Transaction confirmation delays have no impact on the effectiveness of the attack.
% Transaction confirmation delays do not affect the attack accuracy. 
This is because the estimate of the transaction confirmation timestamp relies on the status query/response traffic generated 
\textbf{after} this transaction has been confirmed on-chain, regardless of how long the confirmation takes.
For unconfirmed transactions or failed transactions,
the ML model will classify the corresponding status query/response packets as negative instances, as the responses contain {\em nil} with a small size, preventing false positives.

\textbf{Robustness to network latency.} 
Our method is robust to network latency in the blockchain P2P network and in the interactions between wallets and RPC services.
These network delays are typically within hundreds of milliseconds \cite{RTT} and has been accounted for by the parameter block time $z$ and the $RTT$ in Equations (2) and (3), which are used to theoretically estimate the interval $I_{c,q}$ measured in $k$ blocks.
To account for potential real-world fluctuations, we use a slightly larger $k$ than the theoretical estimate in the attack. 
The search window spans $k$ blocks (e.g., around $36$ seconds in Ethereum), far exceeding the typical network latency in practice.
Moreover, our dataset is collected from geographically distributed routers and consists of thousands of transactions, covering typical network jitters.
%The consistent results across attack locations in 
\autoref{table::accuracyLocation} demonstrates that the network latency has little impact on our attack.

% \red{
% \textbf{Computational cost of identifying and controlling the routing path.} 
% The domain names and IPs of a RPC service is either publicly known or can be obtained by inspecting network packets. 
% Given the IPs of RPC services and wallet users of interest, it is lightweight for an router owner and an ISP to log the RPC traffic that passes through routers under their control.
% For example, in our experiments, a server capability of only 1 vCPU, 2GB RAM can log the traffic of 30 wallet users efficiently.
% Regarding routing control, we emphasize that we do not assume attackers capability of routing control, as the RPC traffic naturally and consistently passes through access routers/ISPs of wallet users and RPC servers. 
% Specific transit routers/ISPs such as border gateways and their operators may also reliably observe the RPC traffic.
% }

% \vspace{-3mm}
% \vspace{-1mm}
\section{Related Work}
% \vspace{-1mm}

In this section, we introduce the related work on deanonymization attacks against blockchains.

% \subsection{Clustering Blockchain Addresses}
% \vspace{1mm}
%\noindent
\textbf{Clustering Blockchain Addresses:} 
Many existing attacks focus on clustering blockchain addresses that belong to the same user or exhibit similar behaviors, by analyzing extensive ledger data such as topological information from transaction flow graphs \cite{zhao2015graph, zheng2020identifying, remy2018tracking, meiklejohn2013fistful,beres2021blockchain,linoy2019exploring, chen2020understanding,dupont2015toward,mastan2018new}.
However, this type of attack cannot reveal the real-world identity of blockchain users.

% and the topological information in transaction flow graphs have been commonly leveraged to group different addresses of a single user \cite{zhao2015graph, zheng2020identifying, remy2018tracking, meiklejohn2013fistful}.
% External public information from off-chain sources such as social media can aid in clustering addresses with similar behaviors or validating the clustering results \cite{liao2016behind,ermilov2017automatic}.
% In addition, the transaction timestamps can be used to group addresses in the same time zone \cite{dupont2015toward}, and network protocols can be exploited to cluster network sessions \cite{mastan2018new}.
% or various types of IP addresses \cite{yang2023evicting} related to a blockchain node.  
% In Ethereum, information from smart contracts can be instrumental in grouping a developer's addresses \cite{linoy2019exploring, chen2020understanding}. 
% Moreover, the transaction graph features can be employed to profile and cluster addresses \cite{beres2021blockchain}. 

% \subsection{Identifying Transaction Source Node}
% \vspace{1mm}
%\noindent
\textbf{Identifying Transaction Source Node:} 
% In such methods,
% %attempt to link a transaction to its source node, which 
% a firstly broadcasts this transaction to the blockchain network.  
These attacks typically leverage the features of transaction propagation order and timing to define heuristic rules or train a machine learning model to identify the source node \cite{biryukov2019deanonymization,biryukov2019transaction,gao2021practical,koshy2014analysis,shen2020transaction,zheng2023ledger}. 
To gather sufficient propagation data and features, they often rely on a super node that maintains thousands of connections to nearly all nodes in the target blockchain, which may not be practical.
%limits their practicality although such access is unrestricted. 
Some work is only effective in particular scenarios \cite{koshy2014analysis, biryukov2019deanonymization, wallace2020can}, such as targeting malformed transactions.
% \cite{koshy2014analysis}, newly added nodes \cite{biryukov2019deanonymization}, or nodes running in software-defined network (SDN) environment \cite{wallace2020can}, 
% limiting their generality.
% In addition, 
% this type of attack only can link transaction with blockchain nodes.
% this type of attack is primarily designed for identifying real-world identities of users who interact with a blockchain network by running a blockchain node. 
%In the context of RPC services, users access the blockchain network through RPC services as intermediaries. 
These methods may link a blockchain address to the IP address of a node belonging to a RPC provider, not the user.
%, rather than the IP address of a user's personal device.
In contrast, our method is general against blockchain RPC users and does not need a super node, offering a practical solution.

% Some other work exploits network attacks to deanonymize bitcoin nodes, but usually has a strong threat assumption. 
% Apostolaki et al. \cite{apostolaki2021perimeter} assume the attacker can control a Autonomous Systems (AS). It first collects all transactions from the target IP address as a potential set. If a transaction is created by the target node, this node will not request this raw transaction from other nodes, while other nodes will more likely request this raw transaction from the target through a GETDATA request. Upon this feature, the attacker can further exclude false result from the potential set and get the final correct results.

% \subsection{Deanonymizing Transaction Initiators}
% \vspace{1mm}
%\noindent
\textbf{Deanonymizing Lightweight Clients:} 
A few deanonymization attacks aim to reveal the real-world identity of a blockchain lightweight client via correlating its IP address with a transaction and the initiator's pseudonym.
Biryukov et al. exploit the address advertisement mechanism and transaction propagation timing from entry nodes to deanonymize Bitcoin lightweight clients \cite{biryukov2014deanonymisation} or users accessing Bitcoin via {\em Tor} \cite{biryukov2015bitcoin}.
However, these vulnerabilities have been addressed \cite{fanti2017deanonymization, getaddrRevision}.
Furthermore, a research has shown that these attack methods in Bitcoin do not apply to Ethereum \cite{klusman2018deanonymisation}.

% \subsection{Attack Surfaces Introduced by Blockchain RPC Services}
% \vspace{1mm}
%\vspace{2mm}
%\noindent
% \vspace{1mm}
\textbf{Attack Surfaces Introduced by Blockchain RPC Services:} 
% The potential attack surfaces incurred by integrating RPC protocols and services into the blockchains have not been thoroughly studied, especially risks to user anonymity.
Only a few studies focus on security issues related to blockchain RPC services, such as DoS attacks \cite{li2021strong, li2021deter}, cryptocurrency stealing \cite{cheng2019towards}, passphrase extraction attacks \cite{wang2018attack}, behavior analysis of malicious users \cite{hara2020profiling}, and handling non-deterministic events and invalid arguments \cite{kim2023etherdiffer}.
We highlight the risks of using public RPC services on user anonymity under
passive network attacks that incur zero transaction fee.\looseness=-1
% which is effective across different mainstream blockchains with zero attack cost.

\textbf{Deanonymizing Blockchain RPC Users:}
There is only one study dedicated to deanonymizing Ethereum RPC users \cite{wang2024deanonymizing}. 
It is an active attack that injects transactions into the Ethereum network via the same router as the victim. It incurs transaction fees and falls short of exposing widespread
flaws in user anonymity across other blockchain networks.
% is not generalized to other blockchains.
In contrast, our methods incur zero transaction fees under passive network attacks without traffic injection, and are applicable to various mainstream blockchains.
In addition, as Solana is relatively new blockchain \cite{yakovenko2018solana,pierro2022can},
only a few work discusses its security problems such as detecting its smart contract vulnerabilities \cite{cui2022vrust, smolka2023fuzz}.
We are pioneering the investigation of risks to Solana's user anonymity.

\textbf{Comparison with Web2 IP deanonymization}.
Both our attack and Web2 IP deanonymization involve deanonymization techniques, but their goals and techniques differ.
In conventional Web2 IP deanonymization, a malicious website observes a user's IP address and seeks to identify the user identity and their physical location \cite{dan2021ip}.
For clarity, we define extended Web2 IP deanonymization as the process of correlating multiple online accounts to the same user, often through shared IPs and behavioral patterns, for purposes such as targeted advertising \cite{wu2021my}.
In contrast, the adversary in our attack observes a blockchain pseudonym in the public ledger and seeks to identify the IP address responsible for the corresponding transactions. If physical location information is needed, established conventional Web2 IP deanonymization techniques can then be applied to the recovered IP.
Due to different application contexts and objectives, the employed attack methods vary.

% A few work focuses the security problems exploiting RPC services, such as DoS attacks \cite{li2021strong,li2021deter}, currency stealing attack \cite{cheng2019towards}, passphrase-extraction attack \cite{wang2018attack}, behavior analysis of malicious users \cite{hara2020profiling} and so on.
% We are the first to exploit the third-party RPC service to deanonymize Ethereum users.

% A blog post \cite{AnonymousBlog1} discusses that RPC providers collect users' IP addresses and Ethereum addresses, which violates anonymity. However, our thread model does not assume that the attacker has privileges like an RPC provider.
% We demonstrate that an attacker who can only monitor network data from a gateway, that a victim user connects to, can successfully deanonymize Ethereum users, under the assumption of benign RPC providers.

% Li, et al \cite{li2021strong,li2021deter} propose some DoS attacks against the third-party RPC services. 
% Cheng, et al \cite{cheng2019towards} study a currency stealing attack due to the unprotected RPC service nodes.
% Wang, et al \cite{wang2018attack} analyse six attacks on RPC APIs, including the passphrase-extraction attack, DoS attack, privacy attack and so on. Each of these attacks has limited effect on Ethereum. 
% Hara, et al \cite{hara2020profiling} deploy honeypots to collect behavior information of malicious users of Ethereum RPC ports, so as to profile attackers.
% \vspace{-2mm}
% \vspace{-2mm}
\section{Conclusion}
% \vspace{-1mm}

In this paper, we propose a novel deanonymization attack named \sys against blockchain RPC users across various blockchain networks.
\sys can accurately link a RPC user's IP address with their blockchain pseudonym, by exploiting the temporal correlation vulnerability in the interactions among wallets, RPC services, and blockchain networks, which collaboratively process a transaction.
\sys only requires a passive network attacker who can monitor network traffic from the victim user and retrieve public ledgers, without incurring transaction fees.
We systematically analyze and explain why our methods are universally applicable across different blockchains and wallets.
Our numerical results and real-world attack results align well, and both consistently demonstrate high deanonymization success rates against normal users.
All results across Ethereum, Bitcoin and Solana can achieve a success rate of over $95\%$.
Although the attacker is assumed to possess network eavesdropping capabilities over encrypted traffic,
the high success rate and zero attack cost (i.e., transaction fee) significantly highlight the risks to blockchain user anonymity posed by blockchain RPC protocols and services.
% although the attacker is assumed to have network eavesdropping capabilities on encrypted traffic.
% Against active users who represent only a small fraction of all blockchain users, 
% (e.g, $0.03\%$ in Ethereum), 
% \sys can link the IP of the monitored user to 17 pseudonym candidates on average.

%which potentially undermine the anti-censorship capability of blockchains.\looseness=-1

% \balance

% \vspace{-2mm}
\section*{Acknowledgment}
% \vspace{-1mm}

This research was supported in part by 
National Key R\&D Program of China (No. 2023YFC3605800), 
National Natural Science Foundation of China (Nos. 62072103, 62232004, 92467205), 
HK RGC Theme-Based Research Scheme (No. T43-513/23-N), 
HK RGC General Research Fund (No. PolyU15220922),
Jiangsu Provincial Key Laboratory of Network and Information Security, 
and Research Institute for Artificial Intelligence of Things, The Hong Kong Polytechnic University.
Any opinions, findings, conclusions, and recommendations in this paper are those of the authors and do not necessarily reflect the views of the funding agencies.\looseness=-1

\bibliographystyle{ACM-Reference-Format}
\bibliography{reference}


\begin{thebibliography}{48}

%%% the Web address.

\ifx \showCODEN    \undefined \def \showCODEN     #1{\unskip}     \fi
\ifx \showISBNx    \undefined \def \showISBNx     #1{\unskip}     \fi
\ifx \showISBNxiii \undefined \def \showISBNxiii  #1{\unskip}     \fi
\ifx \showISSN     \undefined \def \showISSN      #1{\unskip}     \fi
\ifx \showLCCN     \undefined \def \showLCCN      #1{\unskip}     \fi
\ifx \shownote     \undefined \def \shownote      #1{#1}          \fi
\ifx \showarticletitle \undefined \def \showarticletitle #1{#1}   \fi
\ifx \showURL      \undefined \def \showURL       {\relax}        \fi
% invisible to TeX
\providecommand\bibfield[2]{#2}
\providecommand\bibinfo[2]{#2}
\providecommand\natexlab[1]{#1}
\providecommand\showeprint[2][]{arXiv:#2}

\bibitem[AWS(2025)]%
        {RTT}
\bibfield{author}{\bibinfo{person}{AWS}.} \bibinfo{year}{2025}\natexlab{}.
\newblock \bibinfo{title}{What is RTT in Networking?}
\newblock
\urldef\tempurl%
\url{https://aws.amazon.com/what-is/rtt-in-networking/#:~:text=A%20good%20round%2Dtrip%20time,able%20to%20access%20the%20service.}
\showURL{%
\tempurl}


\bibitem[Bailey et~al\mbox{.}(2012)]%
        {bailey2012menlo}
\bibfield{author}{\bibinfo{person}{Michael Bailey}, \bibinfo{person}{David Dittrich}, \bibinfo{person}{Erin Kenneally}, {and} \bibinfo{person}{Doug Maughan}.} \bibinfo{year}{2012}\natexlab{}.
\newblock \showarticletitle{The menlo report}.
\newblock \bibinfo{journal}{\emph{IEEE Security \& Privacy}} \bibinfo{volume}{10}, \bibinfo{number}{2} (\bibinfo{year}{2012}), \bibinfo{pages}{71--75}.
\newblock


\bibitem[B{\'e}res et~al\mbox{.}(2021)]%
        {beres2021blockchain}
\bibfield{author}{\bibinfo{person}{Ferenc B{\'e}res}, \bibinfo{person}{Istv{\'a}n~A Seres}, \bibinfo{person}{Andr{\'a}s~A Bencz{\'u}r}, {and} \bibinfo{person}{Mikerah Quintyne-Collins}.} \bibinfo{year}{2021}\natexlab{}.
\newblock \showarticletitle{Blockchain is watching you: Profiling and deanonymizing ethereum users}. In \bibinfo{booktitle}{\emph{2021 IEEE International Conference on Decentralized Applications and Infrastructures (DAPPS)}}. IEEE, \bibinfo{pages}{69--78}.
\newblock


\bibitem[Biryukov et~al\mbox{.}(2014)]%
        {biryukov2014deanonymisation}
\bibfield{author}{\bibinfo{person}{Alex Biryukov}, \bibinfo{person}{Dmitry Khovratovich}, {and} \bibinfo{person}{Ivan Pustogarov}.} \bibinfo{year}{2014}\natexlab{}.
\newblock \showarticletitle{Deanonymisation of clients in bitcoin p2p network}. In \bibinfo{booktitle}{\emph{ACM SIGSAC conference on computer and communications security}}. \bibinfo{pages}{15--29}.
\newblock


\bibitem[Biryukov and Pustogarov(2015)]%
        {biryukov2015bitcoin}
\bibfield{author}{\bibinfo{person}{Alex Biryukov} {and} \bibinfo{person}{Ivan Pustogarov}.} \bibinfo{year}{2015}\natexlab{}.
\newblock \showarticletitle{Bitcoin over Tor isn't a good idea}. In \bibinfo{booktitle}{\emph{IEEE Symposium on Security and Privacy}}. \bibinfo{pages}{122--134}.
\newblock


\bibitem[Biryukov and Tikhomirov(2019a)]%
        {biryukov2019deanonymization}
\bibfield{author}{\bibinfo{person}{Alex Biryukov} {and} \bibinfo{person}{Sergei Tikhomirov}.} \bibinfo{year}{2019}\natexlab{a}.
\newblock \showarticletitle{Deanonymization and linkability of cryptocurrency transactions based on network analysis}. In \bibinfo{booktitle}{\emph{IEEE European symposium on security and privacy (EuroS\&P)}}.
\newblock


\bibitem[Biryukov and Tikhomirov(2019b)]%
        {biryukov2019transaction}
\bibfield{author}{\bibinfo{person}{Alex Biryukov} {and} \bibinfo{person}{Sergei Tikhomirov}.} \bibinfo{year}{2019}\natexlab{b}.
\newblock \showarticletitle{Transaction clustering using network traffic analysis for bitcoin and derived blockchains}. In \bibinfo{booktitle}{\emph{IEEE INFOCOM 2019-IEEE Conference on Computer Communications Workshops (INFOCOM WKSHPS)}}. IEEE, \bibinfo{pages}{204--209}.
\newblock


\bibitem[Bitcoin(2015)]%
        {getaddrRevision}
\bibfield{author}{\bibinfo{person}{Bitcoin}.} \bibinfo{year}{2015}\natexlab{}.
\newblock \bibinfo{title}{Ignore getaddr messages on Outbound connections}.
\newblock
\urldef\tempurl%
\url{https://github.com/bitcoin/bitcoin/commit/200f29363b06cbe8e35dfc5e1b818a0dd96c8474}
\showURL{%
\tempurl}


\bibitem[Chen et~al\mbox{.}(2020)]%
        {chen2020understanding}
\bibfield{author}{\bibinfo{person}{Ting Chen}, \bibinfo{person}{Zihao Li}, \bibinfo{person}{Yuxiao Zhu}, \bibinfo{person}{Jiachi Chen}, \bibinfo{person}{Xiapu Luo}, \bibinfo{person}{John Chi-Shing Lui}, \bibinfo{person}{Xiaodong Lin}, {and} \bibinfo{person}{Xiaosong Zhang}.} \bibinfo{year}{2020}\natexlab{}.
\newblock \showarticletitle{Understanding ethereum via graph analysis}.
\newblock \bibinfo{journal}{\emph{ACM Transactions on Internet Technology (TOIT)}}  \bibinfo{volume}{20} (\bibinfo{year}{2020}), \bibinfo{pages}{1--32}.
\newblock


\bibitem[Cheng et~al\mbox{.}(2019)]%
        {cheng2019towards}
\bibfield{author}{\bibinfo{person}{Zhen Cheng}, \bibinfo{person}{Xinrui Hou}, \bibinfo{person}{Runhuai Li}, \bibinfo{person}{Yajin Zhou}, \bibinfo{person}{Xiapu Luo}, \bibinfo{person}{Jinku Li}, {and} \bibinfo{person}{Kui Ren}.} \bibinfo{year}{2019}\natexlab{}.
\newblock \showarticletitle{Towards a First Step to Understand the Cryptocurrency Stealing Attack on Ethereum.}. In \bibinfo{booktitle}{\emph{RAID}}, Vol.~\bibinfo{volume}{2019}. \bibinfo{pages}{47--60}.
\newblock


\bibitem[Cui et~al\mbox{.}(2022)]%
        {cui2022vrust}
\bibfield{author}{\bibinfo{person}{Siwei Cui}, \bibinfo{person}{Gang Zhao}, \bibinfo{person}{Yifei Gao}, \bibinfo{person}{Tien Tavu}, {and} \bibinfo{person}{Jeff Huang}.} \bibinfo{year}{2022}\natexlab{}.
\newblock \showarticletitle{VRust: Automated vulnerability detection for solana smart contracts}. In \bibinfo{booktitle}{\emph{Proceedings of the 2022 ACM SIGSAC Conference on Computer and Communications Security}}. \bibinfo{pages}{639--652}.
\newblock


\bibitem[Dan et~al\mbox{.}(2021)]%
        {dan2021ip}
\bibfield{author}{\bibinfo{person}{Ovidiu Dan}, \bibinfo{person}{Vaibhav Parikh}, {and} \bibinfo{person}{Brian~D Davison}.} \bibinfo{year}{2021}\natexlab{}.
\newblock \showarticletitle{IP geolocation through reverse DNS}.
\newblock \bibinfo{journal}{\emph{ACM Transactions on Internet Technology (TOIT)}} \bibinfo{volume}{22}, \bibinfo{number}{1} (\bibinfo{year}{2021}), \bibinfo{pages}{1--29}.
\newblock


\bibitem[DuPont and Squicciarini(2015)]%
        {dupont2015toward}
\bibfield{author}{\bibinfo{person}{Jules DuPont} {and} \bibinfo{person}{Anna~Cinzia Squicciarini}.} \bibinfo{year}{2015}\natexlab{}.
\newblock \showarticletitle{Toward de-anonymizing bitcoin by mapping users location}. In \bibinfo{booktitle}{\emph{Proceedings of the 5th ACM Conference on Data and Application Security and Privacy}}. \bibinfo{pages}{139--141}.
\newblock


\bibitem[Eskandari et~al\mbox{.}(2019)]%
        {eskandari2019sok}
\bibfield{author}{\bibinfo{person}{Shayan Eskandari}, \bibinfo{person}{Seyedehmahsa Moosavi}, {and} \bibinfo{person}{Jeremy Clark}.} \bibinfo{year}{2019}\natexlab{}.
\newblock \showarticletitle{Sok: Transparent dishonesty: front-running attacks on blockchain}. In \bibinfo{booktitle}{\emph{International Conference on Financial Cryptography and Data Security}}. Springer, \bibinfo{pages}{170--189}.
\newblock


\bibitem[Esmail and Fathallah(2012)]%
        {esmail2012physical}
\bibfield{author}{\bibinfo{person}{Maged~Abdullah Esmail} {and} \bibinfo{person}{Habib Fathallah}.} \bibinfo{year}{2012}\natexlab{}.
\newblock \showarticletitle{Physical layer monitoring techniques for TDM-passive optical networks: A survey}.
\newblock \bibinfo{journal}{\emph{IEEE Communications Surveys \& Tutorials}} \bibinfo{volume}{15}, \bibinfo{number}{2} (\bibinfo{year}{2012}), \bibinfo{pages}{943--958}.
\newblock


\bibitem[Fanti and Viswanath(2017)]%
        {fanti2017deanonymization}
\bibfield{author}{\bibinfo{person}{Giulia Fanti} {and} \bibinfo{person}{Pramod Viswanath}.} \bibinfo{year}{2017}\natexlab{}.
\newblock \showarticletitle{Deanonymization in the bitcoin P2P network}.
\newblock \bibinfo{journal}{\emph{Advances in Neural Information Processing Systems}} (\bibinfo{year}{2017}).
\newblock


\bibitem[Gan et~al\mbox{.}(2022)]%
        {gan2022understanding}
\bibfield{author}{\bibinfo{person}{Rundong Gan}, \bibinfo{person}{Le Wang}, \bibinfo{person}{Xiangyu Ruan}, {and} \bibinfo{person}{Xiaodong Lin}.} \bibinfo{year}{2022}\natexlab{}.
\newblock \showarticletitle{Understanding flash-loan-based wash trading}. In \bibinfo{booktitle}{\emph{Proceedings of the 4th ACM Conference on Advances in Financial Technologies}}. \bibinfo{pages}{74--88}.
\newblock


\bibitem[Gao et~al\mbox{.}(2021)]%
        {gao2021practical}
\bibfield{author}{\bibinfo{person}{Yue Gao}, \bibinfo{person}{Jinqiao Shi}, \bibinfo{person}{Xuebin Wang}, \bibinfo{person}{Ruisheng Shi}, \bibinfo{person}{Zelin Yin}, {and} \bibinfo{person}{Yanyan Yang}.} \bibinfo{year}{2021}\natexlab{}.
\newblock \showarticletitle{Practical Deanonymization Attack in Ethereum Based on P2P Network Analysis}. In \bibinfo{booktitle}{\emph{2021 IEEE Intl Conf on Parallel \& Distributed Processing with Applications, Big Data \& Cloud Computing, Sustainable Computing \& Communications, Social Computing \& Networking (ISPA/BDCloud/SocialCom/SustainCom)}}. IEEE, \bibinfo{pages}{1402--1409}.
\newblock


\bibitem[Hara et~al\mbox{.}(2020)]%
        {hara2020profiling}
\bibfield{author}{\bibinfo{person}{Kazuki Hara}, \bibinfo{person}{Teppei Sato}, \bibinfo{person}{Mitsuyoshi Imamura}, {and} \bibinfo{person}{Kazumasa Omote}.} \bibinfo{year}{2020}\natexlab{}.
\newblock \showarticletitle{Profiling of Malicious Users Targeting Ethereum’s RPC Port Using Simple Honeypots}. In \bibinfo{booktitle}{\emph{2020 IEEE International Conference on Blockchain (Blockchain)}}.
\newblock


\bibitem[Heilman et~al\mbox{.}(2015)]%
        {heilman2015eclipse}
\bibfield{author}{\bibinfo{person}{Ethan Heilman}, \bibinfo{person}{Alison Kendler}, \bibinfo{person}{Aviv Zohar}, {and} \bibinfo{person}{Sharon Goldberg}.} \bibinfo{year}{2015}\natexlab{}.
\newblock \showarticletitle{Eclipse attacks on Bitcoin’s peer-to-peer network}. In \bibinfo{booktitle}{\emph{24th USENIX Security Symposium (USENIX Security 15)}}. \bibinfo{pages}{129--144}.
\newblock


\bibitem[Infura(2022)]%
        {infura2}
\bibfield{author}{\bibinfo{person}{Infura}.} \bibinfo{year}{2022}\natexlab{}.
\newblock \bibinfo{title}{How Infura Helps MetaMask Scale at the Speed of Web3 Growth}.
\newblock
\urldef\tempurl%
\url{https://www.infura.io/use-cases/developer-stories/metamask}
\showURL{%
\tempurl}


\bibitem[Kim and Hwang(2023)]%
        {kim2023etherdiffer}
\bibfield{author}{\bibinfo{person}{Shinhae Kim} {and} \bibinfo{person}{Sungjae Hwang}.} \bibinfo{year}{2023}\natexlab{}.
\newblock \showarticletitle{Etherdiffer: Differential testing on rpc services of ethereum nodes}. In \bibinfo{booktitle}{\emph{Proceedings of the 31st ACM Joint European Software Engineering Conference and Symposium on the Foundations of Software Engineering}}. \bibinfo{pages}{1333--1344}.
\newblock


\bibitem[Klusman and Dijkhuizen(2018)]%
        {klusman2018deanonymisation}
\bibfield{author}{\bibinfo{person}{Robin Klusman} {and} \bibinfo{person}{Tim Dijkhuizen}.} \bibinfo{year}{2018}\natexlab{}.
\newblock \bibinfo{title}{Deanonymisation in ethereum using existing methods for bitcoin}.
\newblock


\bibitem[Koshy et~al\mbox{.}(2014)]%
        {koshy2014analysis}
\bibfield{author}{\bibinfo{person}{Philip Koshy}, \bibinfo{person}{Diana Koshy}, {and} \bibinfo{person}{Patrick McDaniel}.} \bibinfo{year}{2014}\natexlab{}.
\newblock \showarticletitle{An analysis of anonymity in bitcoin using p2p network traffic}. In \bibinfo{booktitle}{\emph{Financial Cryptography and Data Security}}. Springer, \bibinfo{pages}{469--485}.
\newblock


\bibitem[Li et~al\mbox{.}(2022)]%
        {li2022foap}
\bibfield{author}{\bibinfo{person}{Jianfeng Li}, \bibinfo{person}{Hao Zhou}, \bibinfo{person}{Shuohan Wu}, \bibinfo{person}{Xiapu Luo}, \bibinfo{person}{Ting Wang}, \bibinfo{person}{Xian Zhan}, {and} \bibinfo{person}{Xiaobo Ma}.} \bibinfo{year}{2022}\natexlab{}.
\newblock \showarticletitle{$\{$FOAP$\}$:$\{$Fine-Grained$\}$$\{$Open-World$\}$ android app fingerprinting}. In \bibinfo{booktitle}{\emph{31st USENIX Security Symposium (USENIX Security 22)}}. \bibinfo{pages}{1579--1596}.
\newblock


\bibitem[Li et~al\mbox{.}(2021a)]%
        {li2021strong}
\bibfield{author}{\bibinfo{person}{Kai Li}, \bibinfo{person}{Jiaqi Chen}, \bibinfo{person}{Xianghong Liu}, \bibinfo{person}{Yuzhe~Richard Tang}, \bibinfo{person}{XiaoFeng Wang}, {and} \bibinfo{person}{Xiapu Luo}.} \bibinfo{year}{2021}\natexlab{a}.
\newblock \showarticletitle{As Strong As Its Weakest Link: How to Break Blockchain DApps at RPC Service.}. In \bibinfo{booktitle}{\emph{NDSS}}.
\newblock


\bibitem[Li et~al\mbox{.}(2021b)]%
        {li2021deter}
\bibfield{author}{\bibinfo{person}{Kai Li}, \bibinfo{person}{Yibo Wang}, {and} \bibinfo{person}{Yuzhe Tang}.} \bibinfo{year}{2021}\natexlab{b}.
\newblock \showarticletitle{Deter: Denial of ethereum txpool services}. In \bibinfo{booktitle}{\emph{Proceedings of the 2021 ACM SIGSAC Conference on Computer and Communications Security}}. \bibinfo{pages}{1645--1667}.
\newblock


\bibitem[Ling et~al\mbox{.}(2019)]%
        {ling2019novel}
\bibfield{author}{\bibinfo{person}{Zhen Ling}, \bibinfo{person}{Junzhou Luo}, \bibinfo{person}{Danni Xu}, \bibinfo{person}{Ming Yang}, {and} \bibinfo{person}{Xinwen Fu}.} \bibinfo{year}{2019}\natexlab{}.
\newblock \showarticletitle{Novel and practical SDN-based traceback technique for malicious traffic over anonymous networks}. In \bibinfo{booktitle}{\emph{IEEE Conference on Computer Communications (INFOCOM)}}. \bibinfo{pages}{1180--1188}.
\newblock


\bibitem[Linoy et~al\mbox{.}(2019)]%
        {linoy2019exploring}
\bibfield{author}{\bibinfo{person}{Shlomi Linoy}, \bibinfo{person}{Natalia Stakhanova}, {and} \bibinfo{person}{Alina Matyukhina}.} \bibinfo{year}{2019}\natexlab{}.
\newblock \showarticletitle{Exploring Ethereum’s blockchain anonymity using smart contract code attribution}. In \bibinfo{booktitle}{\emph{15th International Conference on Network and Service Management (CNSM)}}. IEEE, \bibinfo{pages}{1--9}.
\newblock


\bibitem[Lopes et~al\mbox{.}(2024)]%
        {lopes2024flow}
\bibfield{author}{\bibinfo{person}{Daniela Lopes}, \bibinfo{person}{Jin-Dong Dong}, \bibinfo{person}{Pedro Medeiros}, \bibinfo{person}{Daniel Castro}, \bibinfo{person}{Diogo Barradas}, \bibinfo{person}{Bernardo Portela}, \bibinfo{person}{Joao Vinagre}, \bibinfo{person}{Bernardo Ferreira}, \bibinfo{person}{Nicolas Christin}, {and} \bibinfo{person}{Nuno Santos}.} \bibinfo{year}{2024}\natexlab{}.
\newblock \showarticletitle{Flow Correlation Attacks on Tor Onion Service Sessions with Sliding Subset Sum.}. In \bibinfo{booktitle}{\emph{NDSS}}.
\newblock


\bibitem[Mastan and Paul(2018)]%
        {mastan2018new}
\bibfield{author}{\bibinfo{person}{Indra~Deep Mastan} {and} \bibinfo{person}{Souradyuti Paul}.} \bibinfo{year}{2018}\natexlab{}.
\newblock \showarticletitle{A new approach to deanonymization of unreachable bitcoin nodes}. In \bibinfo{booktitle}{\emph{Cryptology and Network Security: 16th International Conference}}. Springer, \bibinfo{pages}{277--298}.
\newblock


\bibitem[Meiklejohn et~al\mbox{.}(2013)]%
        {meiklejohn2013fistful}
\bibfield{author}{\bibinfo{person}{Sarah Meiklejohn}, \bibinfo{person}{Marjori Pomarole}, \bibinfo{person}{Grant Jordan}, \bibinfo{person}{Kirill Levchenko}, \bibinfo{person}{Damon McCoy}, \bibinfo{person}{Geoffrey~M Voelker}, {and} \bibinfo{person}{Stefan Savage}.} \bibinfo{year}{2013}\natexlab{}.
\newblock \showarticletitle{A fistful of bitcoins: characterizing payments among men with no names}. In \bibinfo{booktitle}{\emph{Proceedings of the 2013 conference on Internet measurement conference}}. \bibinfo{pages}{127--140}.
\newblock


\bibitem[Mitseva and Panchenko(2024)]%
        {mitseva2024stop}
\bibfield{author}{\bibinfo{person}{Asya Mitseva} {and} \bibinfo{person}{Andriy Panchenko}.} \bibinfo{year}{2024}\natexlab{}.
\newblock \showarticletitle{Stop, don't click here anymore: boosting website fingerprinting by considering sets of subpages}. In \bibinfo{booktitle}{\emph{33rd USENIX Security Symposium (USENIX Security 24)}}. \bibinfo{pages}{4139--4156}.
\newblock


\bibitem[Oh et~al\mbox{.}(2023)]%
        {oh2023appsniffer}
\bibfield{author}{\bibinfo{person}{Sanghak Oh}, \bibinfo{person}{Minwook Lee}, \bibinfo{person}{Hyunwoo Lee}, \bibinfo{person}{Elisa Bertino}, {and} \bibinfo{person}{Hyoungshick Kim}.} \bibinfo{year}{2023}\natexlab{}.
\newblock \showarticletitle{AppSniffer: Towards robust mobile app fingerprinting against VPN}. In \bibinfo{booktitle}{\emph{Proceedings of the ACM Web Conference 2023}}. \bibinfo{pages}{2318--2328}.
\newblock


\bibitem[Pierro and Tonelli(2022)]%
        {pierro2022can}
\bibfield{author}{\bibinfo{person}{Giuseppe~Antonio Pierro} {and} \bibinfo{person}{Roberto Tonelli}.} \bibinfo{year}{2022}\natexlab{}.
\newblock \showarticletitle{Can solana be the solution to the blockchain scalability problem?}. In \bibinfo{booktitle}{\emph{2022 IEEE International Conference on Software Analysis, Evolution and Reengineering (SANER)}}. IEEE, \bibinfo{pages}{1219--1226}.
\newblock


\bibitem[Remy et~al\mbox{.}(2018)]%
        {remy2018tracking}
\bibfield{author}{\bibinfo{person}{Cazabet Remy}, \bibinfo{person}{Baccour Rym}, {and} \bibinfo{person}{Latapy Matthieu}.} \bibinfo{year}{2018}\natexlab{}.
\newblock \showarticletitle{Tracking bitcoin users activity using community detection on a network of weak signals}. In \bibinfo{booktitle}{\emph{Complex Networks \& Their Applications VI: Proceedings of Complex Networks 2017 (The Sixth International Conference on Complex Networks and Their Applications)}}. Springer, \bibinfo{pages}{166--177}.
\newblock


\bibitem[Shen et~al\mbox{.}(2020)]%
        {shen2020transaction}
\bibfield{author}{\bibinfo{person}{Meng Shen}, \bibinfo{person}{Junxian Duan}, \bibinfo{person}{Ning Shang}, {and} \bibinfo{person}{Liehuang Zhu}.} \bibinfo{year}{2020}\natexlab{}.
\newblock \showarticletitle{Transaction deanonymization in large-scale bitcoin systems via propagation pattern analysis}. In \bibinfo{booktitle}{\emph{International Conference on Security and Privacy in Digital Economy}}. Springer, \bibinfo{pages}{661--675}.
\newblock


\bibitem[Shen et~al\mbox{.}(2023)]%
        {shensubverting}
\bibfield{author}{\bibinfo{person}{Meng Shen}, \bibinfo{person}{Kexin Ji}, \bibinfo{person}{Zhenbo Gao}, \bibinfo{person}{Qi Li}, \bibinfo{person}{Liehuang Zhu}, {and} \bibinfo{person}{Ke Xu}.} \bibinfo{year}{2023}\natexlab{}.
\newblock \showarticletitle{Subverting Website Fingerprinting Defenses with Robust Traffic Representation}. In \bibinfo{booktitle}{\emph{32nd USENIX Security Symposium (USENIX Security)}}.
\newblock


\bibitem[Smolka et~al\mbox{.}(2023)]%
        {smolka2023fuzz}
\bibfield{author}{\bibinfo{person}{Sven Smolka}, \bibinfo{person}{Jens-Rene Giesen}, \bibinfo{person}{Pascal Winkler}, \bibinfo{person}{Oussama Draissi}, \bibinfo{person}{Lucas Davi}, \bibinfo{person}{Ghassan Karame}, {and} \bibinfo{person}{Klaus Pohl}.} \bibinfo{year}{2023}\natexlab{}.
\newblock \showarticletitle{Fuzz on the beach: Fuzzing solana smart contracts}. In \bibinfo{booktitle}{\emph{Proceedings of the 2023 ACM SIGSAC Conference on Computer and Communications Security}}. \bibinfo{pages}{1197--1211}.
\newblock


\bibitem[Tran et~al\mbox{.}(2020)]%
        {tran2020stealthier}
\bibfield{author}{\bibinfo{person}{Muoi Tran}, \bibinfo{person}{Inho Choi}, \bibinfo{person}{Gi~Jun Moon}, \bibinfo{person}{Anh~V Vu}, {and} \bibinfo{person}{Min~Suk Kang}.} \bibinfo{year}{2020}\natexlab{}.
\newblock \showarticletitle{A stealthier partitioning attack against bitcoin peer-to-peer network}. In \bibinfo{booktitle}{\emph{IEEE Symposium on Security and Privacy (SP)}}. \bibinfo{pages}{894--909}.
\newblock


\bibitem[Wallace and Scott-Hayward(2020)]%
        {wallace2020can}
\bibfield{author}{\bibinfo{person}{Victoria Wallace} {and} \bibinfo{person}{Sandra Scott-Hayward}.} \bibinfo{year}{2020}\natexlab{}.
\newblock \showarticletitle{Can SDN deanonymize Bitcoin users?}. In \bibinfo{booktitle}{\emph{ICC 2020-2020 IEEE International Conference on Communications (ICC)}}. IEEE, \bibinfo{pages}{1--7}.
\newblock


\bibitem[Wang et~al\mbox{.}(2024)]%
        {wang2024deanonymizing}
\bibfield{author}{\bibinfo{person}{Shan Wang}, \bibinfo{person}{Ming Yang}, \bibinfo{person}{Wenxuan Dai}, \bibinfo{person}{Yu Liu}, \bibinfo{person}{Yue Zhang}, {and} \bibinfo{person}{Xinwen Fu}.} \bibinfo{year}{2024}\natexlab{}.
\newblock \showarticletitle{Deanonymizing Ethereum Users behind Third-Party RPC Services}. In \bibinfo{booktitle}{\emph{IEEE Conference on Computer Communications (INFOCOM)}}.
\newblock


\bibitem[Wang et~al\mbox{.}(2018)]%
        {wang2018attack}
\bibfield{author}{\bibinfo{person}{Xu Wang}, \bibinfo{person}{Xuan Zha}, \bibinfo{person}{Guangsheng Yu}, \bibinfo{person}{Wei Ni}, \bibinfo{person}{Ren~Ping Liu}, \bibinfo{person}{Y~Jay Guo}, \bibinfo{person}{Xinxin Niu}, {and} \bibinfo{person}{Kangfeng Zheng}.} \bibinfo{year}{2018}\natexlab{}.
\newblock \showarticletitle{Attack and defence of ethereum remote apis}. In \bibinfo{booktitle}{\emph{2018 IEEE Globecom Workshops (GC Wkshps)}}. IEEE, \bibinfo{pages}{1--6}.
\newblock


\bibitem[Wu et~al\mbox{.}(2021)]%
        {wu2021my}
\bibfield{author}{\bibinfo{person}{Tianqi Wu}, \bibinfo{person}{Yubo Song}, \bibinfo{person}{Fan Zhang}, \bibinfo{person}{Shang Gao}, {and} \bibinfo{person}{Bin Chen}.} \bibinfo{year}{2021}\natexlab{}.
\newblock \showarticletitle{My site knows where you are: a novel browser fingerprint to track user position}. In \bibinfo{booktitle}{\emph{ICC 2021-IEEE International Conference on Communications}}. IEEE, \bibinfo{pages}{1--6}.
\newblock


\bibitem[Yakovenko(2018)]%
        {yakovenko2018solana}
\bibfield{author}{\bibinfo{person}{Anatoly Yakovenko}.} \bibinfo{year}{2018}\natexlab{}.
\newblock \showarticletitle{Solana: A new architecture for a high performance blockchain v0. 8.13}.
\newblock \bibinfo{journal}{\emph{Whitepaper}} (\bibinfo{year}{2018}).
\newblock


\bibitem[Zhao and Guan(2015)]%
        {zhao2015graph}
\bibfield{author}{\bibinfo{person}{Chen Zhao} {and} \bibinfo{person}{Yong Guan}.} \bibinfo{year}{2015}\natexlab{}.
\newblock \showarticletitle{A graph-based investigation of bitcoin transactions}. In \bibinfo{booktitle}{\emph{Advances in Digital Forensics XI: 11th IFIP WG 11.9 International Conference, Orlando, FL, USA, January 26-28, 2015, Revised Selected Papers 11}}. Springer, \bibinfo{pages}{79--95}.
\newblock


\bibitem[Zheng et~al\mbox{.}(2020)]%
        {zheng2020identifying}
\bibfield{author}{\bibinfo{person}{Baokun Zheng}, \bibinfo{person}{Liehuang Zhu}, \bibinfo{person}{Meng Shen}, \bibinfo{person}{Xiaojiang Du}, {and} \bibinfo{person}{Mohsen Guizani}.} \bibinfo{year}{2020}\natexlab{}.
\newblock \showarticletitle{Identifying the vulnerabilities of bitcoin anonymous mechanism based on address clustering}.
\newblock \bibinfo{journal}{\emph{Science China Information Sciences}}  \bibinfo{volume}{63} (\bibinfo{year}{2020}).
\newblock


\bibitem[Zheng et~al\mbox{.}(2023)]%
        {zheng2023ledger}
\bibfield{author}{\bibinfo{person}{Che Zheng}, \bibinfo{person}{Shen Meng}, \bibinfo{person}{Duan Junxian}, {and} \bibinfo{person}{Zhu Liehuang}.} \bibinfo{year}{2023}\natexlab{}.
\newblock \showarticletitle{From Ledger to P2P Network: De-anonymization on Bitcoin Using Cross-Layer Analysis}. In \bibinfo{booktitle}{\emph{International Symposium on Advanced Parallel Processing Technologies}}. Springer, \bibinfo{pages}{390--416}.
\newblock


\end{thebibliography}


\appendix

% \appendices

% \clearpage 
% \newpage
% \vspace{3mm}

{\centering\section*{\huge{Appendix}}}
\vspace{1mm}

\noindent
In the main text above, we primarily present analysis, figures and data related to Ethereum. 
In this Appendix section, we present those in other blockchains,
and other details ignored by the main text.\looseness=-1

\vspace{1mm}
\section{TCP Packet Size and Sequence Features across Blockchains and Wallets}
\label{appe::tcpsize}
% \vspace{-1mm}

\begin{figure}[h]
\centering
\includegraphics[width=1.0\columnwidth]{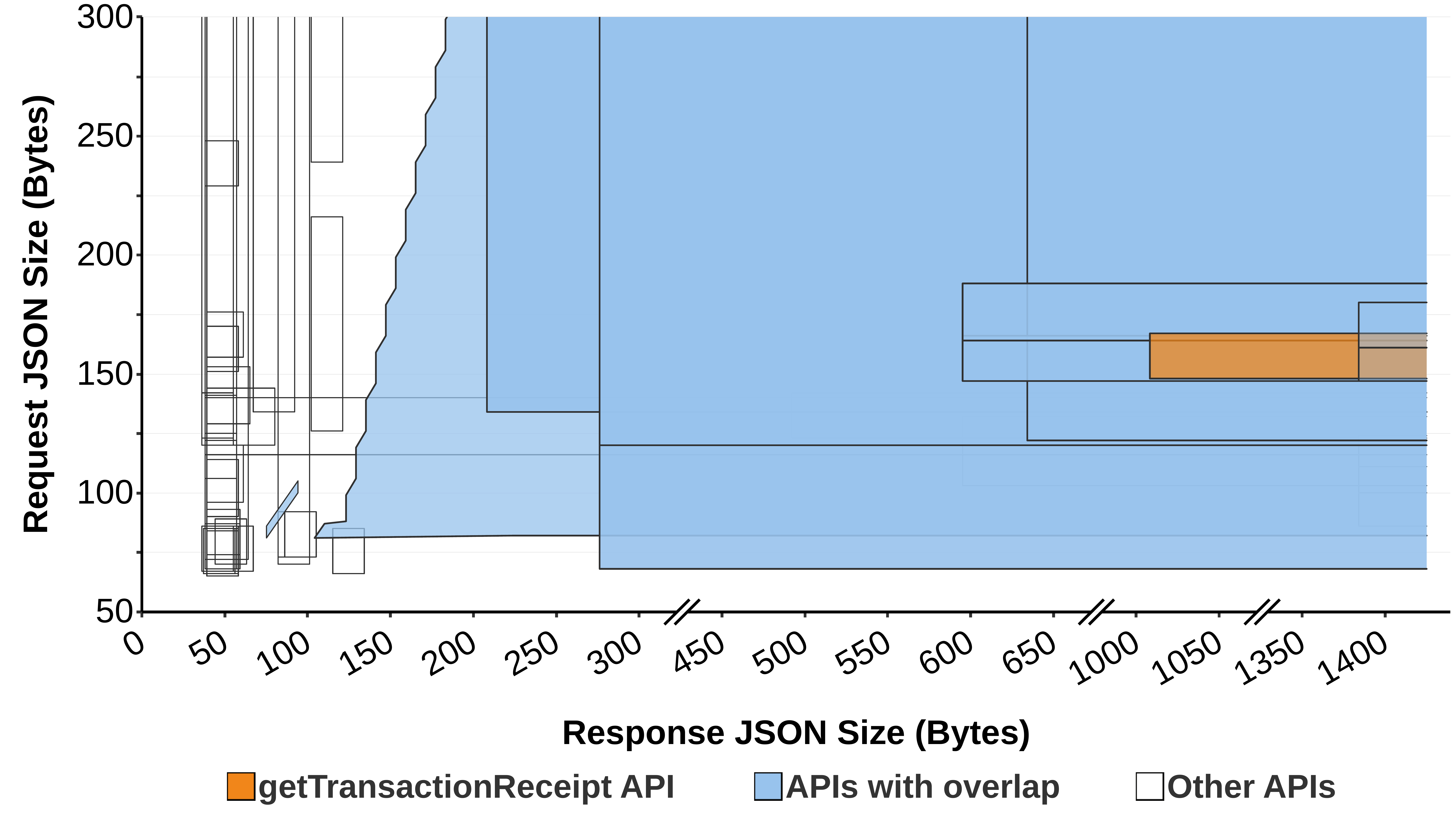}
\vspace{-5mm}
\caption{Size ranges of JSON-RPC request/response in ETH}
\label{fig::JSONETH} 
\vspace{-2mm}
\end{figure}

\autoref{fig::JSONETH} shows the size range of request and response JSON objects in Ethereum RPC protocol. There are 44 RPC APIs in total. 9 out of these 44 APIs have size overlaps with our target API \texttt{eth\_getTransactionReceipt}.

\begin{figure}[h]
\centering
\includegraphics[width=1.0\columnwidth]{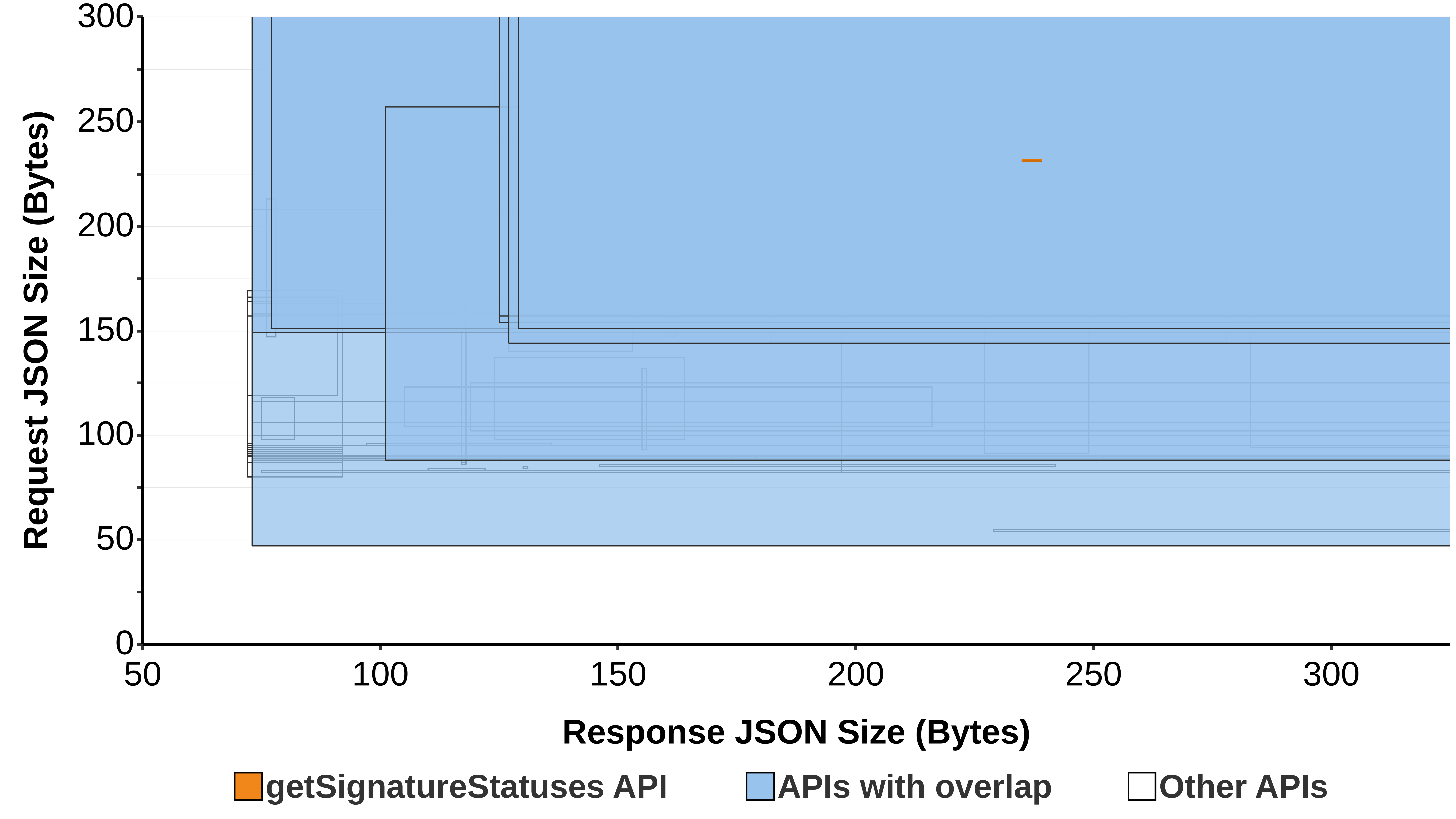}
\vspace{-3mm}
\caption{Size ranges of JSON-RPC request/response in SOL}
\label{fig::solanaJSON} 
\vspace{-2mm}
\end{figure}

\autoref{fig::solanaJSON} shows the size range of request and response JSON objects in Solana RPC protocol. There are 54 RPC APIs in total. 13 out of these 54 APIs have size overlaps with our target API \texttt{getSignatureStatuses}.
\autoref{table::overlapAPI_solana} shows the results considering the practical API usage in the wallet.

\begin{figure}[h]
\centering
\includegraphics[width=1.0\columnwidth]{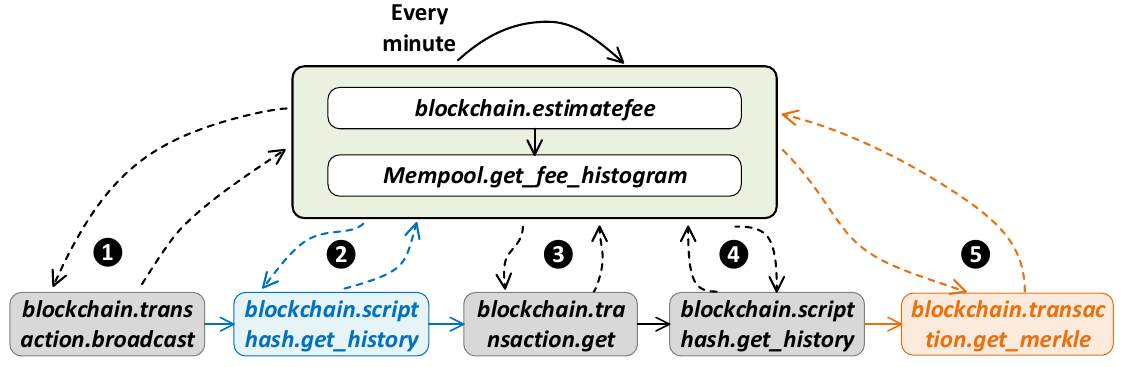}
% % \vspace{-3mm}
\caption{Bitcoin RPC API calls sequence in {\em Electrum}}
\label{fig::BTCAPISequence} 
\end{figure}

\autoref{fig::BTCAPISequence} shows the API call sequence in Bitcoin wallet {\em Electrum}. It first sends out a transaction by calling API \texttt{blockchain.transaction.broadcast}. 
Then it queries the confirmed and unconfirmed transactions by API \texttt{blockchain.scripthash.get\_history}.
When its initiated transaction is confirmed, the wallet queries its status by API \texttt{blockchain.transaction.get\_merkle} for checking its validity. 
{\em Electrum} also periodically calls APIs for estimating transaction fees every minute.

\begin{figure}[h]
\centering
\includegraphics[width=1.0\columnwidth]{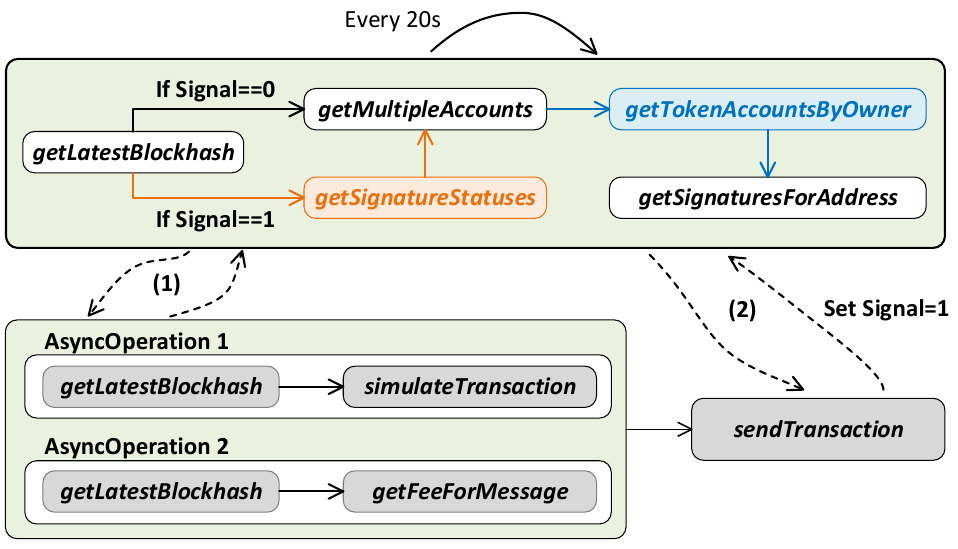}
% \vspace{-3mm}
\caption{Solana RPC API calls sequence in {\em Torus}}
\label{fig::SolanaAPISequence} 
% \vspace{-2mm}
\end{figure}

% \usepackage{multirow}
\begin{table}[ht!]
\caption{TCP segment size ranges considering practical API usage in the Solana wallet named {\em Torus}. The RPC service is {\em QuickNode}, the protocol is HTTP/2, and the target API is \texttt{getSignatureStatuses}.}
\label{table::overlapAPI_solana}
\centering
\scriptsize
\setlength\tabcolsep{0.5pt}
\renewcommand{\arraystretch}{1}
% \begin{tabular}{c|cccccc|lcc}
\resizebox{\linewidth}{!}{
\begin{tabular}{cclcc}
\toprule[1.5pt]
\multicolumn{2}{c}{\textbf{Target TCP Size}}                                                 & \multirow{2}{*}{\textbf{Overlapped APIs in Theory}} & \multicolumn{2}{c}{\textbf{Noise TCP Size}}                                                                                                             \\ \cline{1-2} \cline{4-5} 
\textbf{Request}                            & \textbf{Response}                               &                                                     & \textbf{Request}                                                            & \textbf{Response}                                                           \\ \hline
\multirow{14}{*}{262} & \multirow{14}{*}{266-269} & \texttt{getAccountInfo}                               & 175-338                                                                     & $\geq$158                                                                   \\
                                            &                                                 & \texttt{getMultipleAccounts}                          & $\geq$182                                                                   & $\geq$160                                                                   \\
                                            &                                                 & \texttt{getTokenAccountsByOwner}                      & 185-459                                                                     & $\geq$156                                                                   \\ \cdashline{3-5}[2pt/2pt] 
                                            &                                                 & \texttt{getTransaction}                               & \multirow{4}{*}{\begin{tabular}[c]{@{}c@{}}Without \\ overlap\end{tabular}} & \multirow{4}{*}{\begin{tabular}[c]{@{}c@{}}Without \\ overlap\end{tabular}} \\
                                            &                                                 & \texttt{getFeeForMessage}                             &                                                                             &                                                                             \\
                                            &                                                 & \texttt{getSignaturesForAddress}                      &                                                                             &                                                                             \\
                                            &                                                 & \texttt{simulateTransaction}                          &                                                                             &                                                                             \\ \cdashline{3-5}[2pt/2pt]  
                                            &                                                 & \texttt{getProgramAccounts}                           & \multirow{7}{*}{Unused}                                                     & \multirow{7}{*}{Unused}                                                     \\
                                            &                                                 & \texttt{getRecentPrioritizationFees}                  &                                                                             &                                                                             \\
                                            &                                                 & \texttt{getTokenAccountsByDelegate}                   &                                                                             &                                                                             \\
                                            &                                                 & \texttt{getVoteAccounts}                              &                                                                             &                                                                             \\
                                            &                                                 & \texttt{getLeaderSchedule}                            &                                                                             &                                                                             \\
                                            &                                                 & \texttt{getBlock}                                     &                                                                             &                                                                             \\
                                            &                                                 & \texttt{getInflationReward}                           &                                                                             &                                                                             \\ \bottomrule[1.5pt]
\end{tabular}
} 
\end{table}

\autoref{fig::SolanaAPISequence} shows the API calls sequence in the Solana wallet named {\em Torus}. It first calls APIs to estimate the transaction fees, and then send a transaction by API \texttt{sendTransaction}. In an asynchronous task, the wallet periodically queries the latest block and account information every $20$ seconds. If a transaction has been initiated by the wallet, it also queries the transaction status by calling \texttt{getSignatureStatuses}. It can be observed that, our target API \texttt{getSignatureStatuses} in orange has different context in the API calls sequence from the noise API such as \texttt{getTokenAccountByOwner} in blue.

\section{Modeling the Attack}
\label{appendix::models}

We provide more details about Equation (\ref{eqn::f}).
If there are one or more transactions from $s_i$ being included in the selected $k$ blocks, the pseudonym $s_i$ will appear in the potential set $\mathcal{S}_j$.
Therefore, the probability (denoted as $y_j$) of $s_i$ appearing in the set $\mathcal{S}_j$ should be,
\begin{equation}
    y_j = 1 - (1 - p_i)^{x_j}.
\label{equ::yj}
\end{equation}

Let $f_i$ denote the probability of pseudonym $s_i$ being excluded after $m$ rounds of intersection. Note: the first intersection is $\mathcal{S}_1 \cap \mathcal{S}_0=\mathcal{S}_1 \cap \mathbb U = \mathcal{S}_1$. Once $s_i$ in $\mathcal{S}_1$ does not appear in any following set in the intersection process, it can be excluded. We have,
\begin{equation}
\begin{aligned}
    f_i &= 1-\prod_{j=2}^{m}y_j.
    % \\ &= 1-\prod_{j=2}^{m}(1 - (1 - b_i)^{x_j}).
\end{aligned}   
\label{equ::fi_appendix}
% \vspace{-2mm}
\end{equation}
Based on Equations (\ref{equ::yj})$\sim$(\ref{equ::fi_appendix}), we derive Equation (\ref{eqn::f}). \looseness=-1

\section{Estimation Methods of False Positive Rate and Success Rate based on Measurements}
\label{sec::estimationOfE(P)}

The blockchain user activity intensity will decide $P_{t^{'}}$ and thus $P$, which is beyond control of the attacker.
Their activity intensity can be described by three variables, including a vector $\boldsymbol{p}$ giving the probabilities of a transaction being initiated by different pseudonyms, and $\boldsymbol{x}$ and $\boldsymbol{n}$ denoting the number of transactions and the number of pseudonyms included in $k$ consecutive blocks respectively.
These variables, $\boldsymbol{p}$, $\boldsymbol{x}$ and $\boldsymbol{n}$, follow standard stochastic processes and are random variables.
The $p_i$ may take any value of variable $\boldsymbol{p}$, and $x_j$ may take any value of variable $\boldsymbol{x}$ where each value corresponds to a different probability.
As the parameter $\boldsymbol{p}$ and $\boldsymbol{x}$ follow stochastic processes, the $f_i$, $P_{t^{'}}$ and $P$ are also random variables, whose each possible value occurs with a distinct probability.\looseness=-1

% In probability theory, the expectation is a measure of the average outcome of a random process over numerous trials. The expectation of a random variable is decided by its possible values and the corresponding probabilities.
In probability theory, the average success rate should be the mathematical expectation of $P$, i.e., $\mathbb{E}[P]$.
To derive $\mathbb{E}[P]$, we first derive the probability density function (PDF) denoted as $F(\cdot)$ for the variables $f_i$ and $P$, 
giving the probability of each possible value.
The average false positive rate, denoted as $R$ where $R = 1-P_{t^{'}}$, which is introduced by non-target users appearing in the final intersection result, can be derived in similar way.

% based on the statistical distributions of random variables $\boldsymbol{p}$, $\boldsymbol{x}$, and $\boldsymbol{n}$. 
% Please find the details about distributions $F(\boldsymbol{p})$, $F(\boldsymbol{x})$, and $F(\boldsymbol{n})$ in \S\ref{appendix::models} in Appendix.

% probability density function
% statistical distribution

$F(f_i)$ can be derived based on $F(p)$ and $F(x)$, according to Equation (\ref{eqn::f}). 
Let $\mathcal{D}$ denote the set of tuples $\langle \widetilde{p},\widetilde{x} \rangle$ where $f(\widetilde{m}, \widetilde{p},\widetilde{x})= \widetilde{f_i}$, given $m$. We have,
\begin{equation}
    F(f_i=\widetilde{f_i},m=\widetilde{m}) = \sum_{\langle \widetilde{p},\widetilde{x} \rangle\in\mathcal{D}}F(\widetilde{p})F(\widetilde{x}),
\label{equ::F(m,f)}
% \vspace{-2mm}
\end{equation}
which gives the probability of $\boldsymbol{f_i} = \widetilde{f_i}$ when $m=\widetilde{m}$. 
Let $\mathcal{M}$ denote the range of $f_i$ given $m$. The expectation of $f_i$ can be derived by,
\begin{equation}
    \mathbb{E}[f_i] = \sum\limits_{f_i\in\mathcal{M}}f_i\cdot F(f_i).
\label{equ::E(f_i)}
\end{equation}
$\mathbb{E}[f_i]$ is the average value of $f_i$ estimated in theory, given $m$.
% The false positive rate denoted as $R$ should be $R = 1-P_n$.
Similarly, we can derive the $\mathbb{E}[R]$ and $\mathbb{E}[P_{t^{'}}]$, where $\mathbb{E}[R] = 1 - \mathbb{E}[P_{t^{'}}]$.

% Then, $F(P)$ is given by statistical distributions of $\boldsymbol{p}$, $\boldsymbol{x}$, and $\boldsymbol{n}$, which describe the realistic blockchain networks.
Similarly, $F(P)$ of the success rate can be given by $F(p)$, $F(n)$ and $F(x)$, according to Equation (\ref{eqn::P_Final}).
Let $\mathcal{C}$ denote the set of tuples $\langle \widetilde{p},\widetilde{x},\widetilde{n} \rangle$ where $P(\widetilde{\alpha}, \widetilde{m}, \widetilde{p},\widetilde{x},\widetilde{n})=\widetilde{P}$, given $\alpha$ and $m$. 
The following equation, 
\begin{small}
\begin{equation}
\begin{aligned}
    F(P=\widetilde{P}, \alpha=\widetilde{\alpha}, m=\widetilde{m}) 
    = \sum_{\langle \widetilde{p}, \widetilde{x},\widetilde{n} \rangle\in\mathcal{C}}F(\widetilde{p})F(\widetilde{x})F(\widetilde{n}), 
\end{aligned}
\label{equ::F(P)}
% \vspace{-2mm}
\end{equation}
\end{small}
gives the probability of $P=\widetilde{P}$ when $\alpha=\widetilde{\alpha}$ and $m = \widetilde{m}$. Let $\mathcal{N}$ denote the range of $P$. The expectation of $P$ can be derived by,
\begin{equation}
    \mathbb{E}[P] = \sum\limits_{P\in\mathcal{N}}P\cdot F(P),
\label{equ::E(P)}
\end{equation}
which gives the average success rate of deanonymization in theory, given $\alpha$ and $m$.\looseness=-1

\section{Measurements of Key Parameters in Ethereum Mainnet}
\label{sec::measureActivityLevel}

To derive the numerical false positive rate and success rate based on measurements, we empirically measure the probability density functions (PDF) of the three key variables, i.e., $F(\boldsymbol{n})$, $F(\boldsymbol{x})$ and $F(p_i)$, 
which characterize the user activity intensity on blockchain networks, as analyzed in \S\ref{sec::estimationOfE(P)}.

\begin{figure}[h!]
\vspace{-2mm}
\begin{minipage}[c]{0.48\columnwidth}
\centering
\includegraphics[width=0.99\textwidth]{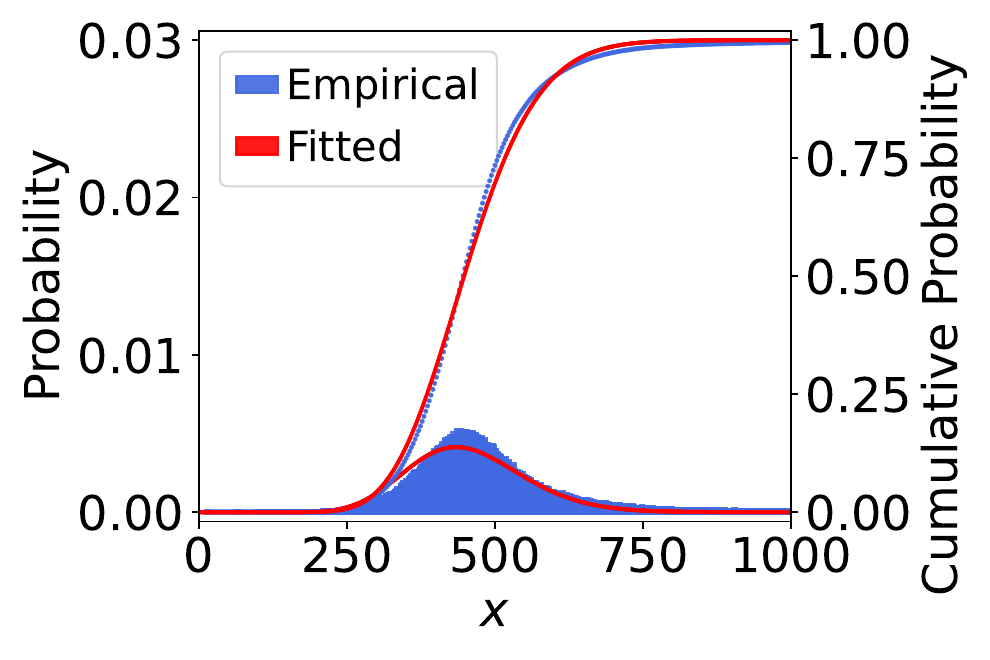}
\vspace{-7mm}
\caption{The number of transactions in $3$ consecutive blocks in Ethereum mainnet}
\label{fig::TxNumDis3blocks} 
\end{minipage}
% \vspace{-3mm}
\hspace{0.05cm}
\begin{minipage}[c]{0.48\columnwidth}
\centering
\includegraphics[width=0.99\textwidth]{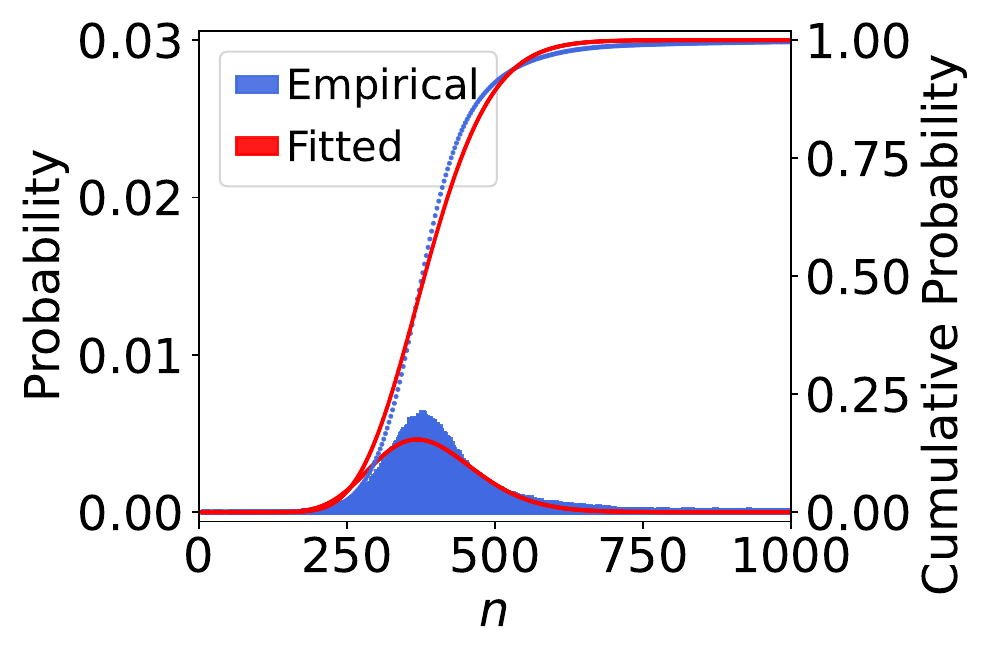}
\vspace{-7mm}
\caption{The number of pseudonyms in $3$ consecutive blocks in Ethereum mainnet}
\label{fig::pseudonymNumDis3blocks}
\end{minipage}
\vspace{-2mm}
\end{figure}

% \vspace{2mm}
%\noindent
\textbf{Measurements of \boldmath{$F(\boldsymbol{x})$} and \boldmath{$F(\boldsymbol{n})$}:} 
The variables $\boldsymbol{x}$ and $\boldsymbol{n}$ follow Poisson distribution as analyzed in \S\ref{sec::attackModel}.
Based on the ledger data, we derive that the average rate of $\boldsymbol{x}$ is $\lambda = 12.68$, and the average rate of $\boldsymbol{n}$ is $\lambda^{'} = 10.80$, when the unit time is set to $1$ seconds.
\autoref{fig::TxNumDis3blocks} and \autoref{fig::pseudonymNumDis3blocks} show that the empirical measurements approximate the fitted Poisson distributions, validating the correctness of our models of $\boldsymbol{x}$ and $\boldsymbol{n}$. 
In Ethereum, 
$3$ consecutive blocks (a period of around 36s) include around $458.93$ transactions and $390.95$ pseudonyms on average.\looseness=-1

%\vspace{2mm}
%\noindent
\textbf{Measurements of \boldmath{$F(p_i)$} and \boldmath{$\lambda_i$}:} 
User transaction behaviors follow the multinomial process as analyzed in \S\ref{sec::attackModel}.
Assume there are a total of $L$ transactions, and $l_i$ of them are initiated by a pseudonym $s_i$. 
We then derive $p_i = \frac{l_i}{L}$.
We calculate $p_i$ for each of the 6,769,835 users in our Ethereum mainnet dataset. 
\autoref{fig::distributionOfb} shows the probability density function of $p_i$. 
Its minimum value is only $0.00000294\%$ while the maximum value reaches $0.64308559\%$, demonstrating significant variability in user activity levels.
Each user is associated with a transacting rate $\lambda_i$, where $\lambda_i = \lambda p_i$. 
\autoref{fig::distributionOfLamda} shows the cumulative proportion of users across different rates.
Based on the threshold $\theta_{\lambda} = 0.00028$ derived in \S\ref{sec::model_Analysis}, the attacker can identify and exclude active users.
Notably, 99.97\% of users have transaction rates below this threshold, indicating that only 0.03\% of users are considered active and thus will be filtered out.
This threshold $\theta_{\lambda} = 0.00028$ corresponds to the $p_i$'s threshold $\theta_{p} =  0.0000215385$.
This means that filtering out the $0.03\%$ active users can effectively narrow the maximum value of $p_i$ from $0.64308559\%$ down to $0.00215385\%$.

% ETH testnet

\begin{figure}[h!]
\vspace{-2mm}
\begin{minipage}[t]{0.48\columnwidth}
\centering
\includegraphics[height=0.64\textwidth]{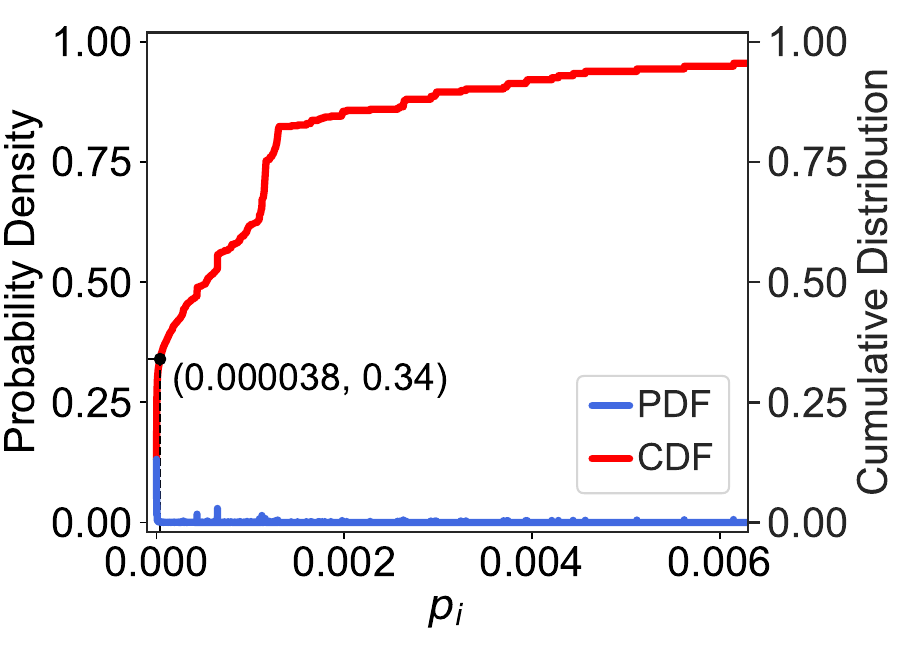}
\vspace{-5mm}
% \caption{Distribution of $b$ in Ethereum testnet}
\caption{Distribution of $p_i$ in Ethereum testnet}
\label{fig::distributionOfbETHtest} 
\end{minipage}
\hspace{0.05cm}
\begin{minipage}[t]{0.48\columnwidth}
\centering
\includegraphics[height=0.64\textwidth]{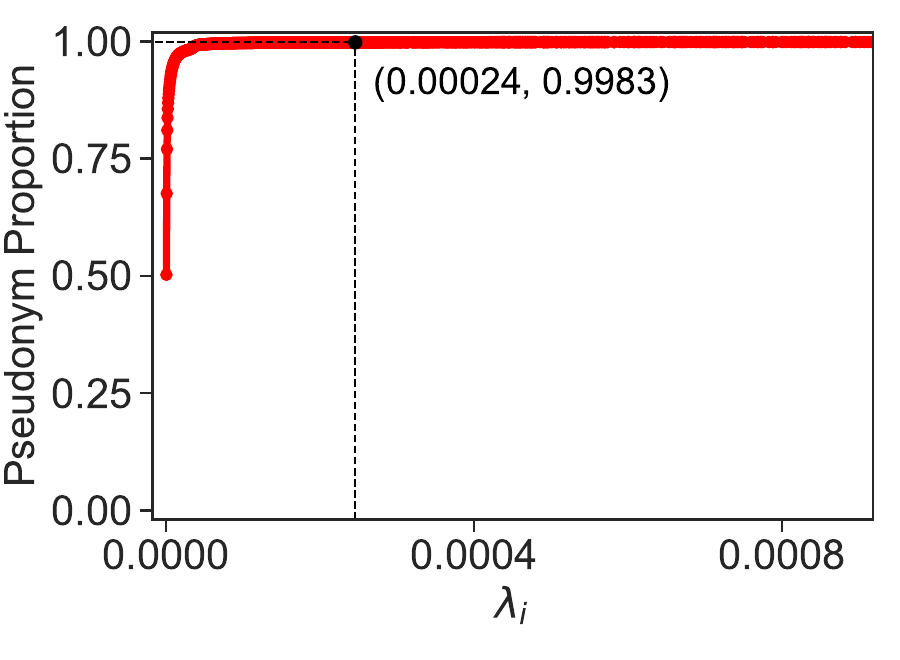}
\vspace{-5mm}
% \caption{Distribution of $f$ in Ethereum testnet}
\caption{The proportion of users within each transacting rate $\lambda_i$ in Ethereum testnet}
\label{fig::F(m,f)ETHtest} 
\end{minipage}
\end{figure}

\begin{figure}[h!]
\begin{minipage}[t]{0.48\columnwidth}
\centering
\includegraphics[height=0.64\textwidth]{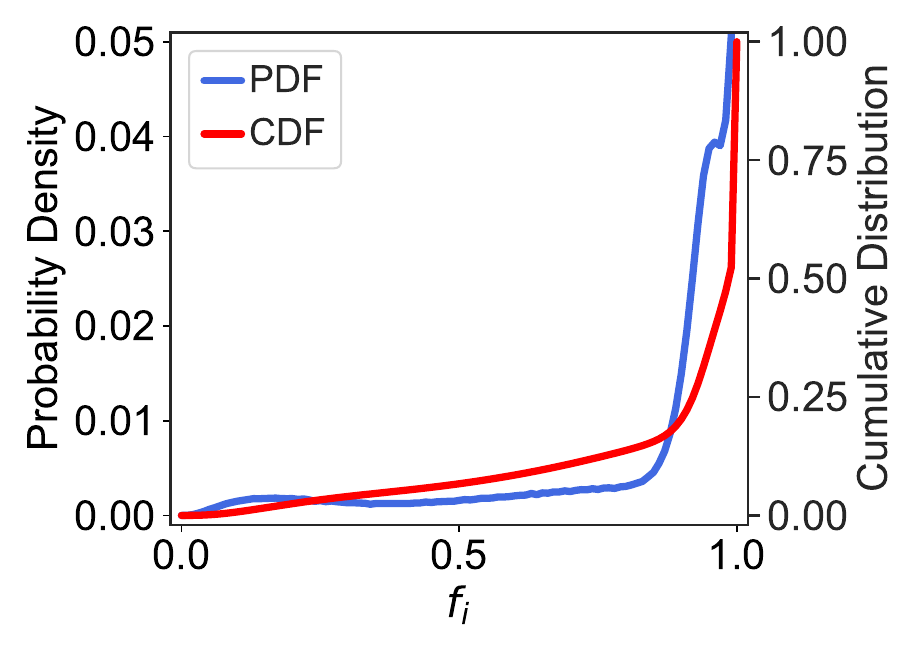}
\vspace{-5mm}
% \caption{Pseudonym proportion of different $b_i$ in Ethereum testnet}
\caption{Distribution of $f_i$ without adopting optimization (m = 3) in Ethereum testnet}
\label{fig::addressTxNumETHtest}  
\end{minipage}
\hspace{0.05cm}
\begin{minipage}[t]{0.48\columnwidth}
\centering
\includegraphics[height=0.64\textwidth]{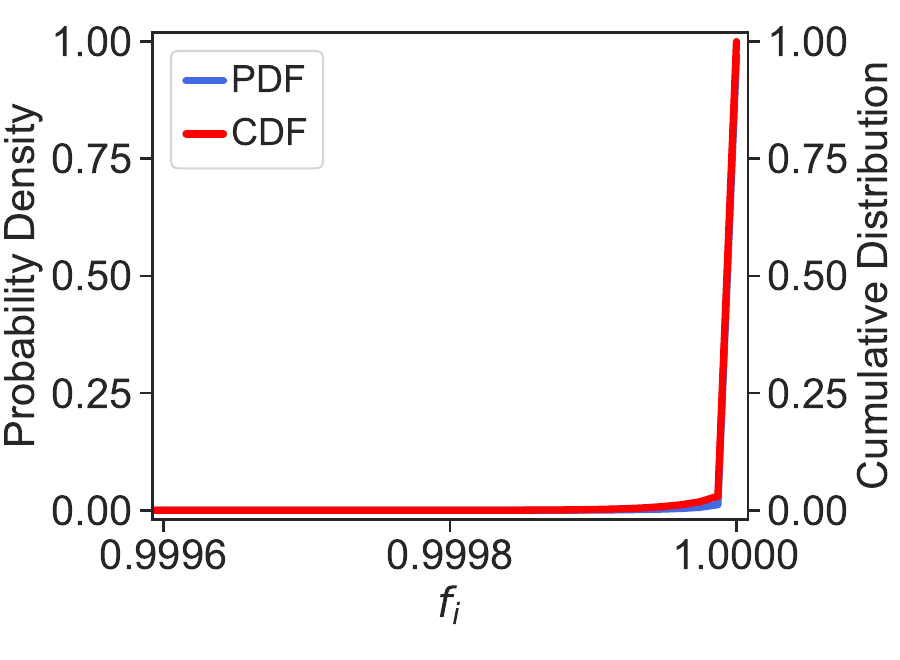}
\vspace{-5mm}
% \caption{Distribution of $f$ after filtering out $1\%$ most active users in Ethereum testnet}
\caption{Distribution of $f_i$ in the optimized identification method (m = 3) in Ethereum testnet}
\label{fig::F(m,f)filter_bETHtest}  
\end{minipage}
\end{figure}

\begin{figure}[h!]
\begin{minipage}[t]{0.48\columnwidth}
\centering
\includegraphics[height=0.64\textwidth]{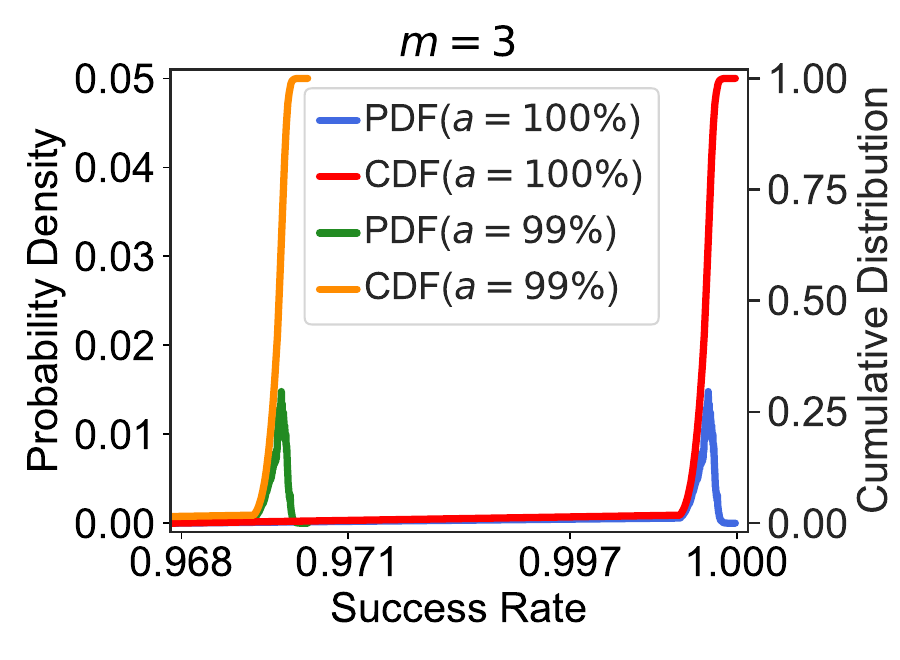}
\vspace{-3mm}
% \caption{Distribution of the identification accuracy $P$ when $m = 2$ in Ethereum testnet}
\caption{Distribution of the success rate $P$ (m = 3) in Ethereum testnet}
\label{fig::P_m=2_ETHtest}  
\end{minipage}
\hspace{0.05cm}
\begin{minipage}[t]{0.48\columnwidth}
\centering
\includegraphics[height=0.64\textwidth]{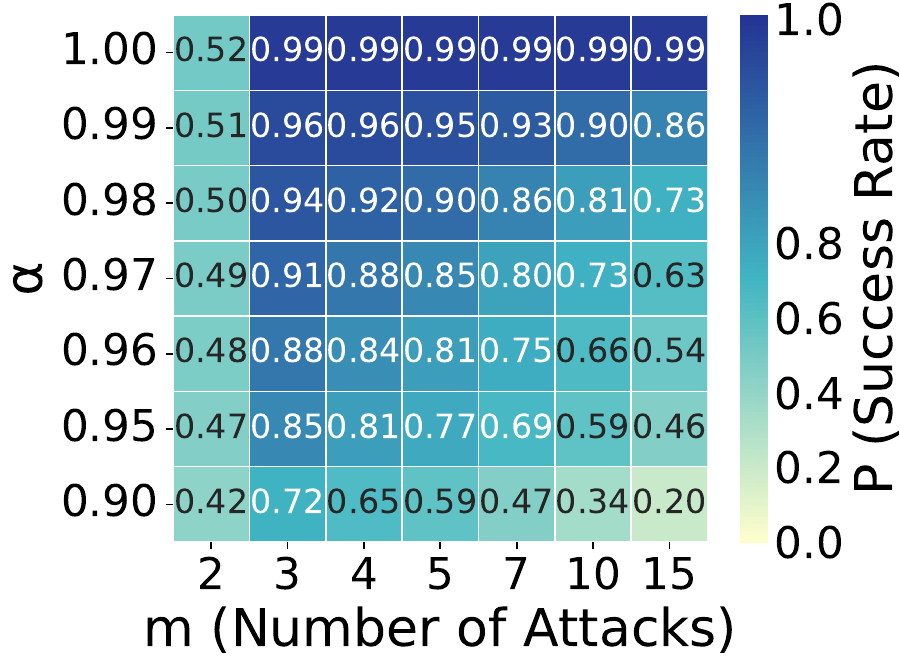}
\vspace{-3mm}
% \caption{Distribution of the identification accuracy $P$ when $m = 3$ in Ethereum testnet}
\caption{The success rate with varying $\alpha$ and $m$ in Ethereum testnet}
\label{fig::P_m=3_ETHtest}  
\end{minipage}
\end{figure}

% BTC mainnet

\begin{figure}[h!]
\begin{minipage}[t]{0.48\columnwidth}
\centering
\includegraphics[height=0.64\textwidth]{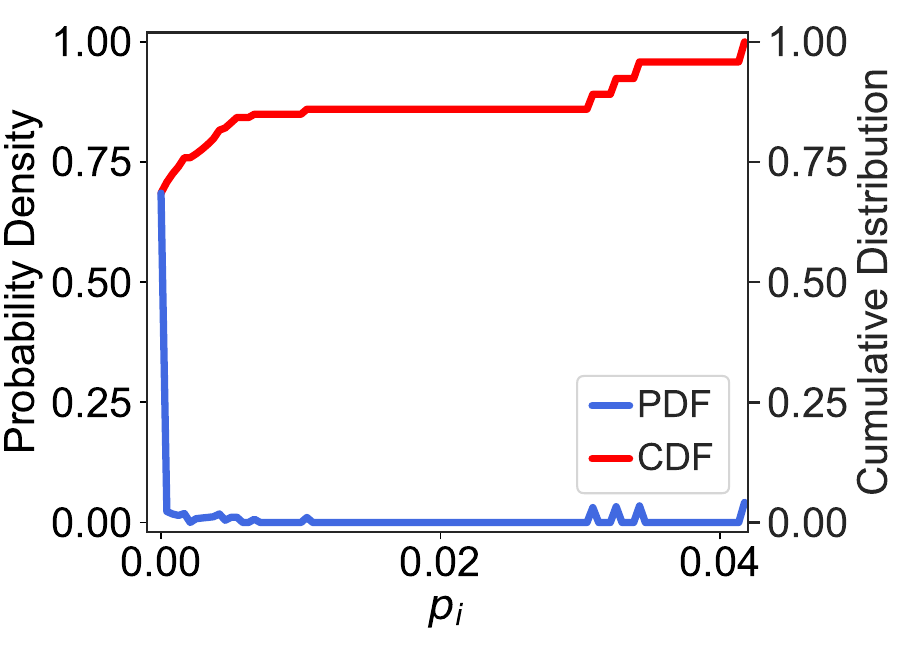}
\vspace{-5mm}
\caption{Distribution of $p_i$ in Bitcoin mainnet}
\label{fig::distributionOfbBTCmain} 
\end{minipage}
\hspace{0.05cm}
% \hspace{0.1cm}
\begin{minipage}[t]{0.48\columnwidth}
\centering
\includegraphics[height=0.64\textwidth]{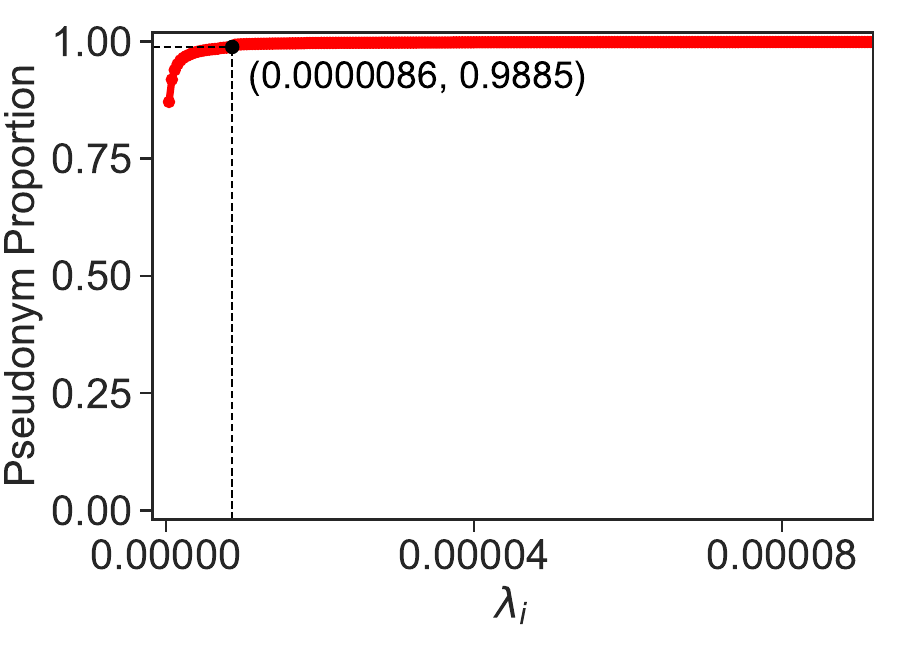}
\vspace{-5mm}
\caption{The proportion of users within each transacting rate $\lambda_i$ in Bitcoin mainnet}
\label{fig::F(m,f)BTCmain} 
\end{minipage}
\end{figure}

\begin{figure}[h!]
\begin{minipage}[t]{0.48\columnwidth}
\centering
\includegraphics[height=0.64\textwidth]{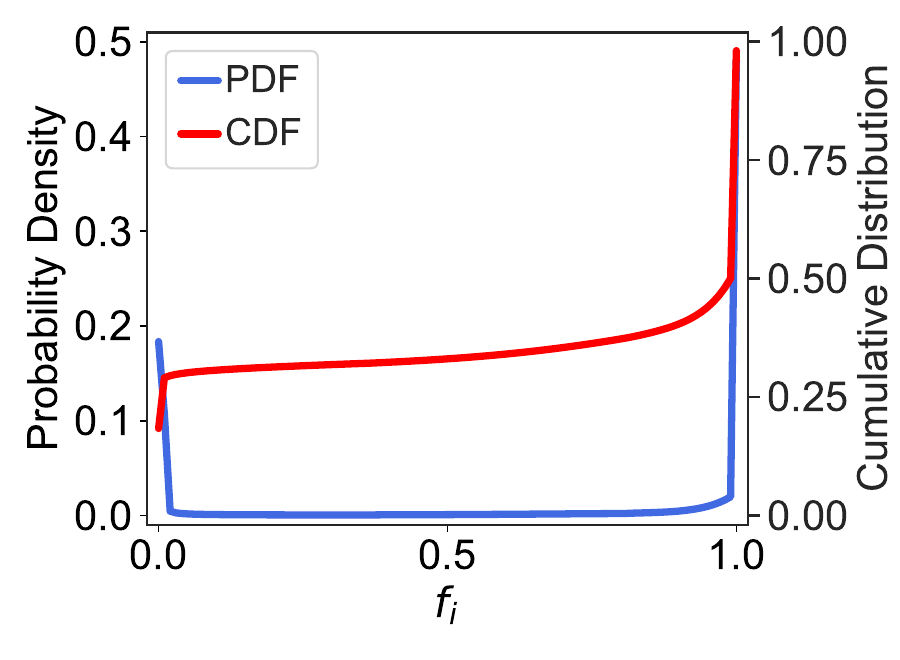}
\vspace{-5mm}
\caption{Distribution of $f_i$ without adopting optimization (m = 4) in Bitcoin mainnet}
\label{fig::addressTxNumBTCmain}  
\end{minipage}
\hspace{0.05cm}
\begin{minipage}[t]{0.48\columnwidth}
\centering
\includegraphics[height=0.64\textwidth]{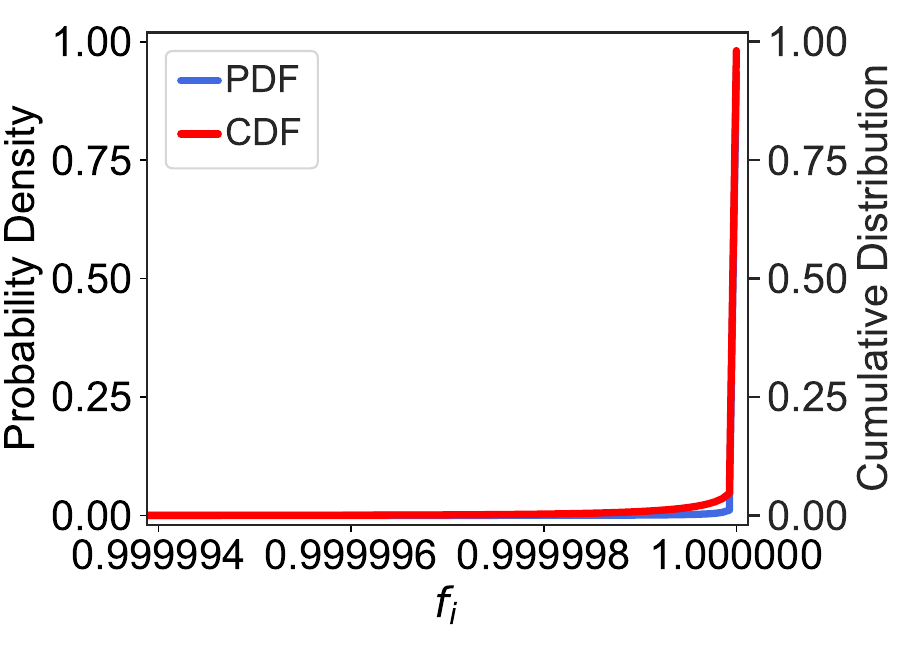}
\vspace{-5mm}
\caption{Distribution of $f_i$ in the optimized identification method (m = 4) in Bitcoin mainnet}
\label{fig::F(m,f)filter_b_BTCmain}  
\end{minipage}
\end{figure}

\begin{figure}[h!]
\begin{minipage}[t]{0.48\columnwidth}
\centering
\includegraphics[height=0.64\textwidth]{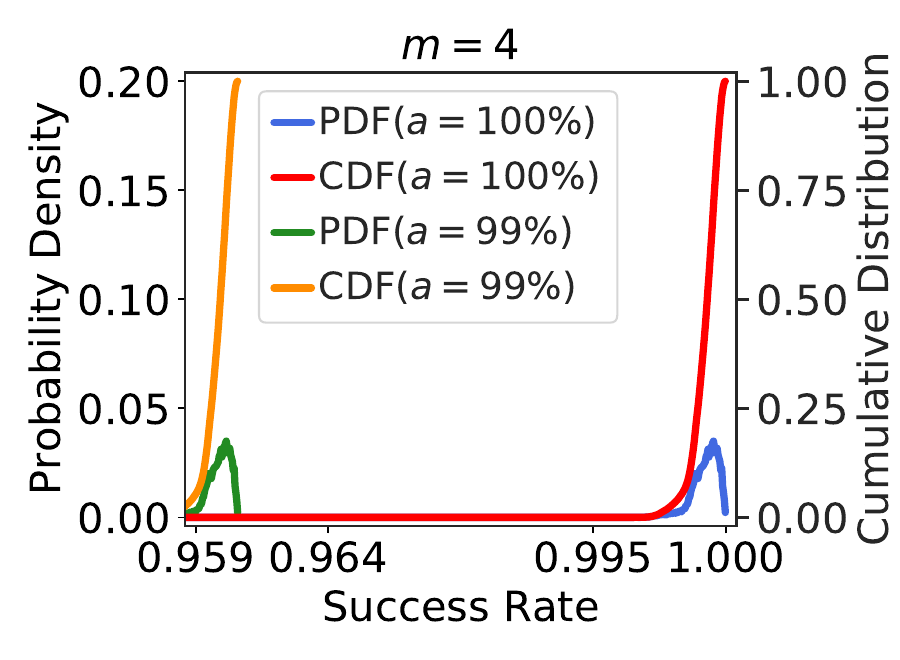}
\vspace{-3mm}
\caption{Distribution of the success rate $P$ (m = 4) in Bitcoin mainnet}
\label{fig::F(P)_m=3_BTCmain}  
\end{minipage}
\hspace{0.05cm}
\begin{minipage}[t]{0.48\columnwidth}
\centering
\includegraphics[height=0.64\textwidth]{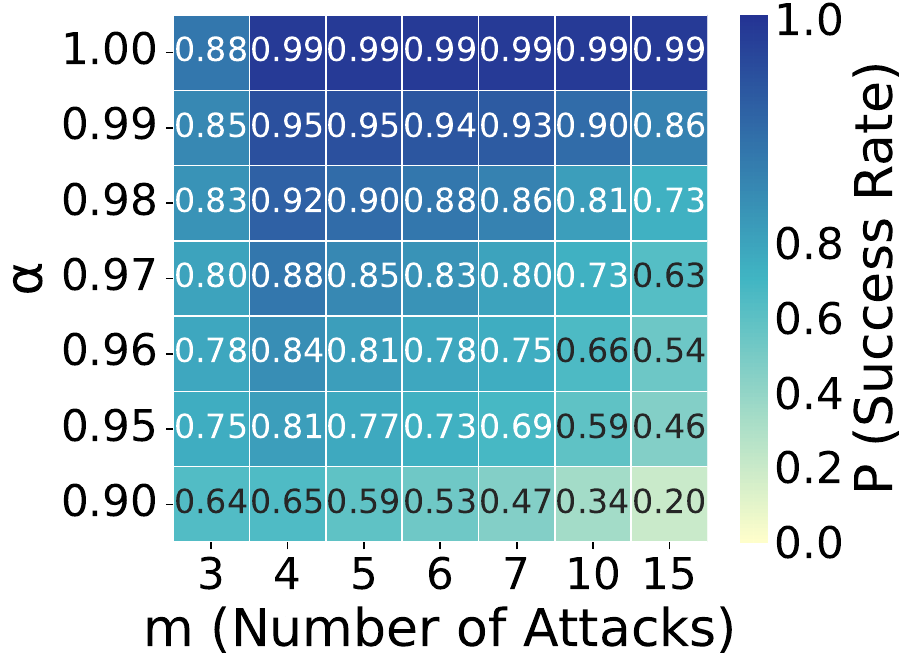}
\vspace{-3mm}
\caption{The success rate with varying $\alpha$ and $m$ in Bitcoin mainnet}
\label{fig::F(P)_m=4_BTCmain}  
\end{minipage}
\end{figure}

% BTC testnet

\begin{figure}[h!]
\begin{minipage}[t]{0.48\columnwidth}
\centering
\includegraphics[height=0.64\textwidth]{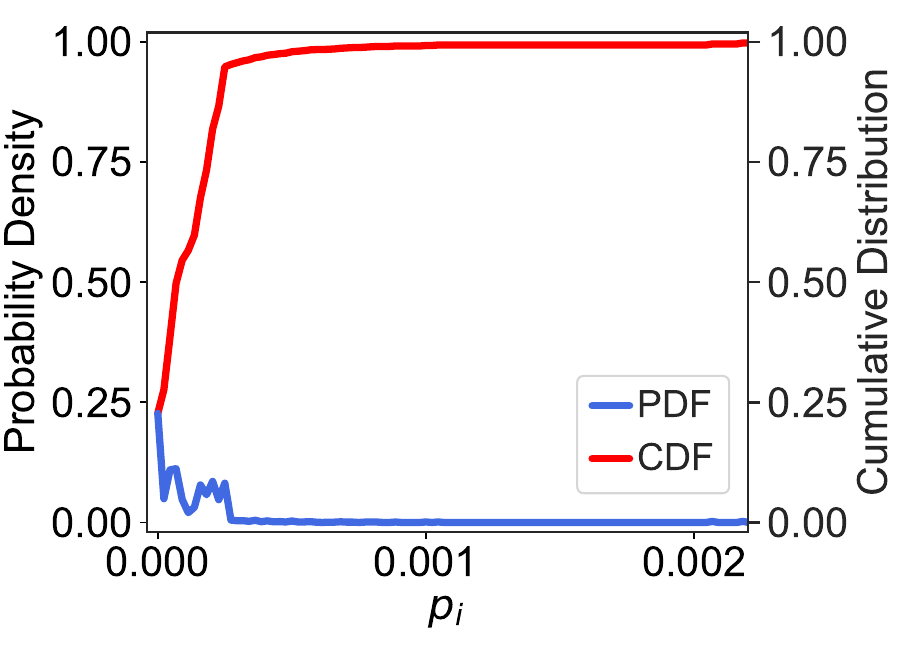}
\vspace{-5mm}
\caption{Distribution of $p_i$ in Bitcoin testnet}
\label{fig::distributionOfb_BTCtest} 
\end{minipage}
\hspace{0.05cm}
% \hspace{0.1cm}
\begin{minipage}[t]{0.48\columnwidth}
\centering
\includegraphics[height=0.64\textwidth]{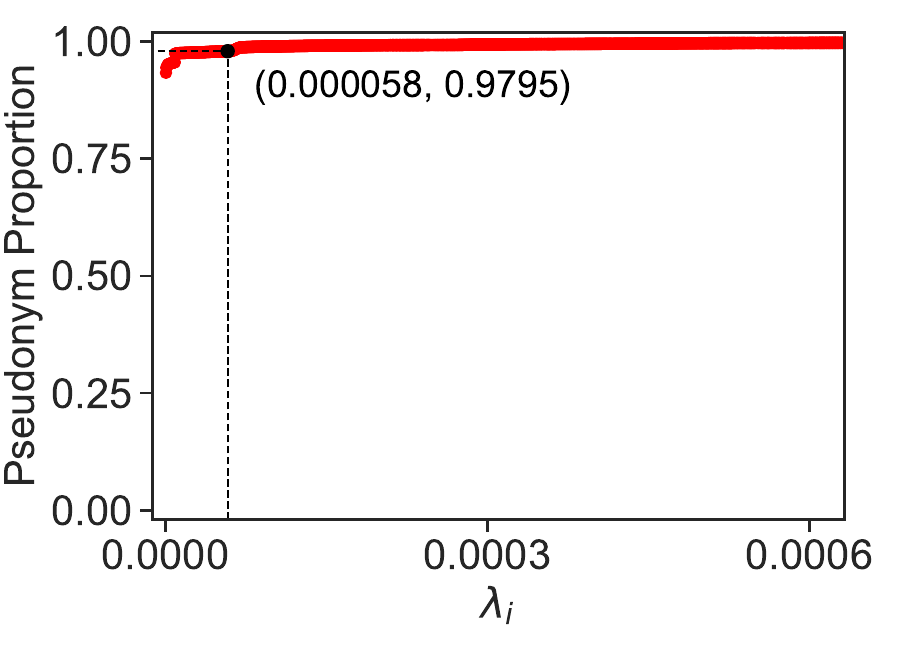}
\vspace{-5mm}
\caption{The proportion of users within each transacting rate $\lambda_i$ in Bitcoin testnet}
\label{fig::F(m,f)_BTCtest} 
\end{minipage}
\end{figure}

\begin{figure}[h!]
\begin{minipage}[t]{0.48\columnwidth}
\centering
\includegraphics[height=0.64\textwidth]{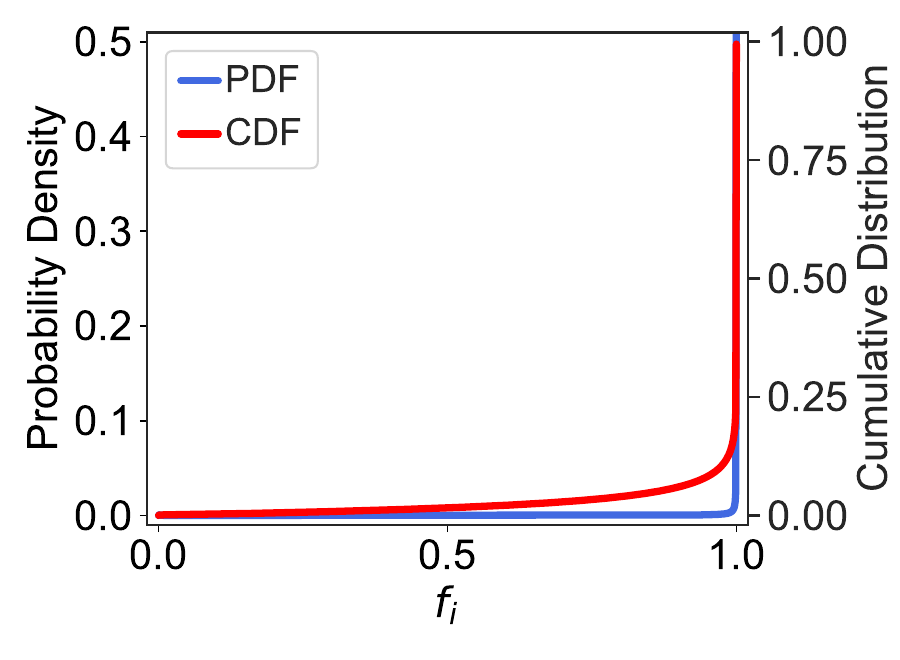}
\vspace{-5mm}
\caption{Distribution of $f_i$ without adopting optimization (m = 4) in Bitcoin testnet}
\label{fig::addressTxNum_BTCtest}  
\end{minipage}
\hspace{0.05cm}
\begin{minipage}[t]{0.48\columnwidth}
\centering
\includegraphics[height=0.64\textwidth]{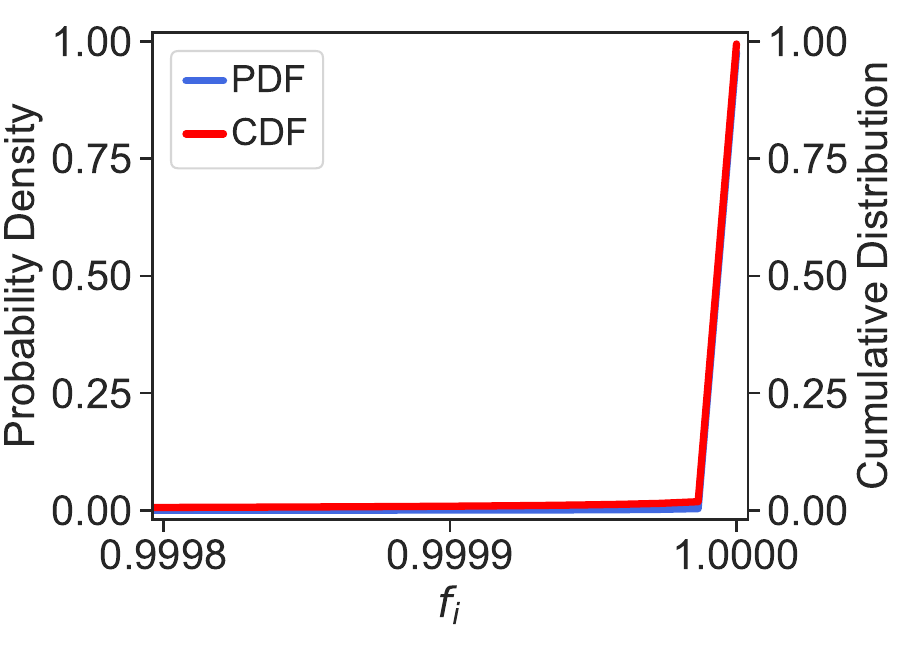}
\vspace{-5mm}
\caption{Distribution of $f_i$ in the optimized identification method (m = 4) in Bitcoin testnet}
\label{fig::F(m,f)filter_b_BTCtest}  
\end{minipage}
\end{figure}

\begin{figure}[h!]
\begin{minipage}[t]{0.48\columnwidth}
\centering
\includegraphics[height=0.64\textwidth]{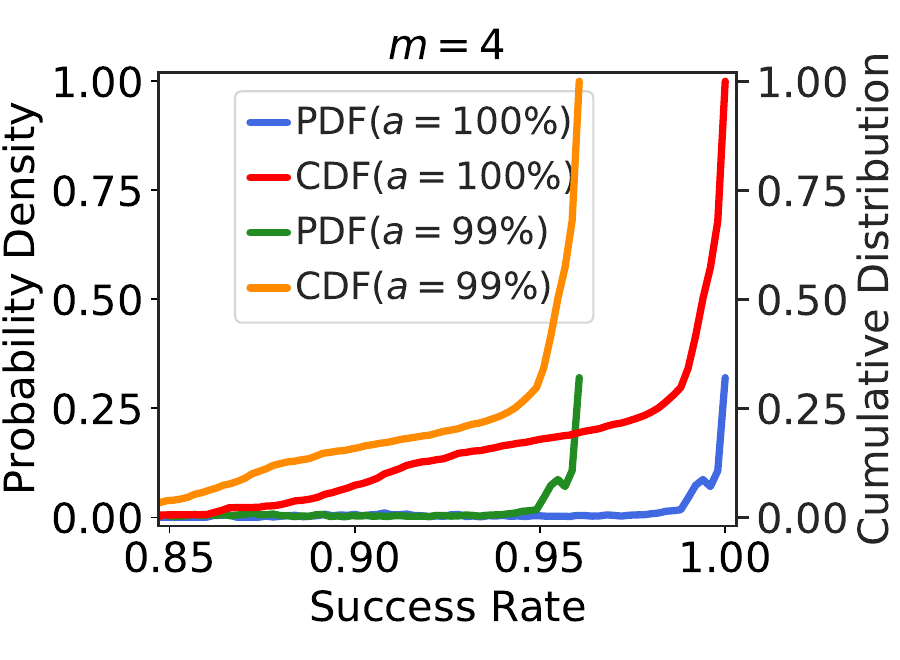}
\vspace{-3mm}
\caption{Distribution of the success rate $P$ (m = 4) in Bitcoin testnet}
\label{fig::F(P)_m=3_BTCtest}  
\end{minipage}
\hspace{0.05cm}
\begin{minipage}[t]{0.48\columnwidth}
\centering
\includegraphics[height=0.64\textwidth]{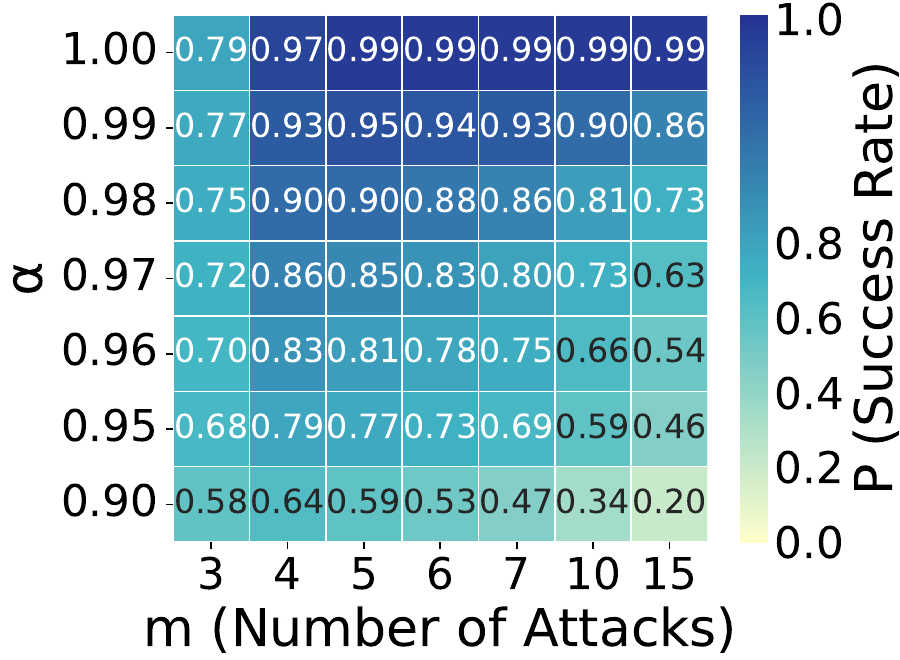}
\vspace{-3mm}
\caption{The success rate with varying $\alpha$ and $m$ in Bitcoin testnet}
\label{fig::F(P)_m=4_BTCtest}  
\end{minipage}
\end{figure}

% SOL mainnet

\begin{figure}[h!]
\begin{minipage}[t]{0.48\columnwidth}
\centering
\includegraphics[height=0.64\textwidth]{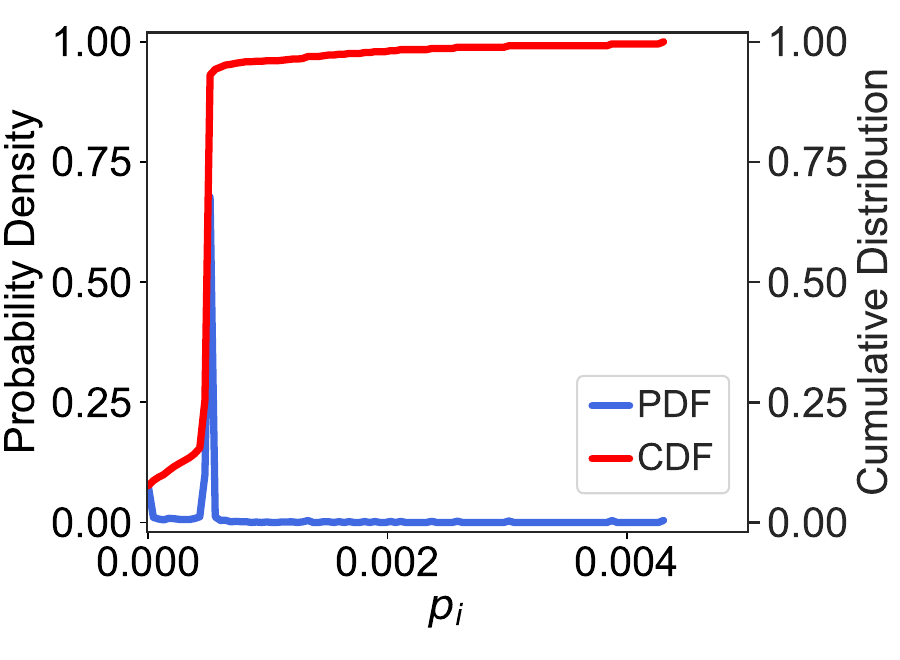}
\vspace{-5mm}
\caption{Distribution of $p_i$ in Solana mainnet}
\label{fig::distributionOfb_SOLmain} 
\end{minipage}
\hspace{0.05cm}
% \hspace{0.1cm}
\begin{minipage}[t]{0.48\columnwidth}
\centering
\includegraphics[height=0.64\textwidth]{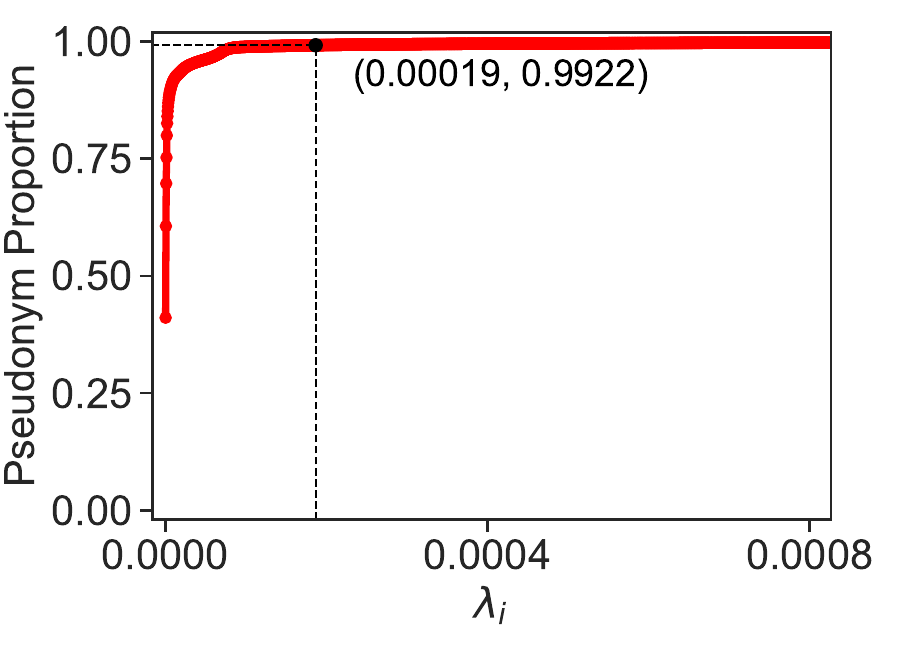}
\vspace{-5mm}
\caption{The proportion of users within each transacting rate $\lambda_i$ in Solana mainnet}
\label{fig::F(m,f)_SOLmain} 
\end{minipage}
\end{figure}

\begin{figure}[h!]
\begin{minipage}[t]{0.48\columnwidth}
\centering
\includegraphics[height=0.64\textwidth]{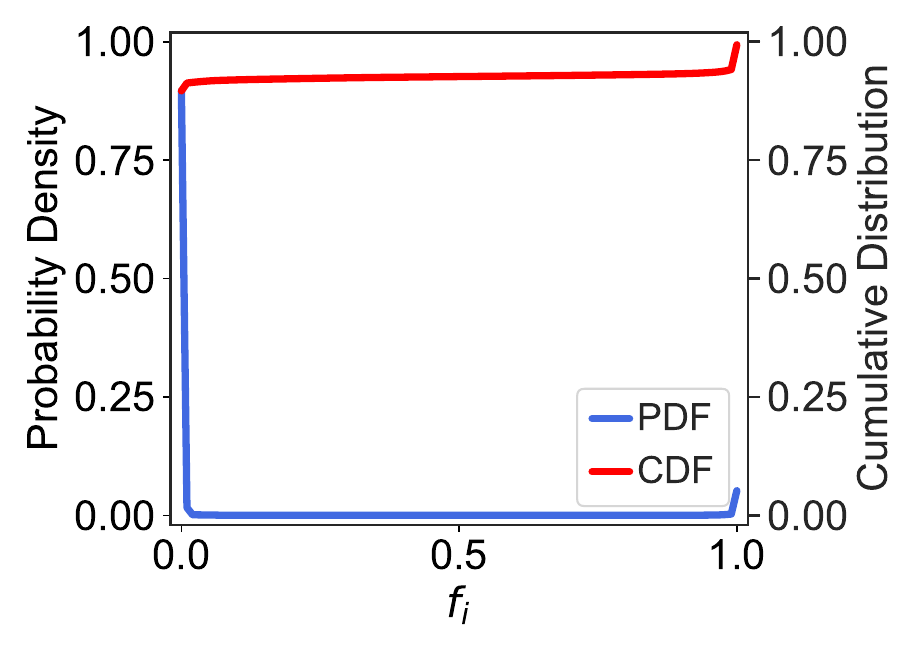}
\vspace{-5mm}
\caption{Distribution of $f_i$ without adopting optimization (m = 4) in Solana mainnet}
\label{fig::addressTxNum_SOLmain}  
\end{minipage}
\hspace{0.05cm}
\begin{minipage}[t]{0.48\columnwidth}
\centering
\includegraphics[height=0.64\textwidth]{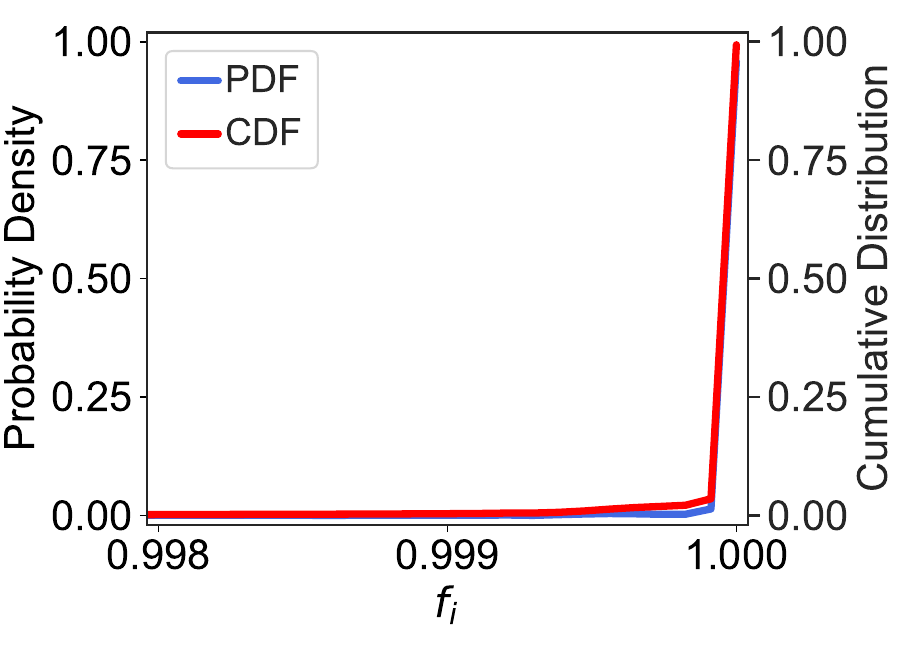}
\vspace{-5mm}
\caption{Distribution of $f_i$ in the optimized identification method (m = 4) in Solana mainnet}
\label{fig::F(m,f)filter_b_SOLmain}  
\end{minipage}
\end{figure}

\begin{figure}[h!]
\begin{minipage}[t]{0.48\columnwidth}
\centering
\includegraphics[height=0.64\textwidth]{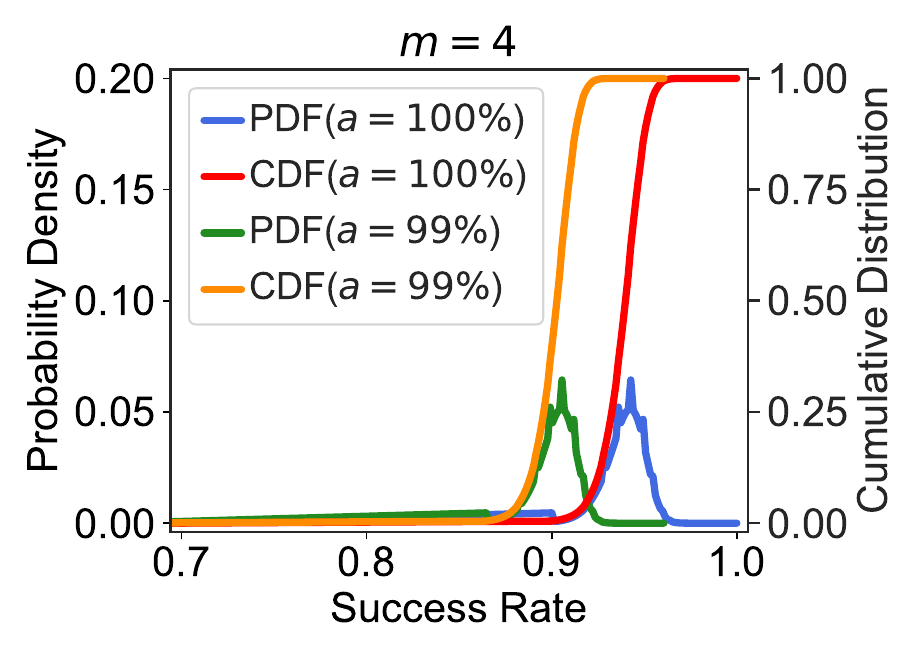}
\vspace{-3mm}
\caption{Distribution of the success rate $P$ (m = 4) in Solana mainnet}
\label{fig::F(P)_m=3_b_SOLmain}  
\end{minipage}
\hspace{0.05cm}
\begin{minipage}[t]{0.48\columnwidth}
\centering
\includegraphics[height=0.64\textwidth]{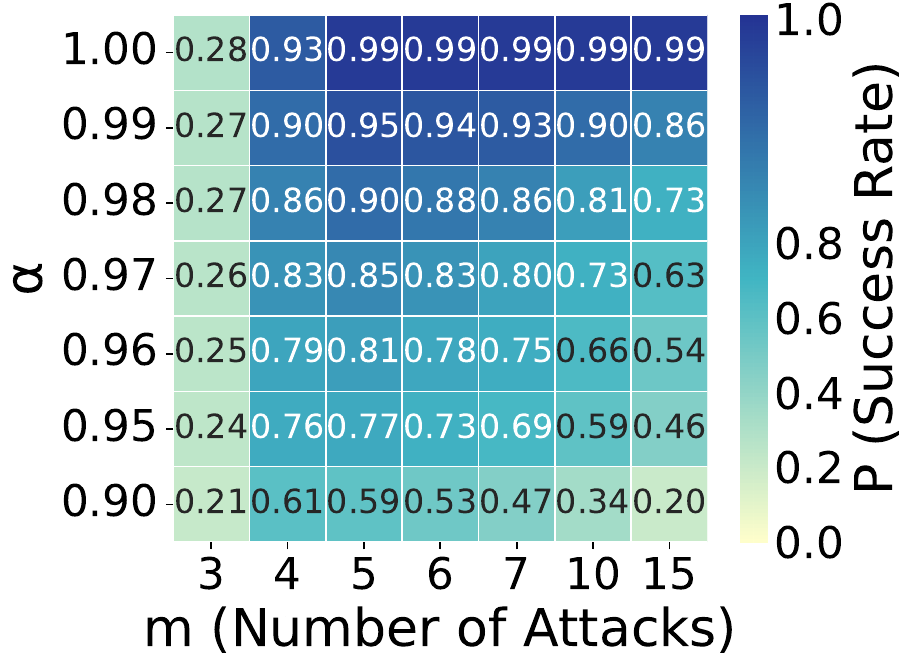}
\vspace{-3mm}
\caption{The success rate with varying $\alpha$ and $m$ in Solana mainnet}
\label{fig::F(P)_m=4_b_SOLmain}  
\end{minipage}
\end{figure}

\vspace{-2mm}

% SOL testnet

\begin{figure}[h!]
\begin{minipage}[t]{0.48\columnwidth}
\centering
\includegraphics[height=0.64\textwidth]{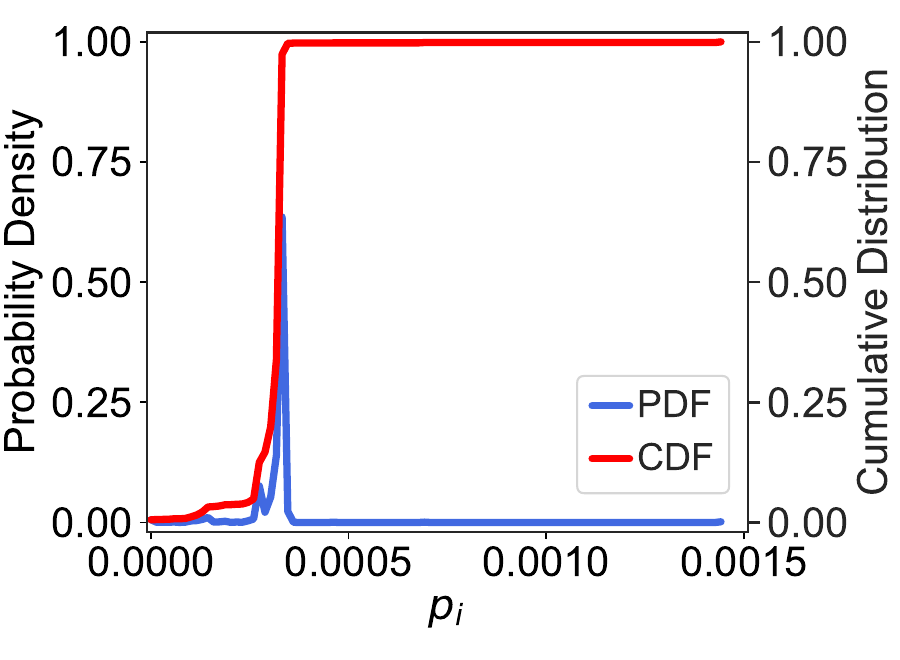}
\vspace{-5mm}
\caption{Distribution of $p_i$ in Solana testnet}
\label{fig::distributionOfb_SOLtest} 
\end{minipage}
\hspace{0.05cm}
% \hspace{0.1cm}
\begin{minipage}[t]{0.48\columnwidth}
\centering
\includegraphics[height=0.64\textwidth]{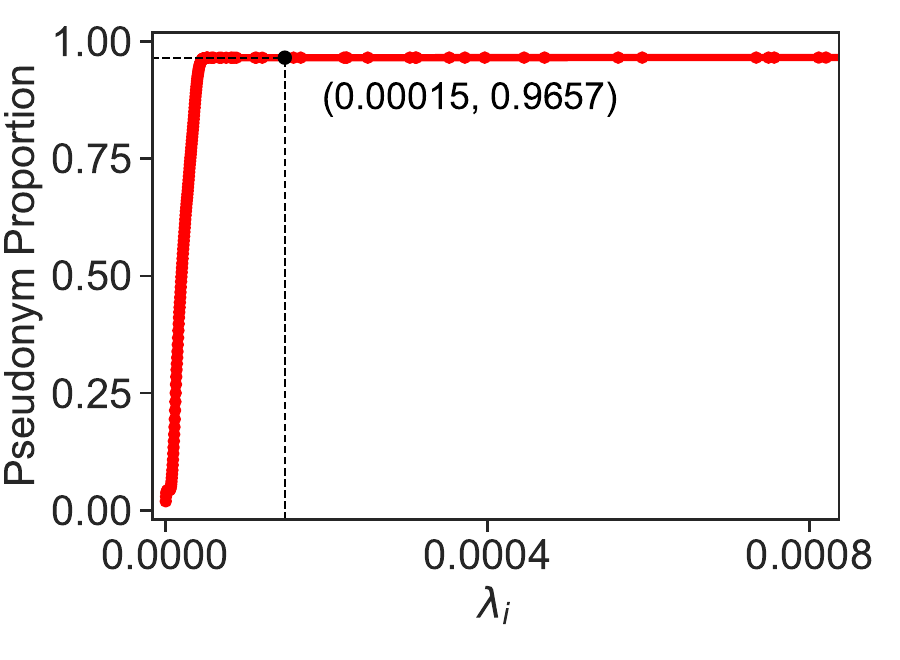}
\vspace{-5mm}
\caption{The proportion of users within each transacting rate $\lambda_i$ in Solana testnet}
\label{fig::F(m,f)_SOLtest} 
\end{minipage}
\end{figure}

\vspace{-2mm}

\begin{figure}[h!]
\begin{minipage}[t]{0.48\columnwidth}
\centering
\includegraphics[height=0.64\textwidth]{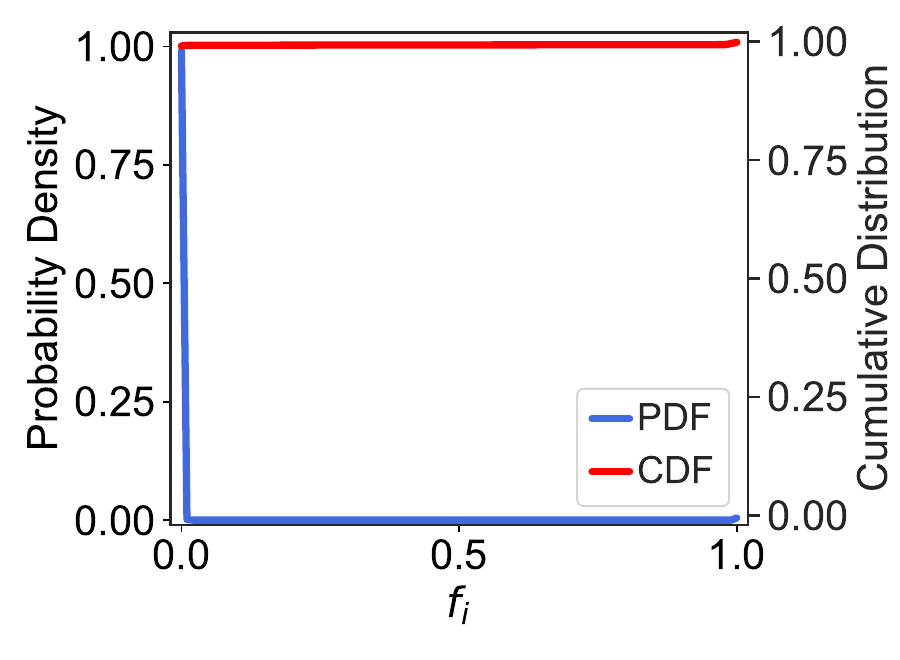}
\vspace{-5mm}
\caption{Distribution of $f_i$ without adopting optimization (m = 4) in Solana testnet}
\label{fig::addressTxNum_SOLtest}  
\end{minipage}
\hspace{0.05cm}
\begin{minipage}[t]{0.48\columnwidth}
\centering
\includegraphics[height=0.64\textwidth]{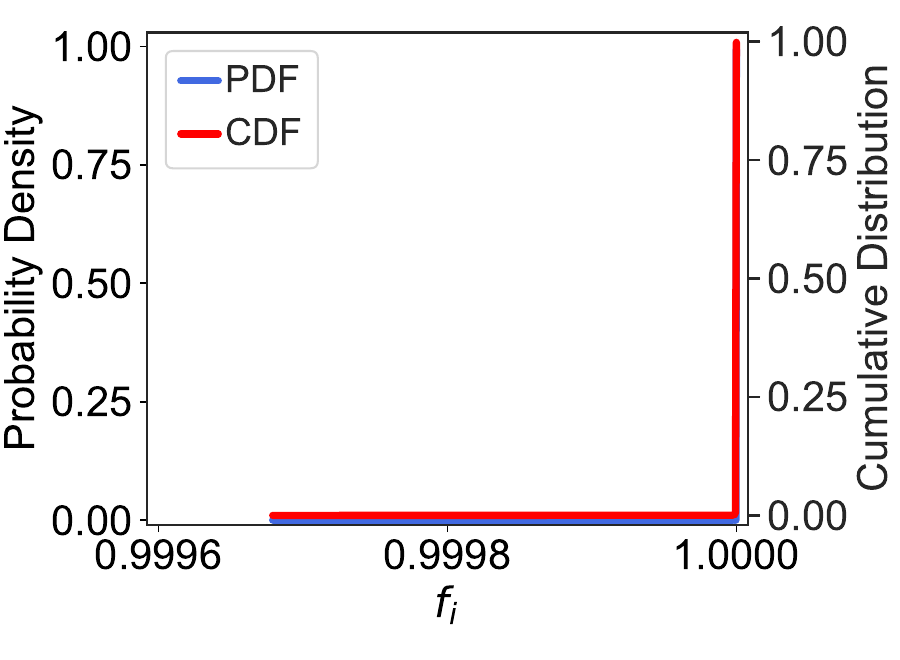}
\vspace{-5mm}
\caption{Distribution of $f_i$ in the optimized identification method (m = 4) in Solana testnet}
\label{fig::F(m,f)filter_b_SOLtest}  
\end{minipage}
\end{figure}

\vspace{-2mm}

\begin{figure}[h!]
\begin{minipage}[t]{0.48\columnwidth}
\centering
\includegraphics[height=0.64\textwidth]{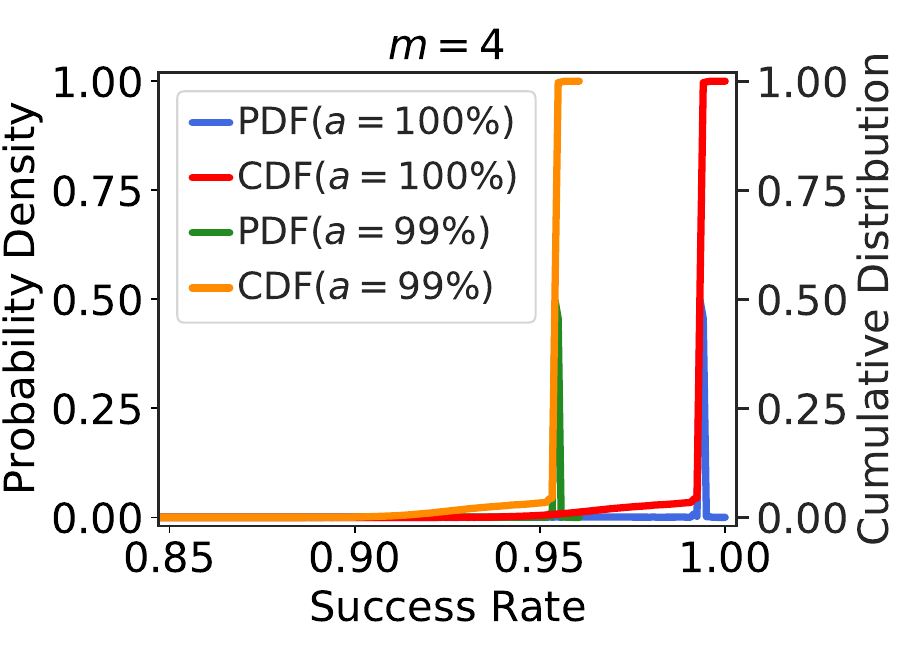}
\vspace{-3mm}
\caption{Distribution of the success rate $P$ (m = 4) in Solana testnet}
\label{fig::F(P)m=3_SOLtest}  
\end{minipage}
\hspace{0.05cm}
\begin{minipage}[t]{0.48\columnwidth}
\centering
\includegraphics[height=0.64\textwidth]{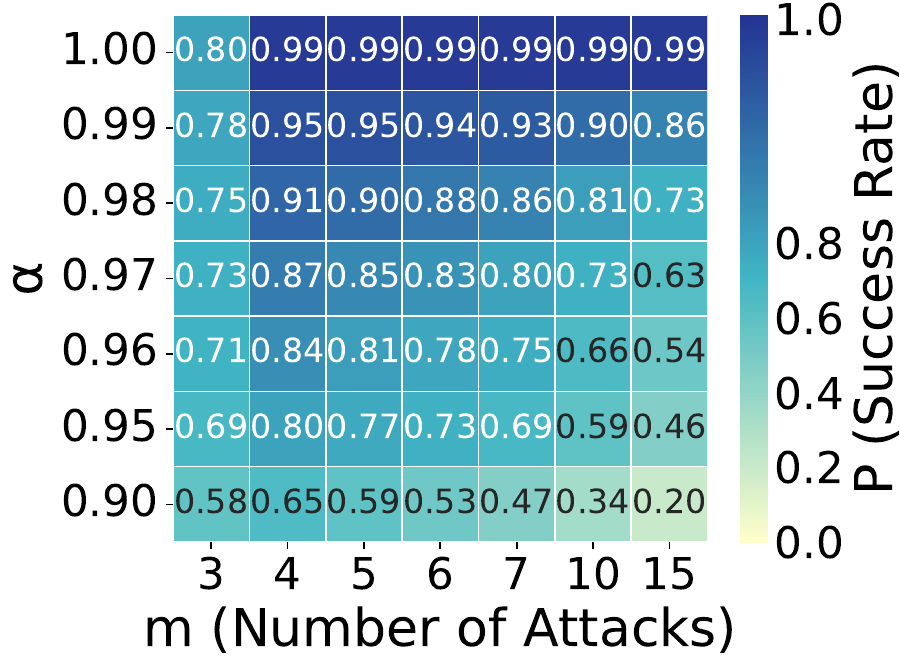}
\vspace{-3mm}
\caption{The success rate with varying $\alpha$ and $m$ in Solana testnet}
\label{fig::F(P)m=4_SOLtest}  
\end{minipage}
\end{figure}

\vspace{-2mm}

% Attack results

\begin{figure}[h!]
\begin{minipage}[t]{0.48\columnwidth}
\centering
\includegraphics[height=0.63\textwidth]{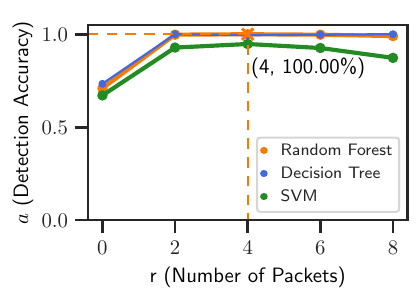}
\vspace{-5mm}
\caption{Detection accuracy of transaction query timestamps in Bitcoin testnet}
\label{fig::mlBTC} 
\end{minipage}
\hspace{0.05cm}
% \hspace{0.1cm}
\begin{minipage}[t]{0.48\columnwidth}
\centering
\includegraphics[height=0.63\textwidth]{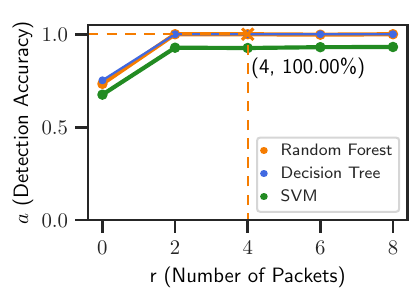}
\vspace{-5mm}
\caption{Detection accuracy of transaction query timestamps in Solana testnet}
\label{fig::mlSOL}  
\end{minipage}
\end{figure}

\begin{figure}[h!]
\begin{minipage}[t]{0.48\columnwidth}
\centering
\includegraphics[height=0.64\textwidth]{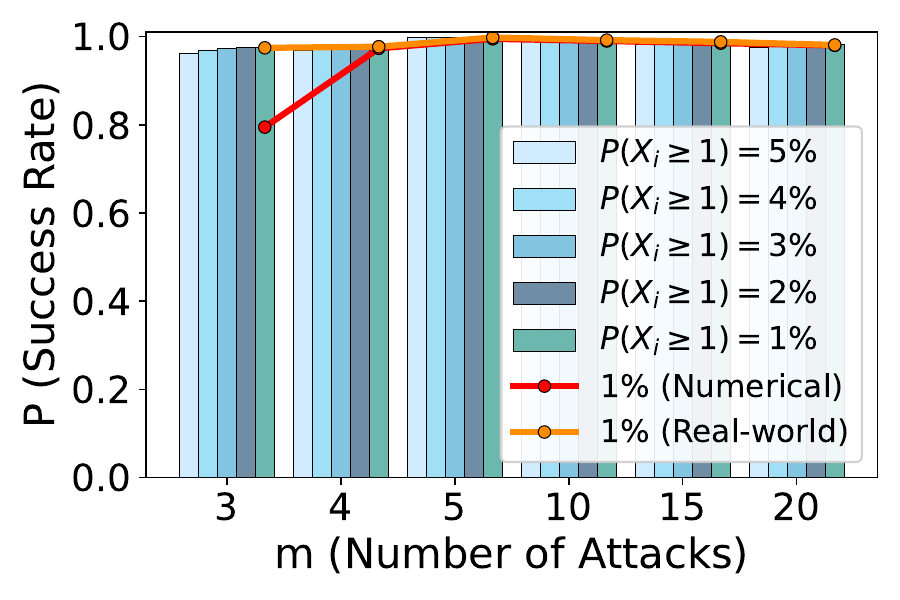}
\vspace{-3mm}
\caption{Success rate of real-world attacks in BTC testnet}
\label{fig::accuracyBTCtest}  
\end{minipage}
\hspace{0.05cm}
\begin{minipage}[t]{0.48\columnwidth}
\centering
\includegraphics[height=0.64\textwidth]{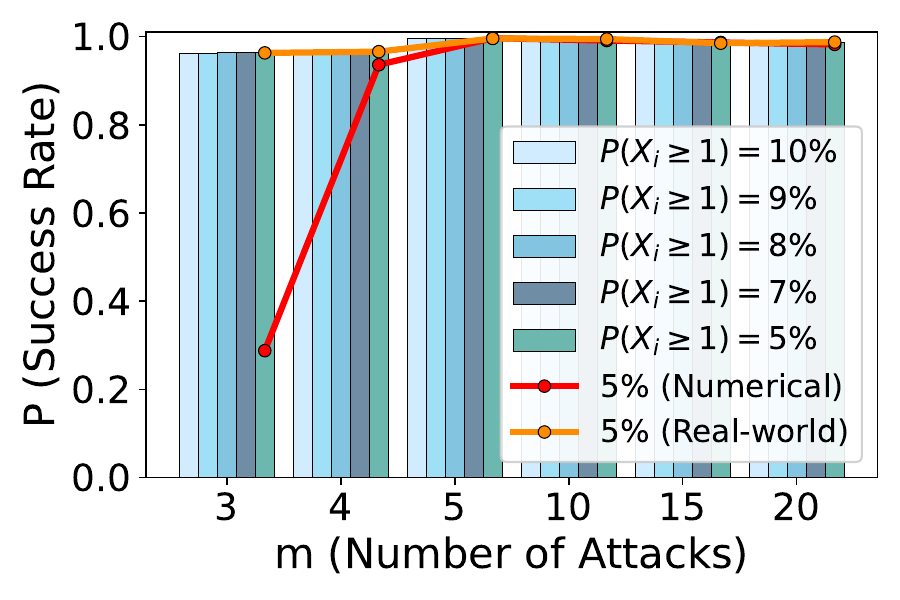}
\vspace{-3mm}
\caption{Success rate of real-world attacks in SOL testnet}
\label{fig::accuracySOLtest}  
\end{minipage}
\vspace{-2mm}
\end{figure}

\vspace{7mm}
\section{Numerical Results based on Empirical Measurements across Blockchains}
\label{sec::measureAppendix}
% \vspace{-1mm}

We provide dataset details and figures demonstrating the main measurements and numerical attack results in different blockchain networks. 
All the datasets in Ethereum testnet, Bitcoin testnet and mainnet are obtained by collecting the corresponding ledger data from August 1st to August 31st 2024. In Solana mainnet and testnet, we use $P(X_i \geq 1) = 5\%$ to derive the threshold $\theta_\lambda$, as their users transact more frequently than other blockchains.
\looseness=-1

\textbf{Ethereum Testnet}. 
% We collect the ledger data from August 1st to August 31st 2024 in Ethereum testnet named {\em Sepolia}.
We study on the Ethereum testnet named {\em Sepolia}.
The dataset contains 196,294 blocks and 17,164,163 transactions. These transactions are initiated by 933,896 user pseudonyms. The volume of the dataset is about $14.88$ GB. 
% The measured $\mu_1 = 5.31$, $\sigma_1=0.20 $, $\mu_2= 5.54$, and $\sigma_2=0.22$. 
The key measurements and numerical attack results are shown in Figures \ref{fig::distributionOfbETHtest}$\sim$\ref{fig::P_m=3_ETHtest}.\looseness=-1

\textbf{Bitcoin Mainnet}.
% We collect the ledger data from August 1st to August 31st 2024 in Bitcoin mainnet. 
The dataset contains 4,430 blocks and 18,543,679 transactions. These transactions are initiated by 4,504,223 users. The volume of the dataset is about $20.35$ GB. 
% The measured $\mu_1 = 7.91$, $\sigma_1= 0.55$, $\mu_2=8.95$, and $\sigma_2=0.41$. 
The key measurements and numerical attack results are shown in Figures \ref{fig::distributionOfbBTCmain}$\sim$\ref{fig::F(P)_m=4_BTCmain}.

\textbf{Bitcoin Testnet}
% We collect the ledger data from August 1st to August 31st 2024 in Bitcoin testnet. 
The dataset contains 30,717 blocks and 16,048,827 transactions. These transactions are initiated by 858,731 users. The volume of the dataset is about $22.89$ GB. 
% The measured $\mu_1 = 4.42$, $\sigma_1= 2.50$, $\mu_2=4.19$, and $\sigma_2=2.20$.
The key measurements and numerical attack results are shown in Figures \ref{fig::distributionOfb_BTCtest}$\sim$\ref{fig::F(P)_m=4_BTCtest}.

\textbf{Solana Mainnet}
We collect the data from August 1st to August 3rd, 2024 in Solana mainnet.
The dataset contains 584,459 blocks and 807,824,762 transactions, which are initiated by 2,389,299 users. The volume of the dataset is about $1.29$ TB.
% The measured $\mu_1 = 8.51$, $\sigma_1=0.13$, $\mu_2=11.44$, and $\sigma_2=0.07$. The probability distribution of key variables 
The key measurements and numerical attack results are shown in Figures \ref{fig::distributionOfb_SOLmain}$\sim$\ref{fig::F(P)_m=4_b_SOLmain}.\looseness=-1

\textbf{Solana Testnet}
We collect the data from August 1st to August 3rd 2024 in Solana testnet. The dataset contains 424,476 blocks and 1,120,429,260 transactions. These transactions are initiated by 97,600 users. The volume of the dataset is about $1.72$ TB.
% The measured $\mu_1 = 8.04$, $\sigma_1=  0.23$, $\mu_2=11.78$, and $\sigma_2=0.16$. The probability distribution of key variables are shown 
The key measurements and numerical attack results are shown in Figures \ref{fig::distributionOfb_SOLtest}$\sim$\ref{fig::F(P)m=4_SOLtest}.\looseness=-1

% \vspace{-1mm}
\section{Real-World Attack Results in BTC and SOL}

We first provide the evaluation of the timestamp detection accuracy of ML models in Bitcoin and Solana in \autoref{fig::mlBTC} and \autoref{fig::mlSOL}. It can be observed that, when we select the random forest algorithm and set $r=4$, the accuracy can reach up to $100\%$ in both networks. 
Next, we present the results of our real-world deanonymization attacks in Bitcoin testnet and Solana testnet in \autoref{fig::accuracyBTCtest} and \autoref{fig::accuracySOLtest}. 
The attack success rate is high. We can derive similar conclusions as those in Ethereum networks.\looseness=-1

% \vspace{-1mm}
% \subsection{Discussion on Robustness to Network Packet Delays and Drops}

% Network delays can introduce variations in packet intervals across different attack routers along the routing path. However, daily network delays are minimal (within 400ms \cite{RTT}) in modern Internet and are negligible compared to the block time.
% As such, they will not impact the selection of $k$, as analyzed in \S\ref{subsec::estimationK}.
% Our ML models are trained based on realistic network packets, including daily delays, and primarily rely on features in packet sizes and sequences. Therefore, they are robust to the minor variations in packet intervals.
% Results in \autoref{table::accuracyLocation} further validate these claims.
% In the rare cases where packets experience significant delays or drops, the packet features will substantially alter. The ML model will classify the affected packet sequence as a negative instance, only resulting in missing one observation opportunity.\looseness=-1

% \vspace{-2mm}
\section{Responses from Affected Vendors}
\label{appendix:vendorResp}

We have reported the vulnerability and attack proposed in this paper to the affected vendors, including 3 blockchains and 9 wallets.
% \autoref{table::repDisclosure} summarizes the responses from vendors to date.
Notably, the Bitcoin RPC protocol designer {\em Electrum} (also a wallet vendor) assigned CVE-2025-43968 to this issue, and awarded us a bug bounty of \$5,000.
They have implemented the suggested countermeasures and released new versions of {\em both the wallet software and RPC server}.
Moreover, they are driving changes at other relevant vendors, including the RPC servers \textbf{{\em Fulcrum}} and \textbf{{\em Electrs}}, to address the issue more broadly.
Sparrow confirmed our attack and treated the vulnerability seriously. 
They are actively working with us on implementing countermeasures continually. 
To date, the related patches can be found in the Sparrow V2.2.0 release note "{\em Optimize and reduce Electrum server RPC calls}" and its V2.2.3 release note of "{\em Improve Electrum server script hash unsubscribe support}". 
Additionally, they have written a mitigation proposal with us together for motivating broad changes at the Electrum-based RPC server side and other similar wallets.
Moreover, they have promised to issue a CVE once the mitigation is fully in place in Sparrow and others.
The Ethereum Foundation awarded us an academic grant for the research on the attack and protocol-level countermeasures (Grant No. FY25-2131; Total Funding: \$47,099). 
The wallet Taho confirmed our attack and requested mitigation strategies from us to fix the issue.
% Moreover, 3 vendors are collaborating with us on implementing and testing the countermeasures, and/or 
% driving changes at other relevant vendors to address the issues more broadly. 

% \vspace{-2mm}
\section{Rule-based Packet Identification Method}
\label{sec::rule_comparison}

To compare the effectiveness of the rule-based method with that of the ML model, we manually define identification rules based on the analysis of the packet sizes and sequences in \S4.2 and \S4.3.
Considering the asynchronous and interleaved nature of periodic and other RPC API calls, we adopt the following heuristic for identification: 
a packet sequence is identified as a match for the target pattern if the sizes corresponding to the target API call and its surrounding API calls appear in the correct {\em relative} order, regardless of the number or type of other intervening API calls.
Specifically, let $\mathbb{T}_0$ and $\mathbb{T}_1$ denote the size range of the request and response packets generated by the target API call, respectively. Similarly, let $\mathbb{P}_0$ and $\mathbb{P}_1$ denote the size range of those packets of the preceding API call, and $\mathbb{F}_0$ and $\mathbb{F}_1$ denote those of the following API call. Let $l_i$ denote the size of the $i^{th}$ packet in the sequence.
Given a sequence of $2r + 2$ packets, if the following rules are satisfied, the $(r+2)^{th}$ packet is identified as the response of the transaction status query.

\begin{itemize}
    \item Rule 1: there exist indices $a$, $b$, $c$ and $d$ such that $a<b<(r+1)<(r+2)<c<d$.
 
    \item Rule 2: $l_{r+1} \in \mathbb{T}_0$ and $l_{r+2} \in \mathbb{T}_1$. 

    \item Rule 3: $l_a \in \mathbb{P}_0$ and $l_b \in \mathbb{P}_1$. 
    
    \item Rule 4: $l_c \in \mathbb{F}_0$ and $l_d \in \mathbb{F}_1$.
    
\end{itemize}
Note that: when $r=0$, the packet sequence contains only two packets. In this case, only {\em Rule 2} is applied, as there are no surrounding packets to evaluate {\em Rule 1}, {\em Rule 3}, or {\em Rule 4}. As shown in \autoref{fig::mlETH}, the ML model outperforms the rule-based method, as it generally provides better adaptability and can assign appropriate weights to different features and values.

% This advantage primarily stems from the fact that, the theoretical packet size ranges are relatively coarse and broad, while in practice, the packet sizes corresponding to an RPC API call tend to concentrate within a narrow range.
% Rule-based approaches are inherently inflexible and depend on manually defined patterns.
% In contrast, the ML model offers better adaptability ability and could assign different weights to different packet size values, leading to improved performance.

\end{sloppypar}
\end{document}